\documentclass[12pt]{article}
\pdfoutput=1

\usepackage{amsmath}
\usepackage{graphicx}
\usepackage[usenames, dvipsnames]{color}
\usepackage{array}
\usepackage{tikz}
\usepackage{color, colortbl}
\definecolor{Gray}{gray}{0.95}
\usepackage[normalem]{ulem}

\usepackage{amssymb}
\usepackage{epsfig}
\usepackage{epstopdf}
\usepackage{latexsym}
\usepackage{color}
\numberwithin{equation}{section}
\usepackage[
      colorlinks=true,
      linkcolor=blue,  
      urlcolor=blue,    
      filecolor=blue,     
      citecolor=red,
linktocpage=true,
      pdfstartview=FitV,
      bookmarksopen=true    
      ]{hyperref}

\begin{document}

\begin{titlepage}

\begin{center}

{\LARGE \textbf{S-folds and $AdS_{3}$ flows from the D3-brane}}

\vspace{60pt}

{\large \bf Adolfo Guarino}$^{\,a,b}$ \,  \large{and} \, {\large \bf Minwoo Suh}$^{\,c}$
		
\vspace{25pt}

{\normalsize  
$^{a}$ Departamento de F\'isica, Universidad de Oviedo,\\
Avda. Federico Garc\'ia Lorca 18, 33007 Oviedo, Spain.}
\\[7mm]

{\normalsize  
$^{b}$ Instituto Universitario de Ciencias y Tecnolog\'ias Espaciales de Asturias (ICTEA) \\
Calle de la Independencia 13, 33004 Oviedo, Spain.}
\\[7mm]

{\normalsize  
$^{c}$ Department of Physics, Kyung Hee University, Seoul 02447, Korea.}
\\[10mm]

\texttt{adolfo.guarino@uniovi.es}  \,\, , \,\, \texttt{minwoosuh1@gmail.com}

\vspace{50pt}

\begin{abstract}

We investigate supersymmetric flows in type IIB supergravity that preserve an $SO(2,2)$ space-time symmetry and asymptote to $AdS_{5} \times S^5$ at both endpoints. The flows are constructed as Janus-type $\mathbb{R} \times  AdS_{3}$ BPS domain-walls in the effective four-dimensional $[SO(1,1)\times{S}O(6)]\ltimes\mathbb{R}^{12}$ gauged maximal supergravity describing the massless sector of type IIB supergravity compactified on $S^1 \times S^{5}$. The compactification includes an S-duality hyperbolic twist along the $S^1$ which, when combined with an appropriate choice of boundary conditions for the running scalars, generates special flows that develop an S-fold regime at their core, thus enhancing the space-time symmetry there to $SO(2,3)$. Via the AdS/CFT correspondence, the flows constructed here are conjectured to describe conformal interfaces in a circle compactification of $\mathcal{N}=4$ $\textrm{SYM}_{4}$.

\end{abstract}

\end{center}

\end{titlepage}

\tableofcontents

\section{Introduction}

Via the AdS/CFT correspondence \cite{Maldacena:1997re}, Janus solutions of supergravity theories \cite{Bak:2003jk} provide a simple and explicit realisation of interfaces in dual field theories \cite{Clark:2004sb, DHoker:2006qeo, Gaiotto:2008sd}. The original type~IIB Janus solutions are characterised by two properties: $i)$ they feature an $AdS_{d-1}$ space-time geometry that asymptotes to an $AdS_{d}$ vacuum on each side of the Janus. $ii)$ the type IIB dilaton field jumps across the interface living at the boundary of space-time and takes different values $\pm \Phi_{0}$ on each side. However, Janus solutions have been generalised to theories without a dilaton field so that they are solely characterised by the first property. Examples are the Janus solutions of \cite{Bobev:2013yra,Anabalon:2022fti} constructed in the four-dimensional $SO(8)$ maximal supergravity that arises upon consistent truncation of eleven-dimensional supergravity on a seven-sphere \cite{deWit:1982bul,deWit:1986oxb}. In this work we will relax the first property and investigate regular Janus-type solutions featuring an $AdS_{d-2}$ space-time geometry that asymptotes to an $AdS_{d}$ geometry on each side of the solution. For the type~IIB scenario investigated in this work one has $\,d=5\,$ whereas $\,d=4\,$ is the relevant value for the M-theory scenario.

The original type IIB Janus solutions are dual to three-dimensional interfaces in $\mathcal{N}=4$ super Yang--Mills (SYM$_{4}$). Remarkably, all possible such interfaces preserving supersymmetry have been classified \cite{DHoker:2006qeo}: when restricted to the largest symmetric cases, these preserve ${\mathcal{N}=1 \,\& \,SU(3)}$, $\mathcal{N}=2 \,\& \, U(2)$ and ${\mathcal{N}=4 \,\& \, SO(4)}$ symmetry\footnote{Our notation $\,\mathcal{N} \,\& \, G_{0}\,$ is adapted to the gravity side of the AdS/CFT correspondence. Namely, we have $\,{G_{0}=SO(\mathcal{N}) \times G_{F}}\,$ where $\,SO(\mathcal{N})\,$ is the $R$-symmetry group of the three-dimensional interface preserving $\mathcal{N}$ supersymmetries and $\,G_{F}\,$ is its flavour symmetry group.}. Following this classification, the solutions with $\mathcal{N}=1 \,\& \,  SU(3)$ and  ${\mathcal{N}=4 \,\& \,  SO(4)}$ symmetry were explicitly constructed in type~IIB supergravity in \cite{DHoker:2006vfr} and \cite{DHoker:2007zhm,DHoker:2007hhe}, respectively. The ${\mathcal{N}=1 \,\& \, SU(3)}$ Janus was alternatively found in five-dimensional gauged supergravity \cite{Clark:2005te, Suh:2011xc} and uplifted to type IIB supergravity \cite{Suh:2011xc}. It is only recently that all supersymmetric Janus solutions have been constructed in the $SO(6)$-gauged $\mathcal{N}\,=\,8$ supergravity in five dimensions, \cite{Gunaydin:1984qu, Gunaydin:1985cu,Pernici:1985ju}, and uplifted to type IIB supergravity in a systematic manner \cite{Bobev:2019jbi, Bobev:2020fon}.

The field theory living on the interface with $\mathcal{N}=4 \,\& \, SO(4)$ symmetry led to the discovery of new 3d $\mathcal{N}\,=\,4$ SCFTs dubbed $T[SU(N)]$ theories in \cite{Gaiotto:2008sd}. They provide the building blocks for 3d $\mathcal{N}=4$ SCFTs. The gravity dual was found to be closely related to the ${\mathcal{N}=4 \,\& \,  SO(4)}$ Janus, but to be a distinct class of solutions \cite{Assel:2011xz, Assel:2012cp}. Furthermore, by gauging the $\,U(N)\times U(N)\,$ global symmetry of 3d $T[SU(N)]$ theories with vector multiplets, yet another new class of strongly coupled 3d SCFTs, named S-folds, was put forward in \cite{Assel:2018vtq} (see also \cite{Terashima:2011qi,Gang:2015wya}). The gravity duals of the S-fold theories were realised by first compactifying one spatial direction -- we denote it $\,\eta\,$ -- into a circle $\,S^1\,$ in a putative $\,AdS_{5}\,$ space-time, and then imposing a non-trivial $SL(2,\mathbb{Z})$ monodromy of hyperbolic type along the circle \cite{Assel:2018vtq, Inverso:2016eet}. The corresponding type IIB backgrounds are of the form $\,{AdS_4 \times S^{1} \times S^{5}}\,$ with a characteristic dilaton profile that is linear on the coordinate $\,\eta\,$ along the $S^1$. As they are the gravity duals of 3d S-fold theories closely related to Janus solutions, they were named J-fold solutions in \cite{Assel:2018vtq}. In \cite{Bobev:2019jbi, Bobev:2020fon}, a systematic method was implemented to obtain J-fold solutions from all the families of Janus solutions in the $\mathcal{N}\,=\,8$ and $SO(6)$-gauged supergravity in five dimensions. These solutions were also uplifted to type IIB supergravity in \cite{Bobev:2019jbi, Bobev:2020fon}. Using the effective five-dimensional approach, new S-fold solutions as well as RG-flows between S-fold CFT's have been investigated in \cite{Arav:2020obl, Arav:2021tpk, Arav:2021gra}.

Interestingly, there is an alternative four-dimensional approach to construct S-fold solutions preceding the five-dimensional method of \cite{Bobev:2019jbi, Bobev:2020fon}. S-folds can be seen as simple $AdS_4$ vacua of the maximal $[SO(1,1)\times{S}O(6)]\ltimes\mathbb{R}^{12}$ dyonically-gauged supergravity in four dimensions that arises from the consistent truncation of type~IIB supergravity on $\,S^{1} \times S^{5}\,$ including an S-duality twist \cite{Inverso:2016eet}. Indeed, the S-fold solutions with ${\mathcal{N}=1 \,\&\, SU(3)}$, \cite{Guarino:2019oct}, $\mathcal{N}=2 \,\&\, U(2)$, \cite{Guarino:2020gfe}, and $\mathcal{N}=4 \,\&\, SO(4)$, \cite{Gallerati:2014xra, Inverso:2016eet}, were originally found in the four-dimensional context and subsequently uplifted to type~IIB supergravity using Exceptional Field Theory techniques \cite{Hohm:2013pua,Hohm:2013uia}. Recently, holographic (flat-sliced) RG-flows between S-fold CFT's have been investigated using the 4D approach \cite{Guarino:2021kyp}. Also the existence of a (non-)supersymmetric conformal manifold of S-fold CFT's has been investigated using this approach \cite{Giambrone:2021zvp,Guarino:2021kyp,Bobev:2021yya,Guarino:2021hrc,Bobev:2021rtg,Giambrone:2021wsm,Guarino:2022tlw}.

In this work we continue the above four-dimensional program and construct supersymmetric domain-wall solutions of the form $\mathbb{R} \times AdS_3$ in the $[SO(1,1)\times{S}O(6)]\ltimes\mathbb{R}^{12}$ dyonically-gauged maximal supergravity in four dimensions. When uplifted to type IIB supergravity, they describe Janus-type solutions of the form
\begin{equation}
\label{AdS3_geometry}
\mathbb{R} \times AdS_3 \times S^{1} \times S^5\ ,
\end{equation}
involving a non-trivial $SL(2,\mathbb{Z})$ monodromy along the $\,S^1\,$. The S-duality twist inducing the monodromy is codified in a parameter $\,c\,$ that is also responsible for the dyonic nature of the gauging in four dimensions \cite{Dall'Agata:2014ita,Inverso:2016eet}.\footnote{As argued in \cite{Dall'Agata:2014ita}, the parameter $\,c\,$ is a discrete (on/off) parameter that can be set to either $1$ or $0$ without loss of generality.} However, and unlike for standard Janus solutions in type IIB \cite{Bobev:2020fon} and M-theory \cite{Bobev:2013yra,Anabalon:2022fti}, this time there is no maximally supersymmetric \textit{round vacuum} of the form $\,AdS_{4} \times S^1 \times S^5\,$ that can serve as an endpoint on each side of the Janus solution.\footnote{There is an $\,AdS_{4} \times S^1 \times S^5\,$ S-fold with $SO(6)$ symmetry featuring a round $\,S^5\,$ internal geometry but it is non-supersymmetric and perturbatively unstable \cite{Guarino:2019oct}.} This implies that there is no maximally supersymemtric $AdS_{4}$ vacuum at the origin of the scalar potential of the effective $[SO(1,1)\times{S}O(6)]\ltimes\mathbb{R}^{12}$ gauged supergravity. As a consequence, the flows we construct in this work are attracted towards a particular non-$AdS_{4}$ background at both endpoints of the domain-walls. Such a non-$AdS_{4}$ behaviour was recognised as the four-dimensional incarnation of \textit{a deformation of} the $\,AdS_5\times{S}^5\,$ background of type IIB supergravity. And the deformation parameter $c$ (the same one codifying the S-duality twist) was interpreted as a source of anisotropy along the compactified $\,S^1\,$ direction in the dual $\,\textrm{SYM}_4\,$ theory \cite{Guarino:2021kyp}. Thus, the seemingly non-conformal asymptotics from a four-dimensional perspective actually correspond to a \textit{deformed} D3-brane with a five-dimensional $\,\mathbb{R} \times AdS_3 \times S^{1}\,$ space-time metric in (\ref{AdS3_geometry}) given by
\begin{equation}
\label{deformed_AdS5}
\begin{array}{rll}
g^2 \, ds_{5}^2 &=& d\lambda^2+f_{1}(\lambda)\,ds_{AdS_3}^2 + f_{2}(\lambda)\,d\eta^2 \ , 
\end{array}
\end{equation} 
in terms of the four-dimensional gauge coupling $\,g\,$ and two functions $\,f_{1,2}(\lambda)\,$ of an slicing coordinate $\,\lambda \in [0,\infty)\,$. The standard D3-brane solution is recovered at $\,c=0\,$ upon setting $\,f_{1}(\lambda)=\cosh^2\lambda\,$ and $\,f_{2}(\lambda)=\sinh^2\lambda\,$ so that (\ref{deformed_AdS5}) reproduces an $\,AdS_{5}\,$ space-time metric with a compactified $S^{1}$ direction parameterised by $\,\eta\,$.

As already stated, the goal of this work is to construct supersymmetric $\, \mathbb{R} \times AdS_{3}\,$ domain-wall solutions in the effective ${[SO(1,1)\times{S}O(6)]\ltimes\mathbb{R}^{12}}$ gauged supergravity that arises upon compactifying type IIB supergravity on $S^{1} \times S^{5}$ including a non-trivial $SL(2,\mathbb{Z})$ monodromy along $S^{1}$. Employing numerical techniques, we find two classes of regular solutions:  
\begin{itemize}

\item The generic Janus-type solutions preserve the $\,SO(2,2)\,$ isometries of the $AdS_{3}$ factor in (\ref{deformed_AdS5}) all along the flow and are attracted towards the deformed D3-brane discussed above. From a 5D perspective, they correspond to flows in the bulk where the space-time geometry along the flows conforms to $\,AdS_{5}-AdS_{3}-AdS_{5}\,$.

\item Upon tuning of the boundary conditions, special Janus-type solutions appear which still approach the deformed D3-brane asymptotically but feature an intermediate S-fold regime. This regime is characterised by a factorised space-time in (\ref{deformed_AdS5}) of the form $AdS_{4} \times S^1$, so there is a symmetry enhancement to $\,SO(2,3)\,$ associated with the $AdS_{4}$ factor. From a 5D perspective, these are flows in the bulk with a space-time geometry that conforms to $\,AdS_{5}-AdS_{3}-AdS_{4}-AdS_{3}-AdS_{5}\,$.

\end{itemize}

\noindent In addition there are also singular solutions (akin to the flows to Hades discussed in \cite{Bobev:2013yra,Anabalon:2022fti}) which diverge at some finite value of the radial direction. All together, these three classes of solutions are the analogue of the Janus solutions in the $SO(8)$ \cite{deWit:1982bul} and $ISO(7)$ \cite{Guarino:2015jca, Guarino:2015qaa, Guarino:2015vca} maximal supergravities constructed in \cite{Bobev:2013yra,Anabalon:2022fti} and \cite{Suh:2018nmp,Karndumri:2021pva}, respectively.

The work is organised as follows. In section~$2$ we review the $\mathbb{Z}_2^{3}$-invariant sector of the dyonically-gauged $[SO(1,1)\times{S}O(6)]\ltimes\mathbb{R}^{12}$ maximal supergravity in order to set up the model and derive the BPS equations. In section~$3$ we first identify the (undeformed, $c=0$) D3-brane solution with a curved worldvolume within our four-dimensional effective model, and then semi-analytically identify the (deformed, $c \neq 0$) D3-brane behaviour that controls the asymptotics at both sides of the Janus-type solutions. Then we numerically construct and characterise such Janus-type solutions paying special attention to the choice of boundary conditions. Upon tuning of the latter, we present examples of Janus-type solutions displaying an $AdS_{4}$ intermediate S-fold regime for all the S-folds discussed in the introduction. Our conclusions and a final discussion are presented in Section~$4$. Some results regarding the type IIB uplift of the Janus-type solutions presented here are collected in the Appendix.

\section{The $\mathbb{Z}_2^3$-invariant truncation}

We consider the $\mathbb{Z}_2^3$-invariant sector of the dyonically-gauged $[SO(1,1)\times{S}O(6)]\ltimes\mathbb{R}^{12}$ maximal supergravity investigated in \cite{Guarino:2020gfe}. This sector has previously been considered in the $ISO(7)$ theory \cite{Guarino:2019snw} as well as in the $SO(8)$ theory \cite{Bobev:2019dik}. It describes a minimal $\mathcal{N}=1$ supergravity coupled to seven chiral fields with scalar components $z_i$ and $i\,=\,1\,,\ldots,7$. In a canonical $\mathcal{N}\,=\,1$ formulation, the Einstein-scalar action is given by
\begin{equation}
S\,=\,\frac{1}{16{\pi}G_4}{\int}d^4x\sqrt{-g}\left[R-K_{z_i\bar{z}_j}dz_i\wedge*d\bar{z}_j-V\,\right].
\end{equation}
The K\"ahler potential describes a $\,[SL(2)/SO(2)]^7\,$ coset space for the scalar geometry and is given by
\begin{equation}
\label{K_potential}
K\,=\,-\sum^7_{i=1}\log\left[-i\left(z_i-\bar{z}_i\right)\right]\,.
\end{equation}
The K\"ahler metric is then defined as
\begin{equation}
\label{K_metric}
K_{z_i\bar{z}_j}\,=\,\partial_{z_i}\partial_{\bar{z}_j}K\,,
\end{equation}
with $K^{z_{i}\bar{z}_{j}}$ denoting its inverse. The holomorphic superpotential is given by
\begin{equation}
\label{Superpotential}
\mathcal{V}\,=\,2g\left[z_1z_5z_6+z_2z_4z_6+z_3z_4z_5+\left(z_1z_4+z_2z_5+z_3z_6\right)z_7\right]+2gc\left(1-z_4z_5z_6z_7\right)\,,
\end{equation}
where $g$ and $c$ are the gauge coupling constant and the electromagnetic deformation parameter, respectively. The scalar potential is obtained from (\ref{K_potential})-(\ref{Superpotential}) as
\begin{equation}
V\,=\,e^K\left[K^{z_i\bar{z}_j}D_{z_i}\mathcal{V} D_{{\bar{z}}_j}\overline{\mathcal{V}}-3\mathcal{V}\overline{\mathcal{V}}\right]\,,
\end{equation}
where the K\"ahler covariant derivative is
\begin{equation}
D_{z_i}\mathcal{V}\,=\,\partial_{z_i}\mathcal{V}+(\partial_{z_i}K)\mathcal{V}\,.
\end{equation}
We also introduce the (complex) gravitino mass term
\begin{equation}
\mathcal{W}\,=\,e^{\frac{K}{2}} \, \mathcal{V}\,,
\end{equation}
which defines a real superpotential
\begin{equation}
W^2\,=\,|\mathcal{W}|^2\,,
\end{equation}
that enters the BPS flow equations (see next section).

\subsection*{Families of $AdS_{4}$ vacua}

Four multi-parametric families of supersymmetric $AdS_4$ vacua have been constructed within this sector of theory. 

\begin{itemize}

\item[I.] A two-parameter family of $\mathcal{N}\,=\,1 \, \&\, U(1)^2$ invariant vacua at
\begin{equation}
\label{solution_family_1}
z_{1}\,=\,z_{2}\,=\,z_{3}\,=\,c\left(-\chi_{1,2,3}+i\frac{\sqrt{5}}{3}\right)\quad , \quad z_4\,=\,z_5\,=\,z_6\,=\,z_7\,=\,\frac{1}{\sqrt{6}}\left(1+i\sqrt{5}\right)\,,
\end{equation}
subject to the constraint
\begin{equation}
\label{sum_chi=0}
\chi_1+\chi_2+\chi_3\,=\,0\,.
\end{equation}
The symmetry enhances to $U(2)$ when the parameters are identified pairwise and to $SU(3)$ when they all vanish \cite{Guarino:2019oct}.

\item[II.] A one-parameter family of $\mathcal{N}\,=\,2 \, \&\, U(1)^2$ invariant vacua at
\begin{equation}
\label{solution_family_2}
z_1\,=\,-\bar{z}_3\,=\,c\left(-\chi+i\frac{1}{\sqrt{2}}\right)\,, \qquad z_2\,=\,ic\,, \qquad z_4\,=\,z_6\,=\,i\,, \qquad z_5\,=\,z_7\,=\,\frac{1}{\sqrt{2}}\left(1+i\right)\,.
\end{equation}
The symmetry enhances to $U(2)$ when $\chi=0$ \cite{Guarino:2020gfe}.

\item[III.] A one-parameter family of $\mathcal{N}\,=\,2 \, \&\, U(1)^2$ invariant vacua at
\begin{equation}
\label{solution_family_3}
z_1\,=\,z_2\,=\, i \, c \frac{\sqrt{\varphi^{2}+1}}{\sqrt{2}}\,, \quad z_3\,=\,ic\,, \quad z_4\,=\,z_5\,=\,\frac{1}{\sqrt{2}}\left(1+i\right) \,, \quad z_6\,=\, - \bar{z}_7\,=\,  \frac{-\varphi+i}{\sqrt{\varphi^{2}+1}}    \,.
\end{equation}
The symmetry enhances to $U(2)$ when $\varphi=0$ \cite{Bobev:2021yya}. At the specific value $\varphi\,=\,0$, the solution (\ref{solution_family_3}) reduces (up to permutation of the complex scalars) to (\ref{solution_family_2}) with $\chi=0$.

\item[IV.]  $\mathcal{N}\,=\,4 \, \&\, SO(4)$ invariant vacuum \cite{Gallerati:2014xra} at
\begin{equation}
\label{solution_family_4}
z_1\,=\,z_2\,=\,z_3\,=\,ic\,, \qquad z_4\,=\,z_5\,=\,z_6\,=\,-\bar{z}_7\,=\,\frac{1}{\sqrt{2}}\left(1+i\right)\,.
\end{equation}
This vacuum also belongs to Family III and is obtained from (\ref{solution_family_3}) upon setting $\varphi=1$. Recently, two axion-like flat deformations breaking supersymmetry have been identified also for this vacuum \cite{Guarino:2021hrc,Guarino:2022tlw}. Higher-dimensional and holographic aspects of these axionic deformations have been further investigated in \cite{Giambrone:2021wsm}.

\end{itemize}

Within a given family, the radius of the corresponding $AdS_4$ vacua, $L$, is always the same and given by
\begin{equation} \label{rads4}
L^2\,=\,\frac{1}{W_0^2}\,=\,-\frac{3}{V_0}\,,
\end{equation}
where $W_0$ and $V_0$ denote values at the vacuum. Note that different values of the free parameters within a given family of $AdS_{4}$ vacua do not change $L$. Moreover, families II, III and IV feature the same value of $L$ and have been shown to be connected by marginal deformations in the two-dimensional conformal manifold of three-dimensional $\mathcal{N}=2$ S-fold CFT's \cite{Bobev:2021yya}.

\subsection*{Further truncations}

In order to construct supersymmetric Janus-type solutions numerically, we will further truncate the $\,\mathbb{Z}_{2}^3\,$ invariant sector described above to simpler subsectors containing a smaller number of scalar fields. We will do it in a minimal manner, namely, for each of the $\mathcal{N}\,=\,1 \, \&\, SU(3)$, ${\mathcal{N}\,=\,2 \, \&\, U(2)}$ and $\mathcal{N}\,=\,4 \, \&\, SO(4)$ $AdS_{4}$ vacua, we will consider the simplest supersymmetric model accommodating that specific $AdS_{4}$ vacuum. This will improve the efficiency of the numerical integration method as there are no other supersymmetric $AdS_{4}$ vacua towards which the flow is attracted.

The set of minimal models we will consider in this work are:
\begin{itemize}

\item[$i)$] The $SU(3)$ invariant sector of the theory to study flows involving the $\,\mathcal{N}\,=\,1 \, \&\, SU(3)\,$ $AdS_{4}$ vacuum.

\item[$ii)$]  An $SU(2)$ invariant sector of the theory to study flows involving the $\,\mathcal{N}\,=\,2 \, \&\, U(2)\,$ $AdS_{4}$ vacuum.

\item[$iii)$] An $SO(3)$ invariant sector of the theory to study flows involving the $\,\mathcal{N}\,=\,4 \, \&\, SO(4)\,$ $AdS_{4}$ vacuum.

\end{itemize}
Needless to say, working with a truncated model necessarily entails a loss of generality in the structure of solutions we will present. We leave a more thorough characterisation/classification of numerical solutions for the future.

\section{Supersymmetric Janus-type solutions}

Janus solutions correspond to $AdS_3$-sliced domain-wall configurations for which the four-dimensional metric takes the form
\begin{equation} 
\label{4met}
ds_{4}^2 = dr^2 + e^{2A(r)}ds^2_{AdS_3} \ ,
\end{equation}
in terms of the $AdS_{3}$ line element $ds^2_{AdS_3}$ and a metric function $A(r)$ that only depends on the radial coordinate $\,r \in (-\infty,\infty)\,$ transverse to the domain-wall. The metric (\ref{4met}) conforms to $AdS_4$ when
\begin{equation}
A(r)\,=\,\log\left(\frac{L}{l}\cosh\left(\frac{r}{L}\right)\right)\,,
\end{equation}
where $L$ and $l$ are the radii of $AdS_4$ and $AdS_3$, respectively.

\subsection*{BPS flow equations}

Let us now solve the supersymmetry variations of fermionic fields on the curved background, as it was done on the Lorentzian $AdS_3$-sliced domain-walls in \cite{Bobev:2013yra, Suh:2018nmp}. The supersymmetric BPS equations for the complex scalar fields read
\begin{equation}
\label{BPS_eqs_scalars}
z_j'\,=\,\left(-i \kappa\frac{e^{-A}}{l}-A'\right)K^{z_j\bar{z}_j}\frac{2}{W}\frac{{\partial}W}{\partial\bar{z}_j}\,,
\end{equation}
whereas for the metric function one finds
\begin{equation} 
\label{BPS_eqs_A}
A'\,=\,\pm\sqrt{W^2-\frac{e^{-2A}}{l^2}}\,,
\end{equation}
in terms of the $AdS_3$ radius $l$ and $\kappa\,=\,\pm1$. The choice $\kappa\,=\,\pm 1$ is not related to the $\pm$ sign in \eqref{BPS_eqs_A}. Reversing the sign of $\kappa$ merely generates a solution reflected along the coordinate $r$.

It will prove convenient to introduce the upper-half plane parameterisation for the $\mathbb{Z}_{2}^{3}$-invariant scalars in the theory, namely,
\begin{equation}
\label{z_upper-half}
z_j\,=\,-\chi_j + i \, e^{-\varphi_j}\,.
\end{equation}
These scalars span an $[SL(2)/SO(2)]^7$ coset geometry and, in terms of their real and imaginary components, the supersymmetric BPS equations take the form
\begin{align} 
\label{BPS_eqs_z_real}
\chi_j'+4 \, e^{-2\varphi_j}\frac{A'}{W}\frac{{\partial}W}{\partial\chi_j}-4\kappa\frac{e^{-A-\varphi_j}}{l}\frac{1}{W}\frac{{\partial}W}{\partial\varphi_j}\,=&\,0\,, \notag \\[2mm]
\varphi_j'+4 \, \frac{A'}{W}\frac{{\partial}W}{\partial\varphi_j}+4\kappa\frac{e^{-A-\varphi_j}}{l}\frac{1}{W}\frac{{\partial}W}{\partial\chi_j}\,=&\,0\,.
\end{align}

\subsection*{Second-order equations of motion}

In our numerical analysis we will look at the second-order equations of motion. The ones for the scalar fields are given by
\begin{align}
\label{2nd_order_EOM_scalars}
e^{-3A}\partial_\mu(e^{3A}g^{\mu\nu}\partial_\nu\varphi_j)-e^{2\varphi_j}g^{\mu\nu}\partial_\mu\chi_j\partial_\nu\chi_j-\frac{\partial{V}}{\partial\varphi_j}\,=\,0\,, \notag \\[2mm]
e^{-3A}\partial_\mu(e^{3A+2\varphi_j}g^{\mu\nu}\partial_\nu\chi_j)-\frac{\partial{V}}{\partial\chi_j}\,=\,0\,,
\end{align}
whereas the Einstein equations yield two additional equations
\begin{align}
\label{2nd_order_EOM_Einstein}
2A''+3(A')^2+\frac{1}{l^2}e^{-2A}\,=\,-\sum_j  \tfrac{1}{2} \big(  (\varphi_j')^2+e^{2\varphi_j}(\chi_j')^2\big)+V\,, \notag \\[2mm]
3 (A')^2+\frac{3}{l^2}e^{-2A}\,=\,\sum_j  \tfrac{1}{2} \big( (\varphi_j')^2+e^{2\varphi_j}(\chi_j')^2\big)+V\,.
\end{align}
We have verified that the equations of motion (\ref{2nd_order_EOM_scalars}) and (\ref{2nd_order_EOM_Einstein}) are verified provided the BPS equations \eqref{BPS_eqs_scalars} and \eqref{BPS_eqs_A} hold.

\subsection*{Numerical methodology}

In order to construct supersymmetric Janus solutions numerically, we closely follow the method employed in \cite{Bobev:2013yra, Suh:2018nmp}. It also proves efficient to solve the differential equations starting from a given $AdS_4$ vacuum and perturbing it to trigger the flow. Moreover we set $l\,=\,1$, $g\,=\,1$, $c\,=\,1$, and $\kappa\,=\,+1$ for all numerical solutions in this work.

As the BPS equation for the warp factor \eqref{BPS_eqs_A} involves a square root, it involves a branch cut. In order to obtain regular Janus solutions in the domain $\,r \in (-\infty , \infty)$, we must choose the positive sign for $\,r>0\,$ and the negative sign for $\,r<0\,$ \cite{Clark:2005te, Suh:2011xc, Bobev:2013yra, Gutperle:2017nwo}. Then, in order to avoid dealing with branch cuts, we solve the second-order equations of motion (\ref{2nd_order_EOM_scalars})-(\ref{2nd_order_EOM_Einstein}) instead of the first-order BPS equations (\ref{BPS_eqs_scalars})-(\ref{BPS_eqs_A}). To numerically solve the second-order equations of motion, we should specify the initial conditions for the scalar fields ($\varphi_{i}$, $\chi_{i}$), the warp factor $A$, and their derivatives. We impose a smoothness condition at the interface
\begin{equation} 
\label{smooth}
A'(0)\,=\,0 \ .
\end{equation}
As a consequence of the smoothness condition in (\ref{smooth}), and once the initial conditions for the scalar fields \mbox{$\{\varphi_{i}(0)$, $\chi_{i}(0)\}$} are fixed, the initial condition for the warp factor $A(0)$ is automatically determined through its BPS equation in \eqref{BPS_eqs_A}. Similarly, \{$\varphi'_{i}(0)$, $\chi'_{i}(0)$\} are also determined through the BPS equations for the scalar fields in \eqref{BPS_eqs_z_real}. Summarising, we have a set of fourteen free parameters specifying the boundary conditions
\begin{equation}
z_{i}(0) = - \chi_{i}(0) + i \, e^{-\varphi_{i}(0)}  \hspace{10mm} \textrm{ with } \hspace{10mm} i = 1, \ldots, 7 \ , 
\end{equation}
and, once they are fixed, the rest of the parameters are also fixed by the smoothness condition \eqref{smooth} and the BPS equations. Finally, once we have obtained a numerical solution of the second-order equations of motion, we numerically verify that it satisfies the BPS equations.

\subsection{D3-brane}

Here we describe the generic non-$AdS_4$ background towards which the regular BPS flows are attracted when reaching the boundary at $r \rightarrow \pm\infty$. As argued in \cite{Guarino:2021kyp}, this non-$AdS_{4}$ configuration uplifts to a deformation of the $AdS_5\times{S}^5$ geometry arising as the near-horizon geometry of a D3-brane in type IIB supergravity. The deformation is induced by the four-dimensional electromagnetic parameter $c$ in (\ref{Superpotential}).

\subsubsection{D3-brane at $c\,=\,0$}
\label{sec:D3-brane_c=0}

\begin{figure}[t]
\begin{center}
\includegraphics[width=0.45\textwidth]{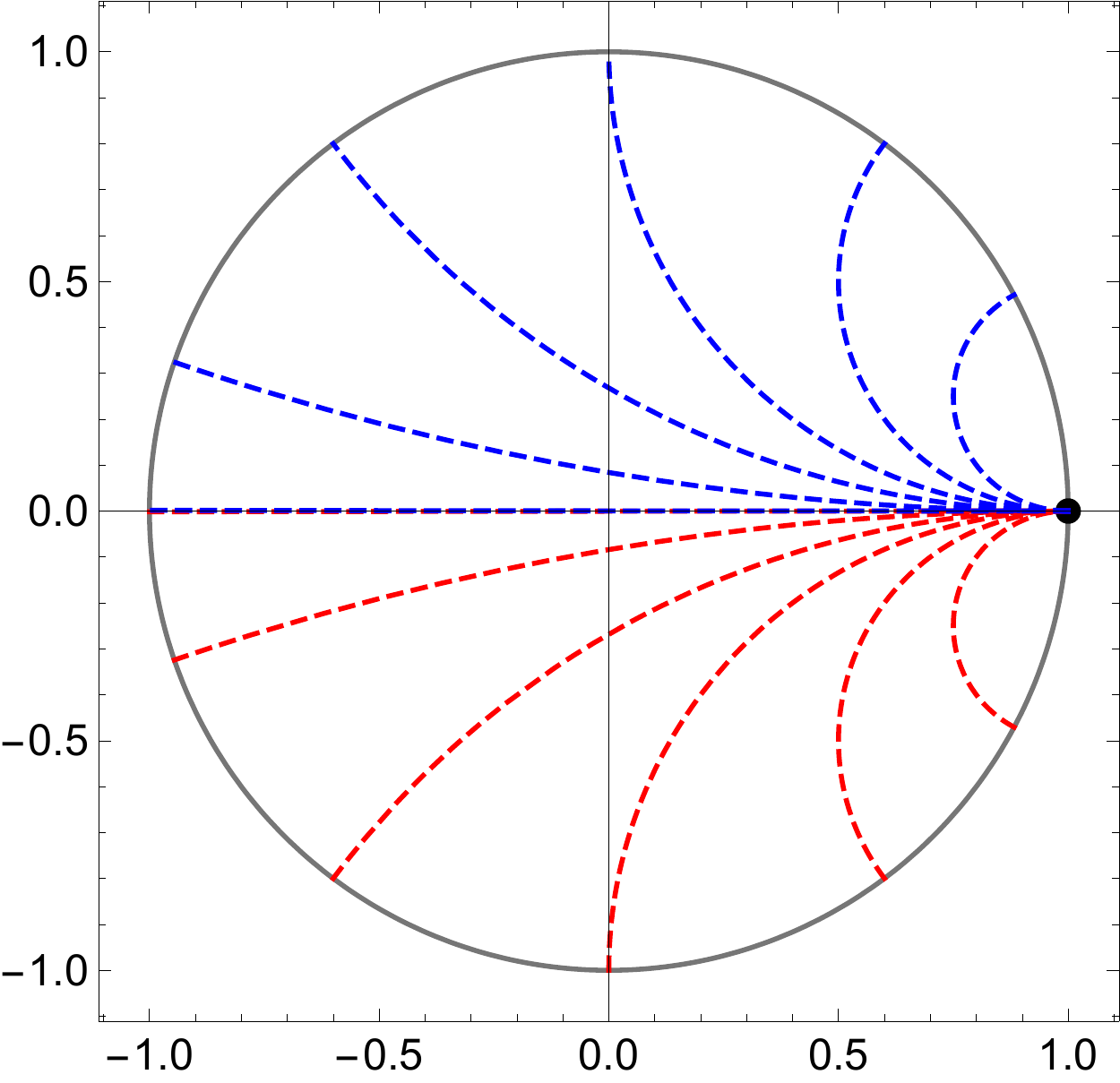} \hspace{10mm} \includegraphics[width=0.45\textwidth]{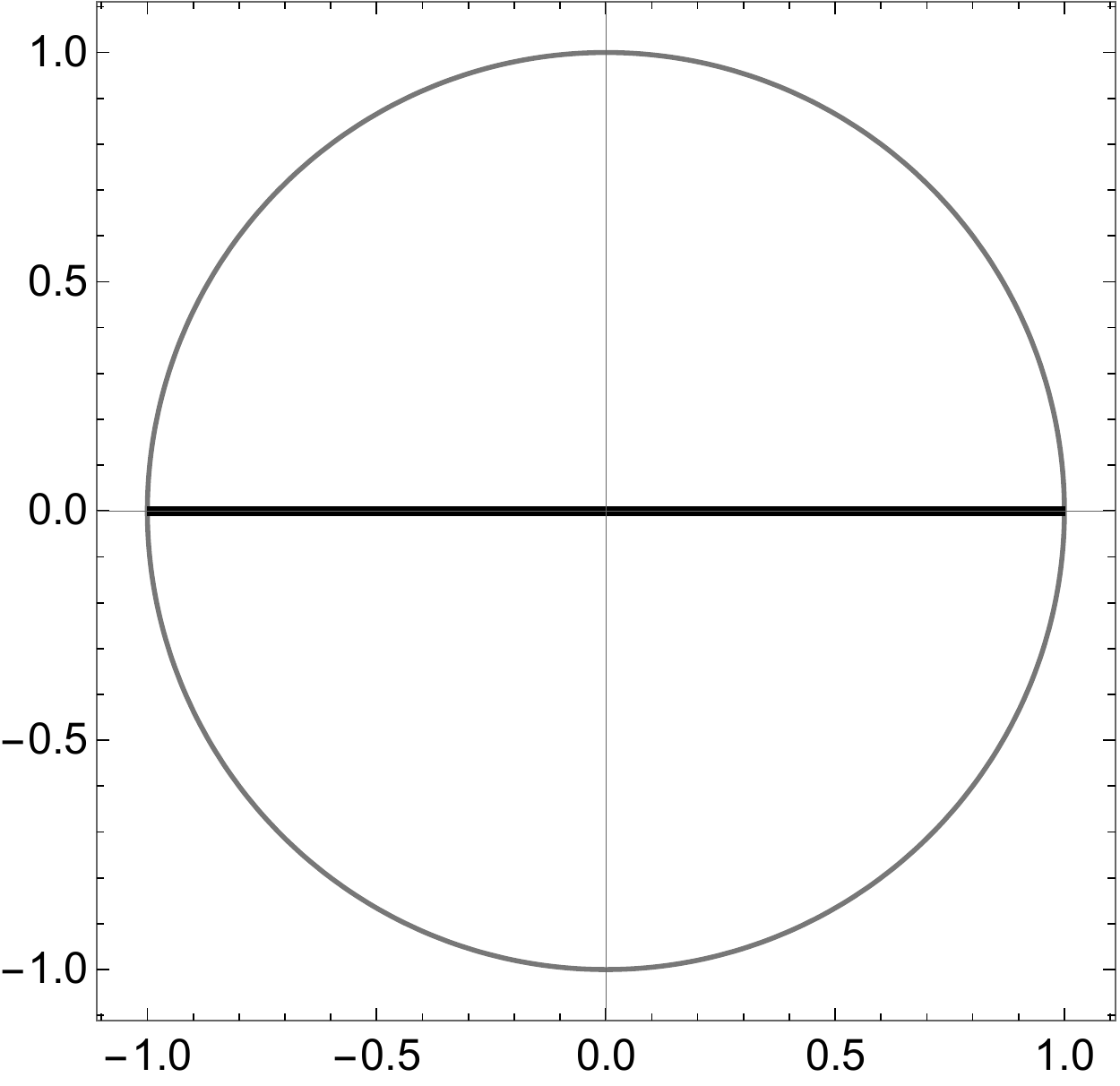}
\put(-370,-15){$\tilde{z}_{1,2,3}$} \put(-115,-15){$\tilde{z}_{4,5,6,7}$}
\caption{{\it Parametric plot of the D3-brane solution at $\,c=0\,$ using the unit-disk parameterisation of the complex scalars $\,\tilde{z}_{1,2,3}\,$ and $\,\tilde{z}_{4,5,6,7}\,$ in (\ref{unit-disk_z123}). On the left plot, the black point at $\,\tilde{z}_{1,2,3} = 1\,$ is reached at the asymptotic values $\, \lambda \rightarrow \infty\,$ (red curves with $\sigma >0$) and $\, \lambda \rightarrow -\infty\,$ (blue curves with $\sigma < 0 $). The singular value $\,\sigma = 0\,$ corresponds to the straight line starting at $\,\tilde{z}_{1,2,3} = -1\,$. On the right plot, different points belonging to the black line in the real axis correspond to different values of the D3-brane constant parameter $\,\Phi_{0}\,$.}}
\label{Fig:D3-brane}
\end{center}
\end{figure}

Let us start by presenting an exact solution to the BPS equations in \eqref{BPS_eqs_scalars} and \eqref{BPS_eqs_A} in the case of $c\,=\,0$. In order to find the solution analytically, we first truncate to the $(SU(3) \cap \mathbb{Z}_{2}^3)$-invariant sector of the theory by performing the scalar identifications
\begin{equation} 
\label{SU(3)-inv_sector}
z_1=z_2=z_3 \equiv z_{1,2,3} \qquad , \qquad  z_4=z_5=z_6=z_7  \equiv z_{4,5,6,7}  \ ,
\end{equation}
and then perform a change of radial coordinate
\begin{equation}
d\lambda\,=\,\frac{g}{\sqrt{2}} \, \textrm{Im}[z_{1,2,3}]^{-\frac{1}{2}} \, dr\,.
\end{equation}
In this manner we obtain an analytic solution with (continuous) $\,U(3)\,$ symmetry of the form
\begin{equation} 
\label{D3-brane_4D_scalars}
z_{1,2,3}  = g^{-1}  \sigma \left(-\frac{1}{3}+i\sinh{\lambda}\right)\qquad , \qquad z_{4,5,6,7}  = i \, e^{-\frac{\Phi_0}{2}} \ ,
\end{equation}
and
\begin{equation} 
\label{D3-brane_4D_metric}
e^{2A}=2 \, g^{-3}\sigma\, \sinh{\lambda} \,\cosh^2\lambda \ ,
\end{equation}
in terms of two arbitrary parameters $(g\,,\,\Phi_0)$ and the scaling parameter $\,\sigma\,$.\footnote{This solution can be generalised to a larger class of solutions solving the second-order equations of motion in (\ref{2nd_order_EOM_scalars}) and (\ref{2nd_order_EOM_Einstein}) but no longer the BPS equations in (\ref{BPS_eqs_scalars}) and (\ref{BPS_eqs_A}). This is given by
\begin{equation} 
\label{D3-brane_4D_scalars_general}
z_{1,2,3} =g^{-1} \left(-\frac{\tilde{\sigma}}{3}+i \, \sigma  \sinh{\lambda}\right)\qquad , \qquad z_{4,5,6,7} =ie^{-\frac{\Phi_0}{2}} \ ,
\end{equation}
and the metric function in (\ref{D3-brane_4D_metric}), so it allows for an additional parameter $\tilde{\sigma}$. Setting $\,\tilde{\sigma}=0\,$ implies $\,\textrm{Re}z_{1,2,3}=0\,$ and the (continuous) symmetry group of the solution is enhanced from $U(3)$ to $SO(6)$. 
} Note that demanding $\,{\textrm{Im}z_{1,2,3} \ge 0}\,$, as imposed by (\ref{z_upper-half}), and $\,e^{2A} \ge 0\,$ requires $\,\lambda \in [0,\infty)\,$ for the choice $\,\sigma>0\,$ and $\,\lambda \in (-\infty,0]\,$ for the choice $\,\sigma<0\,$. Moreover, taking $\,\sigma \rightarrow 0\,$ renders the solution in (\ref{D3-brane_4D_scalars}) pathological as $\,z_{1,2,3} \rightarrow 0\,$ in this limit. Using the (finite) unit-disk parameterisation of the complex scalars, namely,
\begin{equation}
\label{unit-disk_z123}
\tilde{z}_{1,2,3} = \frac{z_{1,2,3}-i}{z_{1,2,3}+i} = 1 - \frac{6}{3+ \sigma \, (3 \sinh\lambda+i)}
\quad , \quad
\tilde{z}_{4,5,6,7} = \frac{z_{4,5,6,7}-i}{z_{4,5,6,7}+i} = \tanh\left(  - \frac{\Phi_{0}}{4}\right)     \ ,
\end{equation}
various plots of the $\,\tilde{z}_{1,2,3}\,$ scalar in (\ref{unit-disk_z123}) are presented in Figure~\ref{Fig:D3-brane} for different values of $\,\sigma\,$. Note that the scalar $\,\tilde{z}_{4,5,6,7}\,$ in (\ref{unit-disk_z123}) covers the real interval $\,[-1,1]\,$ in Figure~\ref{Fig:D3-brane} as a function of the constant parameter $\,\Phi_{0}\,$ setting the asymptotic value of the type IIB dilaton field.

This four-dimensional solution uplifts to a D3-brane in ten-dimensional type IIB supergravity (see Appendix~\ref{app:uplift} for more details on the geometry of the uplift). More concretely the metric reads
\begin{equation}
\label{D3-brane_10D_1}
\begin{array}{rll}
g^2  \,  ds_{10}^2 &=& d\lambda^2+\cosh^2\lambda\,ds_{AdS_3}^2+\sigma^2\sinh^2\lambda\,d\eta^2+\left(ds_{\mathbb{CP}_2}^2+\eta_1\otimes\eta_1\right) \ , 
\end{array}
\end{equation}
with $\sigma>0$ and the coordinate $\,\lambda \in [0,\infty)\,$ specifying an $\,AdS_{3} \times S^{1}\,$ slicing of $\,AdS_{5}\,$ that parameterises the curved worldvolume of the D3-brane. The rest of the type IIB fields are given by
\begin{equation}
\label{D3-brane_10D_2}
\begin{array}{rll}
m_{\alpha\beta} & = & \left(\begin{array}{ll}
 e^{-\Phi_0} & 0 \\
 0 & e^{\Phi_0}
\end{array}
\right)
\qquad , \qquad
\mathbb{B}^{\alpha}  \,= \, 0 
\qquad , \qquad
\widetilde{F}_5 = 4 \, g \left(1+*\right) \,  vol_{\mathbb{CP}_2}\wedge \eta_1  \ .
\end{array}
\end{equation}
From (\ref{D3-brane_10D_1}) one observes that the parameter $\,\sigma\,$ can be reabsorbed in a redefinition of the coordinate $\,\eta\,$ thus rendering the limit $\,\sigma \rightarrow 0\,$ pathological this time from a higher-dimensional perspective. In addition, having $\,\textrm{Re}z_{1,2,3} = -\tfrac{1}{3} g^{-1}  \sigma \neq 0\,$ formally induces a monodromy (see (\ref{U(1)-fibration})) on the $\,S^1\,$ in $\,S^{5} = \mathbb{CP}_{2} \rtimes S^{1}\,$ when moving along the compactified $\,\eta\,$ direction. This results in the global breaking of the $\,S^{5}\,$ isometries from $\,SO(6)\,$ to $\,U(3)\,$ as discussed in \cite{Guarino:2021kyp}. Unlike for the curved-sliced D3-brane discussed here, the flat-sliced D3-brane in \cite{Guarino:2021kyp} corresponds to a four-dimensional analytic flow solution with $\,\textrm{Re}z_{1,2,3} = 0\,$ and, therefore, features a larger $\,SO(6)\,$ symmetry.

\subsubsection{D3-brane at $c\,\neq\,0$}

Let us study a perturbation of the analytic solution at $\,c=0\,$ in (\ref{D3-brane_4D_scalars})-(\ref{D3-brane_4D_metric}) with $\,U(3)\,$ symmetry. We will redefine the radial coordinate as
\begin{equation}
\rho = \sinh\lambda \ ,
\end{equation}
so that $\,\rho \in [0,\infty)\,$ and the solution in (\ref{D3-brane_4D_scalars})-(\ref{D3-brane_4D_metric}) at $\,c=0\,$ takes the form
\begin{equation}
\label{D3-brane_4D_c_0}
z_{1,2,3} = g^{-1}  \sigma \left(-\tfrac{1}{3}+i \rho\right)
\hspace{8mm} , \hspace{8mm}
z_{4,5,6,7}  = i \, e^{-\frac{\Phi_0}{2}}
\hspace{8mm} , \hspace{8mm}
e^{2A}=2 \, g^{-3}\sigma\, \rho \, (1+\rho^2) \ .
\end{equation}
Using the new radial coordinate $\,\rho\,$ it is possible to analytically solve the BPS equations (\ref{BPS_eqs_scalars})-(\ref{BPS_eqs_A}) order by order as a power series in the deformation parameter $\,c\,$ (which can be set to any desired numerical value without loss of generality). At linear order in $\,c\,$, and further expanding around the region at $\,\rho\rightarrow \infty$, the scalar fields conform to the power series
\begin{equation}
\label{scalars_UV_expansion}
\begin{array}{rll}
z_{1,2,3}  & = &  - \frac{1}{3} \, g^{-1} \,  \sigma \, \left[  1 + c \, \left( \lambda_0 - 2  \lambda_2   + 2 \, g^{-2} \sigma^{-1} \, \sinh\Phi_0  - (\lambda_{2}+4 \lambda_{4})  \left( \dfrac{1}{\rho^2} - \dfrac{1}{\rho^4} + \ldots \right)  \right)  \right] \\[4mm] 
&& +  \,  i \,  g^{-1} \,  \sigma\, \rho \left[  1 + c \, \left( \lambda_{0} + \dfrac{\lambda_{2}}{\rho^2}  + \dfrac{\lambda_{4}}{\rho^4} + \ldots  \right)  \right] \ ,  \\[8mm]
z_{4,5,6,7}   & = &   c \, e^{-\frac{\Phi_0}{2}} \left[ \dfrac{g^{-2} \sigma^{-1} \, \cosh\Phi_0}{\rho} - 2 \kappa_{4} \left(  \dfrac{1}{\rho^3} - \dfrac{1}{\rho^5} + \ldots  \right)  \right]  \\[4mm]
&& + \, i \, e^{-\frac{\Phi_0}{2}}  \, \left[ 1 + c \left(  \kappa_{0} + \dfrac{\kappa_{4}}{\rho^{4}} + \ldots \right)    \right] \ ,
\end{array}
\end{equation}
in terms of five integration constants $\,\{  \lambda_{0} \,,\, \lambda_{2} \,,\, \lambda_{4} \,,\, \kappa_{0} \,,\, \kappa_{4}\}\,$. The scale factor approaches $\,\rho \rightarrow \infty\,$ as
\begin{equation}
\label{metric_UV_expansion}
e^{2A} = 2 \, g^{-3}\sigma\, \rho \, (1+\rho^2) \, \left[   1 + c \, \left( \lambda_{0} + 4 \lambda_{2} - \dfrac{\lambda_{2}}{\rho^2} + \dfrac{4 \, \lambda_{2} + \lambda_{4}}{3 \, \rho^4}   + \ldots \right) \right] \ .
\end{equation}
In the following we will construct numerical solutions to the BPS equations (\ref{BPS_eqs_scalars})-(\ref{BPS_eqs_A}) with $\,c\neq 0\,$. Such solutions will generically approach the $\,\rho \rightarrow \infty\,$ region as in (\ref{scalars_UV_expansion})-(\ref{metric_UV_expansion}) with specific values of the integration constants $\,\{  \lambda_{0} \,,\, \lambda_{2} \,,\, \lambda_{4} \,,\, \kappa_{0} \,,\, \kappa_{4}\}\,$ as well as $\,\Phi_{0}\,$. Note that $\,\textrm{Re}z_{4,5,6,7} \neq 0\,$ at linear order in $\,c\,$ but dies off as $\,\rho^{-1}\,$ switching off the type IIB two-form potentials in (\ref{mathbbB})-(\ref{mathfrakb1b2}). As a result, the deformed D3-brane solution in (\ref{scalars_UV_expansion})-(\ref{metric_UV_expansion}) breaks the $\,U(3)\,$ symmetry of (\ref{D3-brane_4D_scalars})-(\ref{D3-brane_4D_metric}) down to $\,SU(3)$. This concludes our characterisation of the D3-brane behaviour in the $\,\rho \rightarrow \infty\,$ region when $\,c \neq 0\,$.

\subsection{$SU(3)$-invariant flows}

The $(SU(3) \cap \mathbb{Z}_{2}^{3})$-invariant sector of the theory is recovered upon the identifications in (\ref{SU(3)-inv_sector}), namely,
\begin{equation} 
\label{SU(3)_sector_scalars}
z_1=z_2=z_3  \qquad , \qquad  z_4=z_5=z_6=z_7 \ .
\end{equation}
This yields an effective two-field model and the holomorphic superpotential in (\ref{Superpotential}) reduces to
\begin{equation}
\label{Superpotential_SU(3)}
\mathcal{V}\,=\,12 \, g \, z_{1} \, z_{4}^2 + 2 \, g \, c \, \left(1-z_{4}^4\right)\,.
\end{equation}
The identifications in (\ref{SU(3)_sector_scalars}) are compatible with the Family I of $AdS_{4}$ vacua presented in (\ref{solution_family_1}). However, the condition (\ref{sum_chi=0}) requires $\,\textrm{Re}z_1 = 0\,$ at the corresponding $AdS_{4}$ vacuum. Still, as we will see (\textit{e.g.} red flow in Figure~\ref{Fig:N=1_plots_2}), a generic Janus-type solution involving such an $AdS_{4}$ vacuum in the intermediate regime will arrive at the asymptotic D3-brane behaviour at $r \rightarrow \pm \infty$ with a non-zero value of the axion $\textrm{Re}z_1$.

\begin{figure}[t!]
\begin{center}
\includegraphics[width=2.8in]{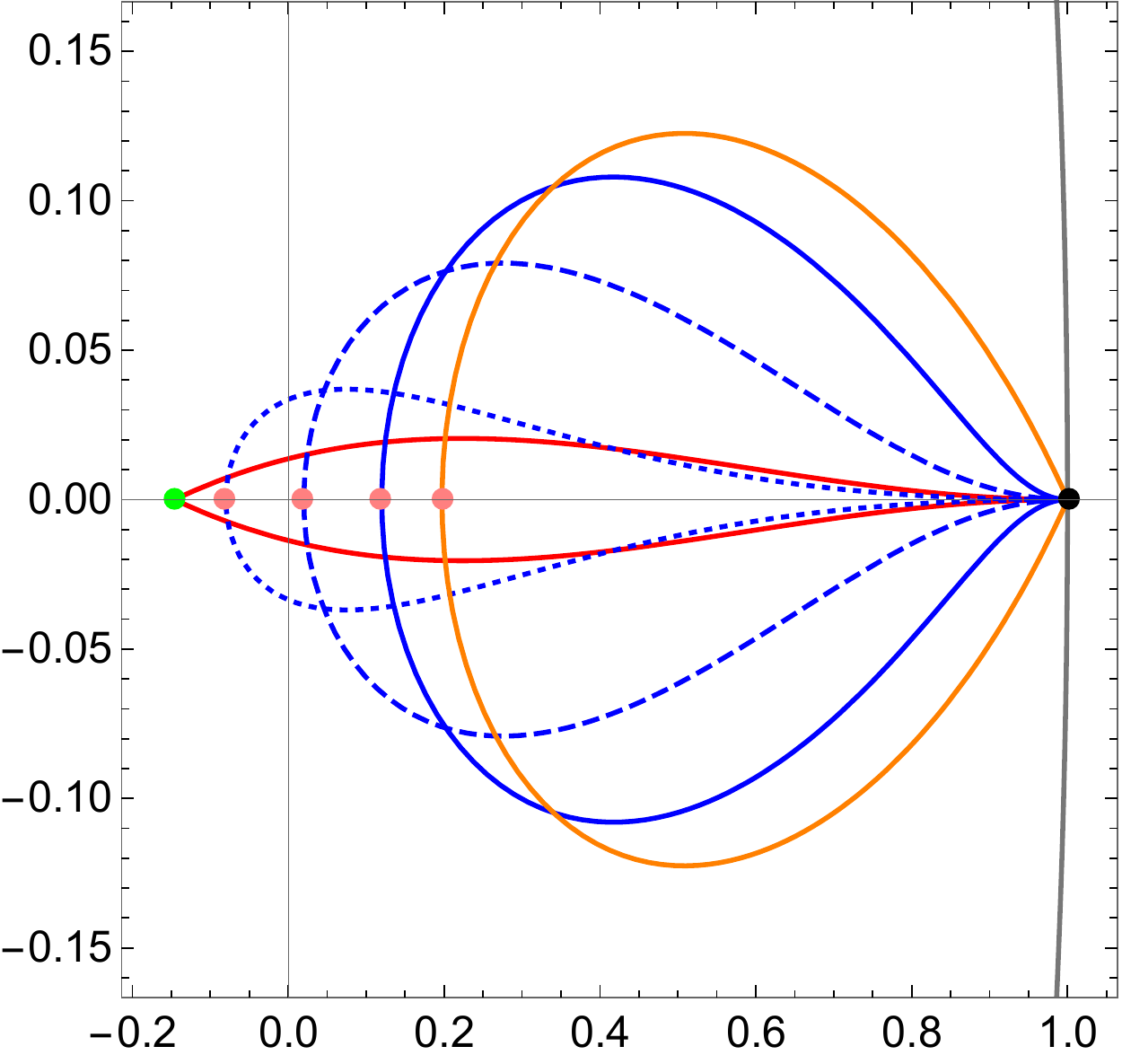} \hspace{10mm} \includegraphics[width=2.8in]{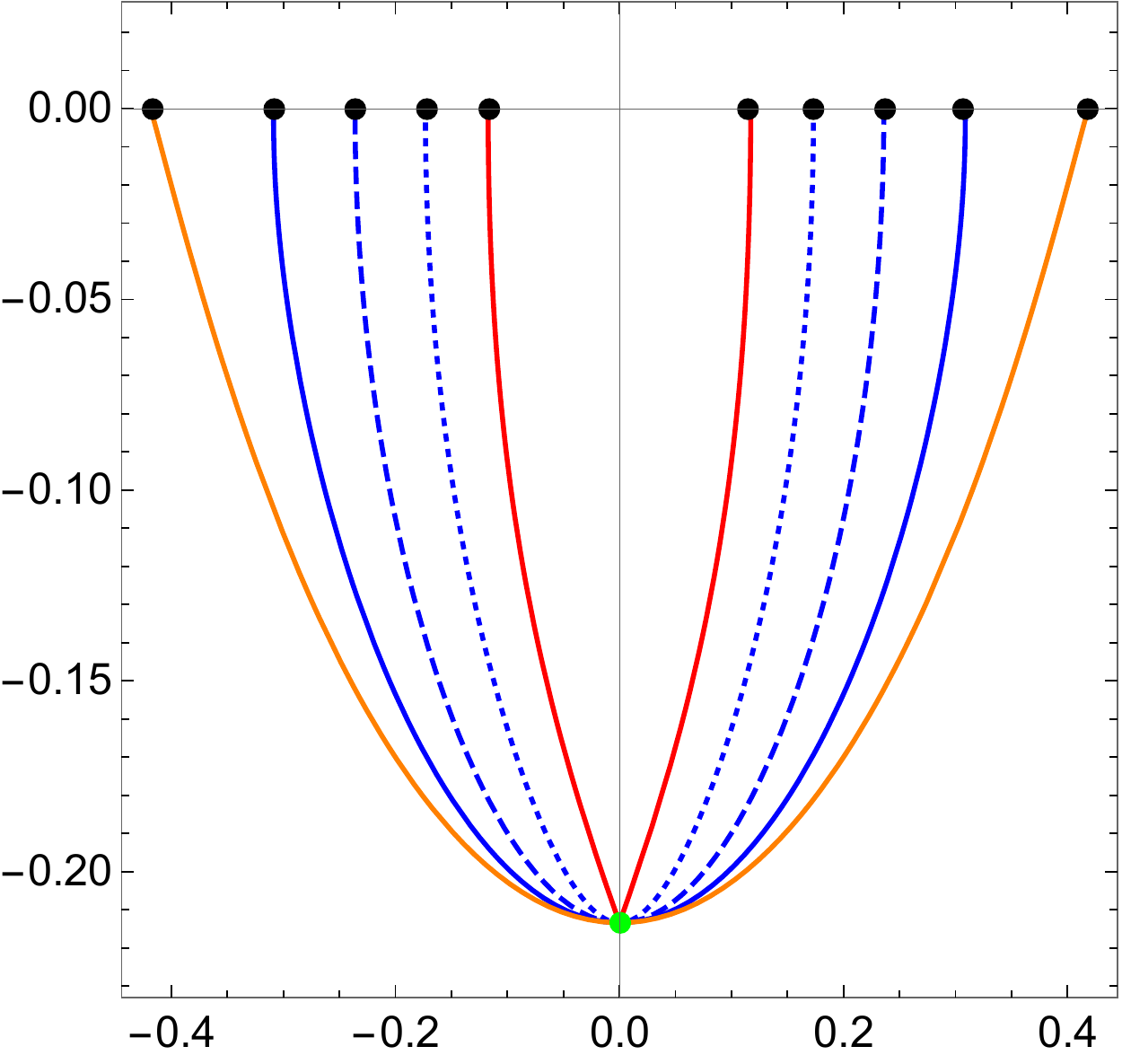}
\put(-335,-15){$\tilde{z}_{1}$}\put(-95,-15){$\tilde{z}_{4}$}
\caption{{\it Samples of regular flows within the $SU(3)$-invariant sector using the boundary conditions (\ref{boundary_SU3_z123}) for the complex scalar fields $\,\tilde{z}_{1}\,$ (left) and $\,\tilde{z}_{4}\,$ (right). The $\mathcal{N}= 1\,\&\,SU(3)$ $AdS_{4}$ vacuum and the D3-brane asymptotics are denoted by green and black dots, respectively. The grey line in the left plot corresponds to the boundary values $\,|\tilde{z}_{1}|=1\,$. The red flow corresponds to the choice $\,\epsilon \approx 0\,$ whereas the orange one corresponds to $\,\epsilon \approx \epsilon_{\textrm{crit}}\,$. Black dots in the right plot are compatible with SO(6) invariance, \textit{i.e.} $\,\tilde{z}_{4}\,$ is real-valued, and correspond to the values $\,\pm\Phi_{0}\,$ of the type IIB dilaton in the D3-brane asymptotics at $\,r \rightarrow \pm \infty\,$.}}
\label{Fig:N=1_plots_1}
\end{center}
\end{figure}

\subsubsection*{Numerical study}

Following the numerical methodology introduced before, a numerical Janus-type flow is constructed upon suitable choice of boundary conditions $\,\lbrace  z_{1}(0) \, , \, z_{4}(0) \rbrace\,$. It proves covenient to switch again to the unit-disk parameterisation for the complex scalars
\begin{equation} 
\label{SU3_unit-disk}
\tilde{z}_{1} = \frac{z_{1}-i}{z_{1}+i} \qquad  , \qquad  \tilde{z}_{4} = \frac{z_{4}-i}{z_{4}+i} \ ,
\end{equation}
and specify the boundary conditions in terms of $\,\lbrace  \tilde{z}_{1}(0) \, , \, \tilde{z}_{4}(0) \rbrace\,$. The location in field space of the ${\mathcal{N}= 1\,\&\,SU(3)}$ $AdS_{4}$ vacuum and the D3-brane asymptotic behaviour (see Figure~\ref{Fig:N=1_plots_1}) suggest two natural choices of boundary conditions.

\begin{figure}[t!]
\begin{center}
\scalebox{0.75}{
\includegraphics[width=2.8in]{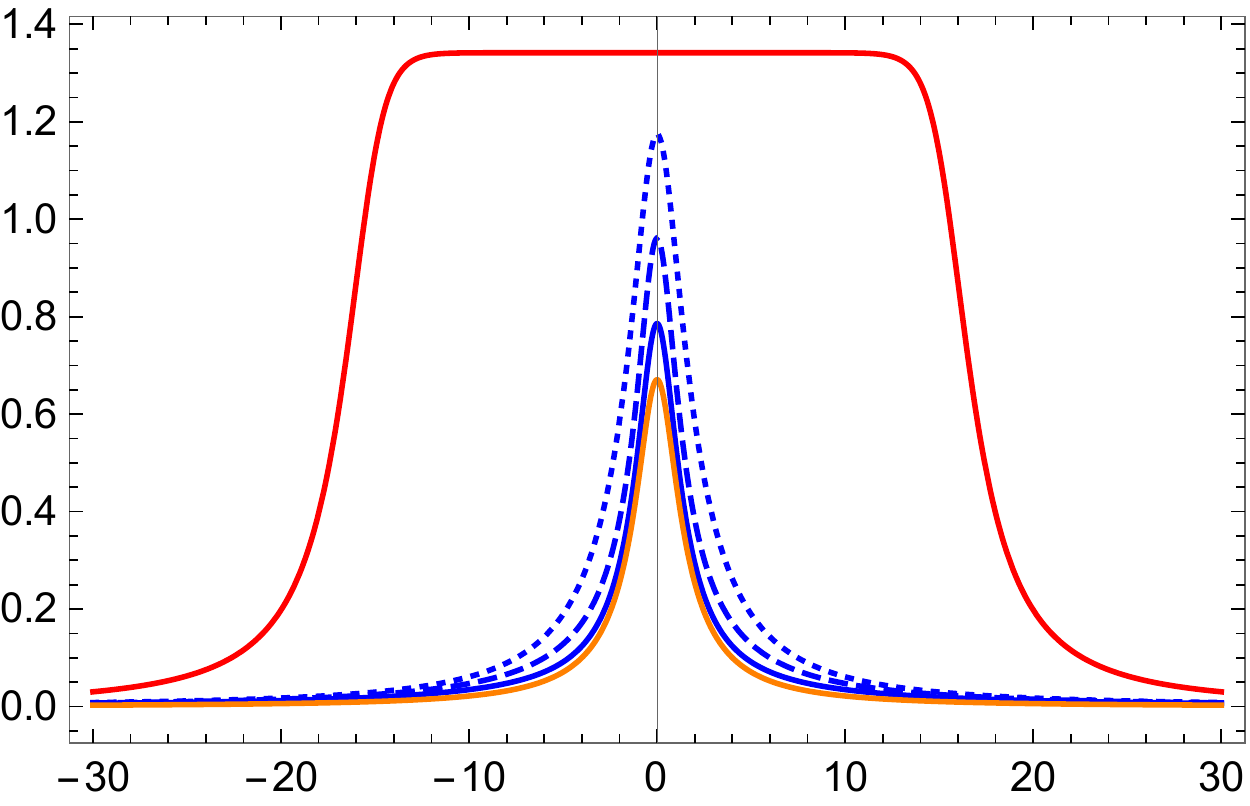}  \hspace{10mm} \includegraphics[width=2.8in]{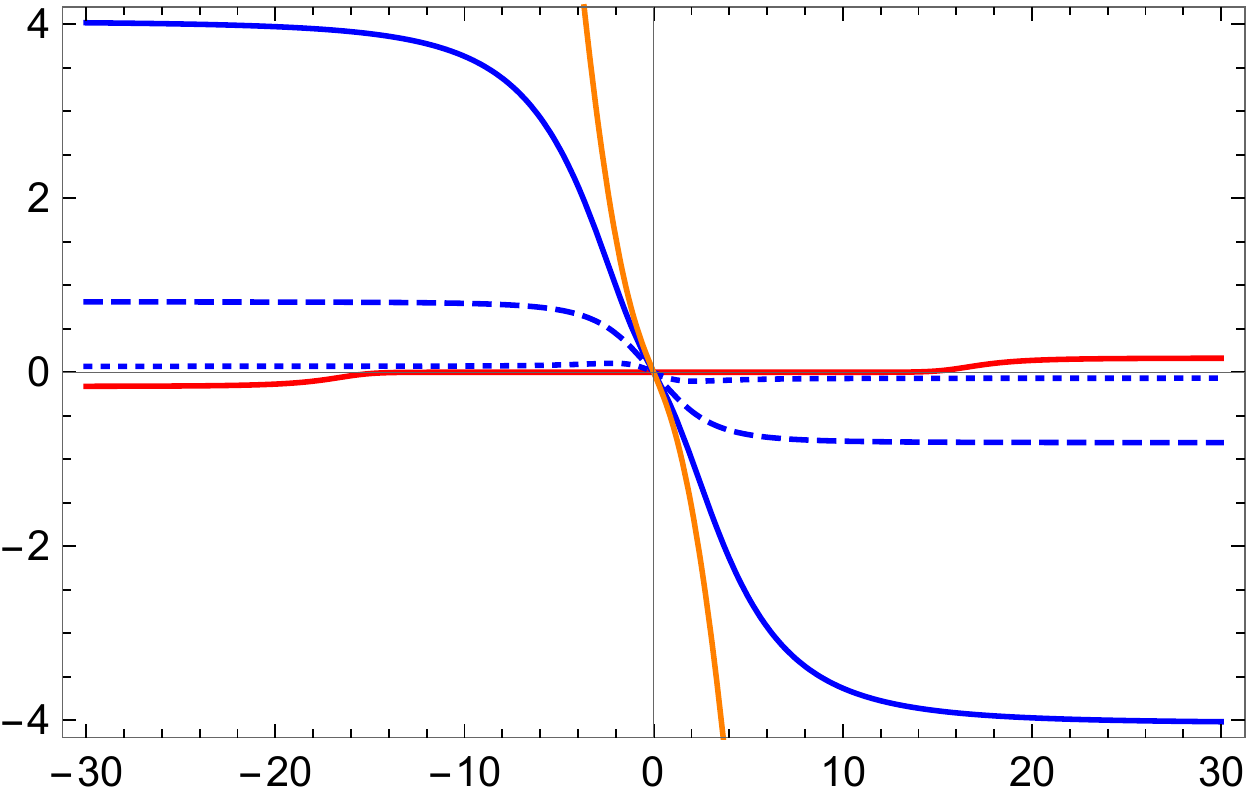} 
\put(-462,50){\rotatebox{90}{$\left(\textrm{Im}z_{1}\right)^{-1}$}}
\put(-218,57){\rotatebox{90}{$\textrm{Re}z_{1}$}}}
\\[4mm]
\scalebox{0.75}{\includegraphics[width=2.8in]{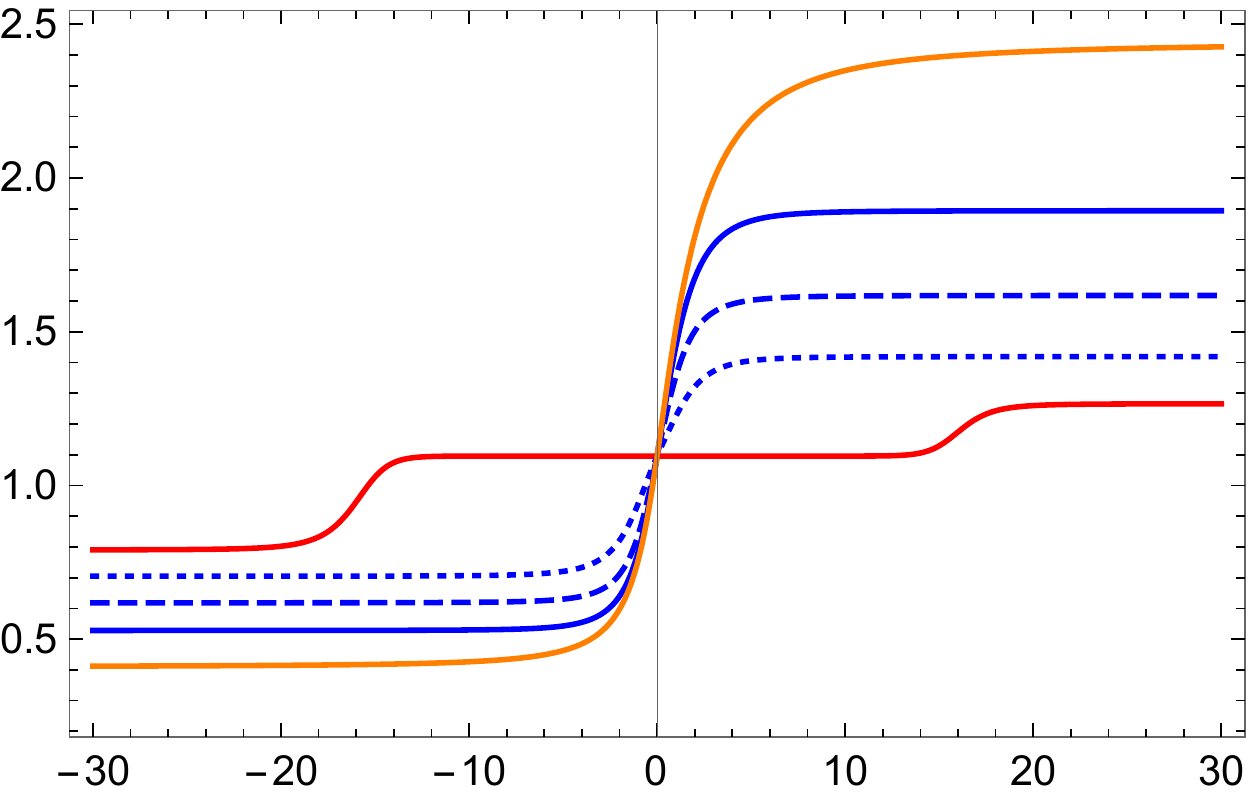}  \hspace{10mm} \includegraphics[width=2.8in]{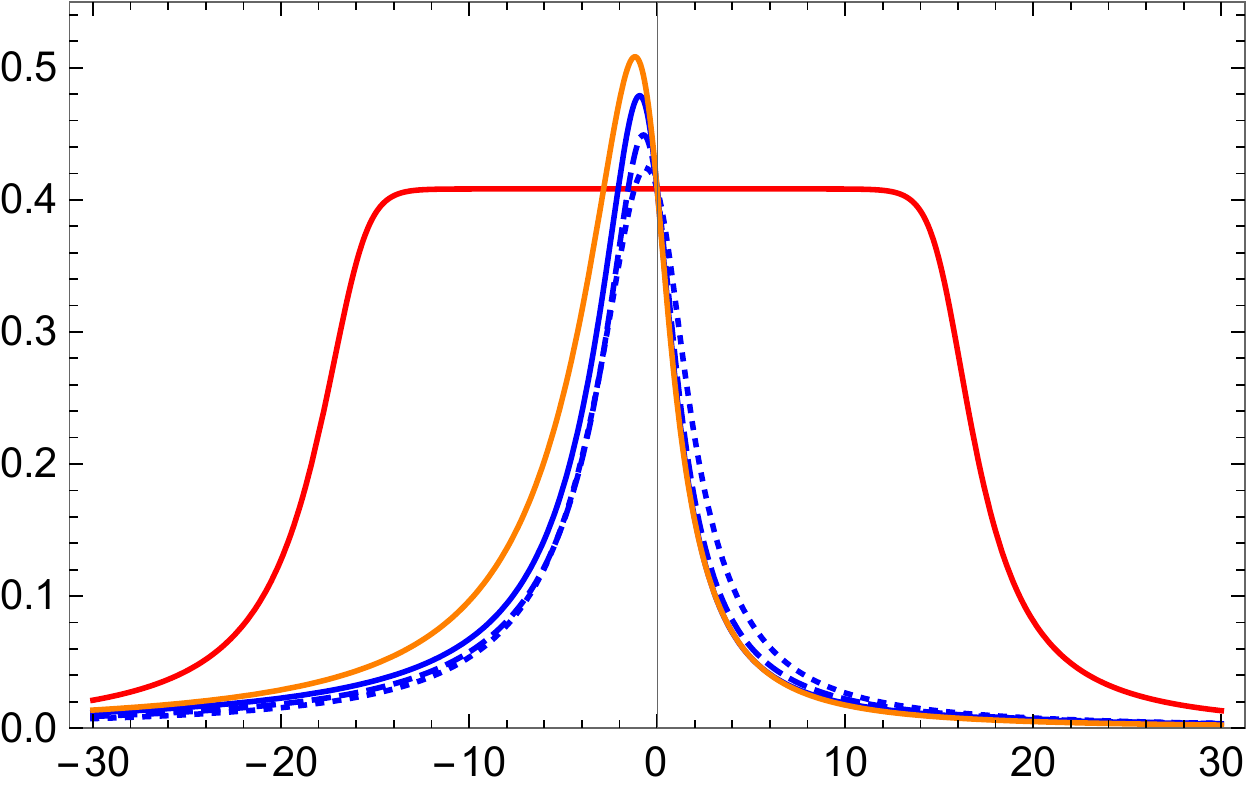} 
\put(-462,50){\rotatebox{90}{$\left(\textrm{Im}z_{4}\right)^{-1}$}}
\put(-218,57){\rotatebox{90}{$\textrm{Re}z_{4}$}}}
\\[4mm]
\scalebox{0.75}{\includegraphics[width=2.8in]{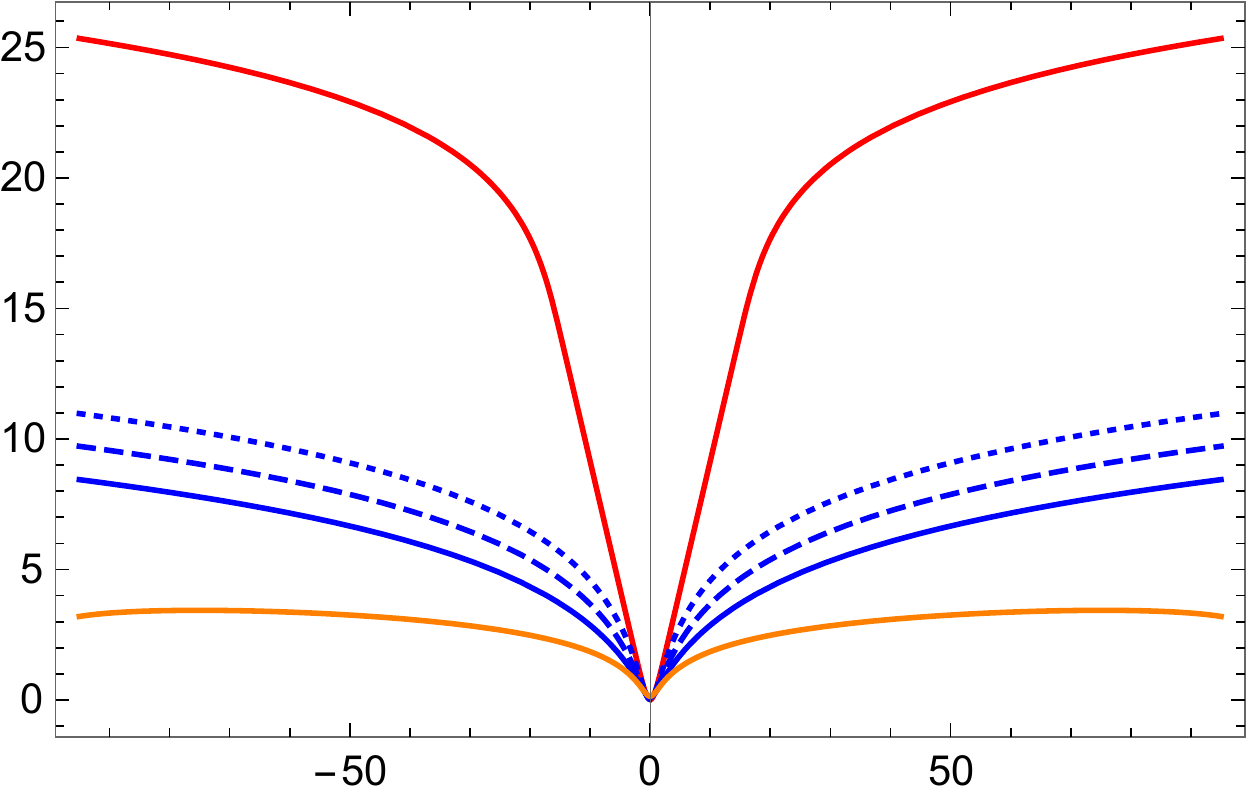}  \hspace{10mm} \includegraphics[width=2.8in]{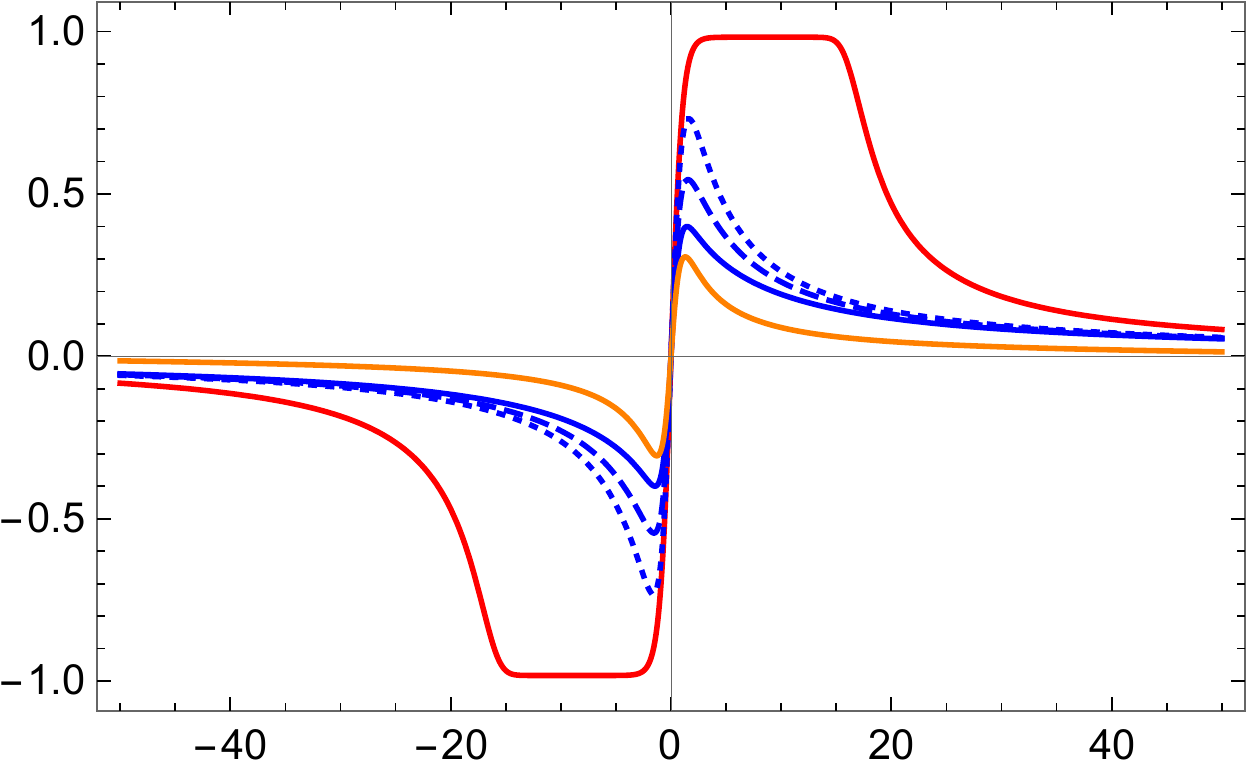} 
\put(-462,65){\rotatebox{0}{$A$}}
\put(-218,65){\rotatebox{0}{$A'$}}
\put(-339,-15){$r$}\put(-96,-15){$r$}}
\caption{{\it Samples of regular flows within the $SU(3)$-invariant sector using the boundary conditions (\ref{boundary_SU3_z123}) for the the complex scalar fields $\,z_{1}\,$ and $\,z_{4}\,$.}}
\label{Fig:N=1_plots_2}
\end{center}
\end{figure}

The first choice is to distribute the turning points of the numerical flows (pink points in Figure~\ref{Fig:N=1_plots_1}) along the $\,\textrm{Re}\tilde{z}_{1}\,$ real axis while keeping $\,\tilde{z}_{4}\,$ fixed at its value in the $AdS_{4}$ vacuum. This is
\begin{equation}
\label{boundary_SU3_z123}
\tilde{z}_{1}(0) =   \tilde{z}_{1}^{(*)} + \epsilon
\qquad , \qquad 
\tilde{z}_{4}(0) =  \tilde{z}_{4}^{(*)}  \ ,
\end{equation}
where $\,\tilde{z}_{1,4}^{(*)}\,$ denote the expectation values of the scalars at the $\,\mathcal{N}= 1\,\&\,SU(3)\,$ $AdS_{4}$ vacuum. Regular flows only exist for $\,0 < \epsilon <  \epsilon_{\textrm{crit}}\,$ with $\,\epsilon_{\textrm{crit}} \approx 0.34315 \,$. Varying the parameter $\,\epsilon\,$ within this range we obtain the numerical flows depicted in Figure~\ref{Fig:N=1_plots_1}. The profiles for the original fields $\,z_{1}\,$ and $\,z_{4}\,$ in the upper half-plane parameterisation are displayed in Figure~\ref{Fig:N=1_plots_2}. From the latter figure, we observe the existence of two limiting/bounding flows. The first one is the flow with $\,\epsilon \approx 0\,$ (red solid lines). This flow develops an $AdS_{4}$ intermediate behaviour with constant scalars around the turning point at $\,r=0\,$ (green point in Figure~\ref{Fig:N=1_plots_1}).  When uplifted to ten dimensions (see (\ref{dlambda_appendix})--(\ref{f1f2_funcs_appendix})), this intermediate behaviour is identified with the $\mathcal{N}=1\,\&\,SU(3)$ type IIB S-fold of \cite{Guarino:2019oct}. The other bounding flow is the critical flow (orange solid lines) corresponding to the largest value  $\,\epsilon \approx \epsilon_{\textrm{crit}}\,$. This critical flow comes along with a very large value of $\,\textrm{Re}z_{1}\,$ around the asymptotic D3-brane regions, as it can be recognised in Figure~\ref{Fig:N=1_plots_2}. All the numerical solutions approach the D3-brane behaviour at the flow endpoints, \textit{i.e.} $\,r \rightarrow \pm \infty\,$, with a different value of $\,\pm\Phi_{0}\,$ depending on the choice of $\,\epsilon\,$ in the boundary conditions (\ref{boundary_SU3_z123}).

The second natural choice of boundary conditions is to distribute the turning points of the numerical flows along the $\,\textrm{Im}\tilde{z}_{4}\,$ imaginary axis while keeping $\,\tilde{z}_{1}\,$ fixed at its value in the $AdS_{4}$ vacuum. This is
\begin{equation}
\label{boundary_SU3_z4567}
\tilde{z}_{1}(0) = \tilde{z}_{1}^{(*)}
\qquad , \qquad 
\tilde{z}_{4}(0) = \tilde{z}_{4}^{(*)} + i \, \epsilon  \ .
\end{equation}
However, we find no regular flows for this choice of boundary conditions. Actually, we cannot find regular flows with boundary conditions different from (\ref{boundary_SU3_z123}).

\begin{figure}[t!]
\begin{center}
\scalebox{0.8}{\includegraphics[width=2.55in]{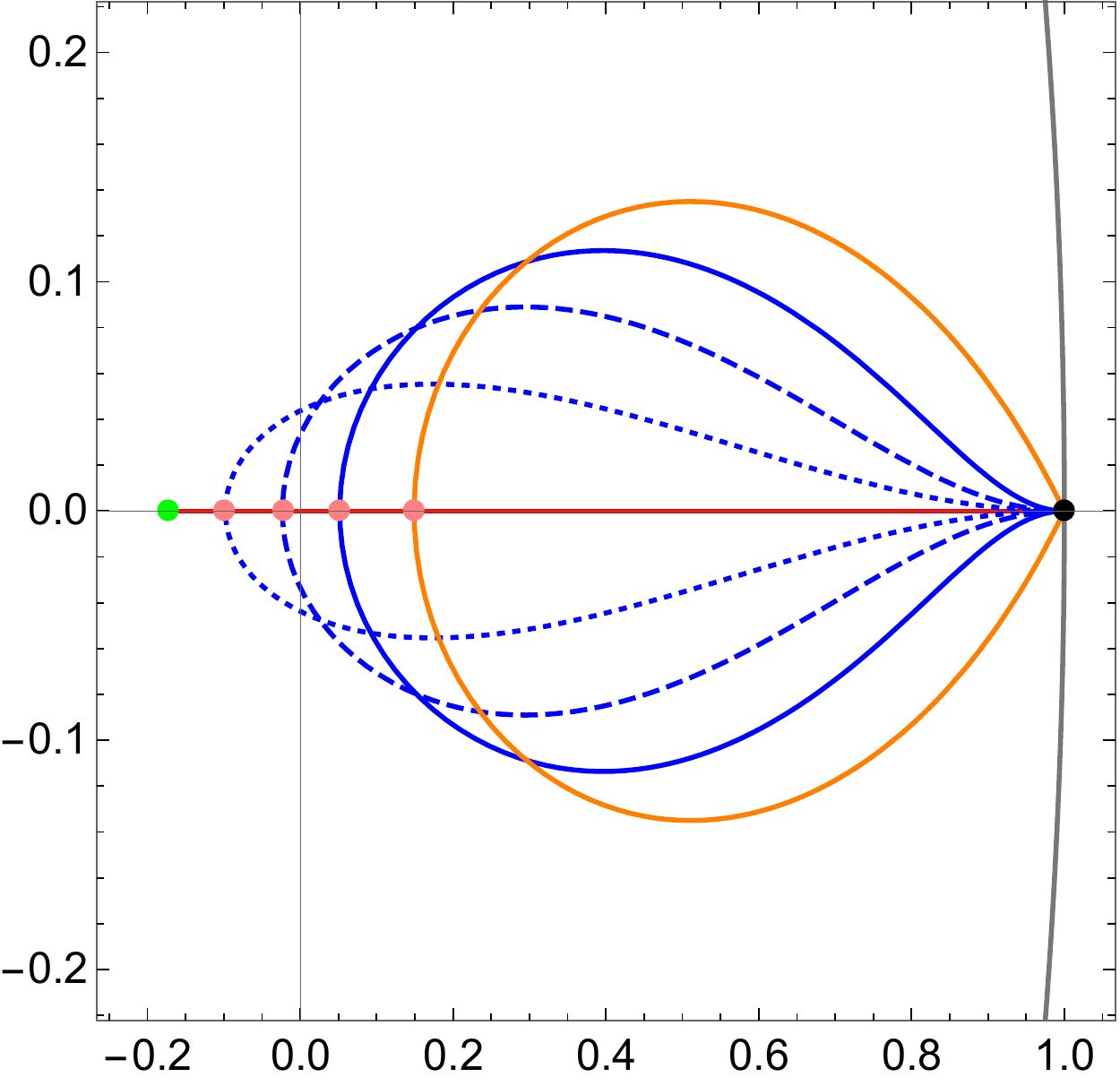} \hspace{12mm} \includegraphics[width=2.6in]{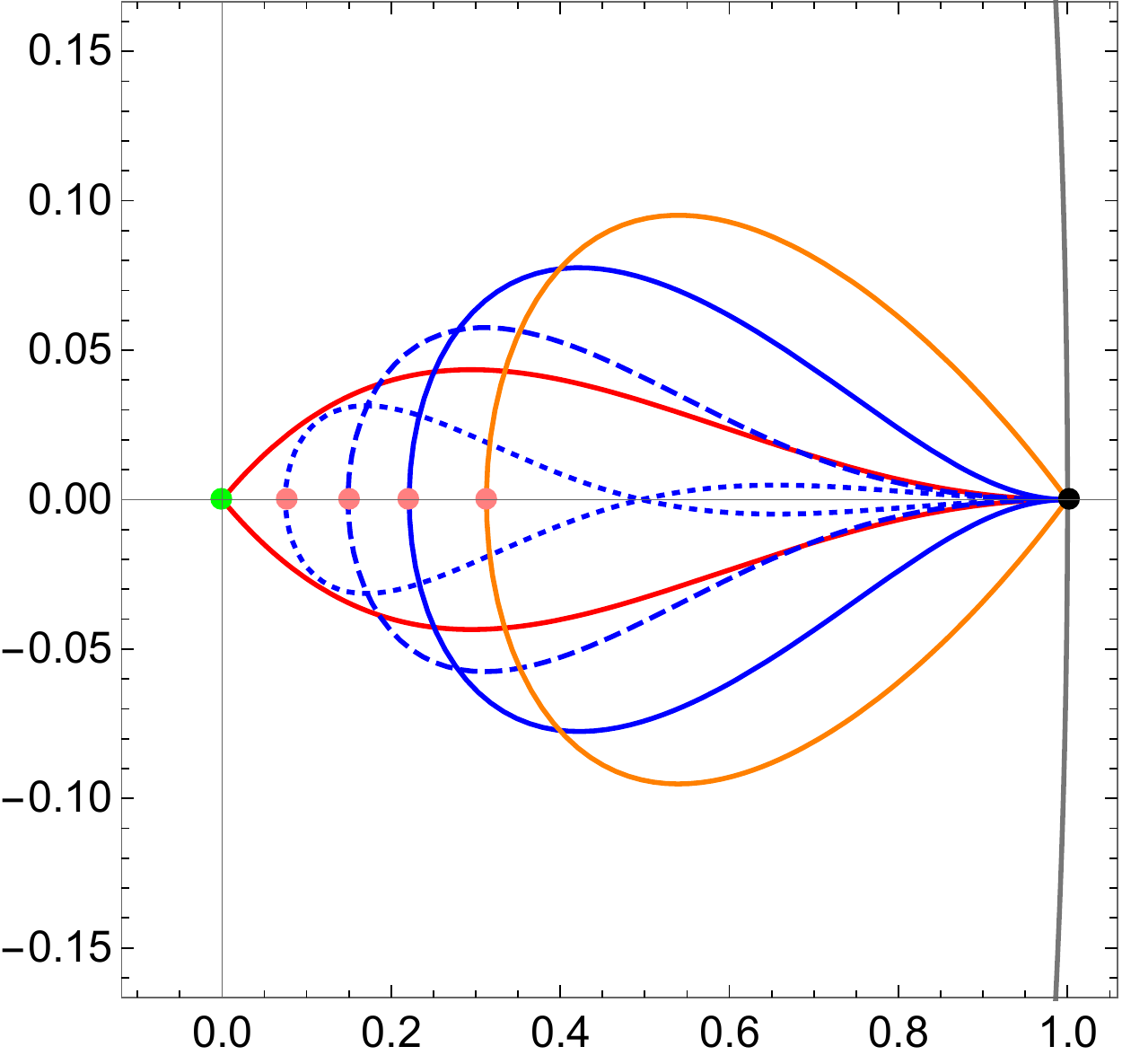}
\put(-315,-15){$\tilde{z}_{1}$}\put(-90,-15){$\tilde{z}_{2}$}} \\[2mm]
\scalebox{0.8}{\includegraphics[width=2.75in]{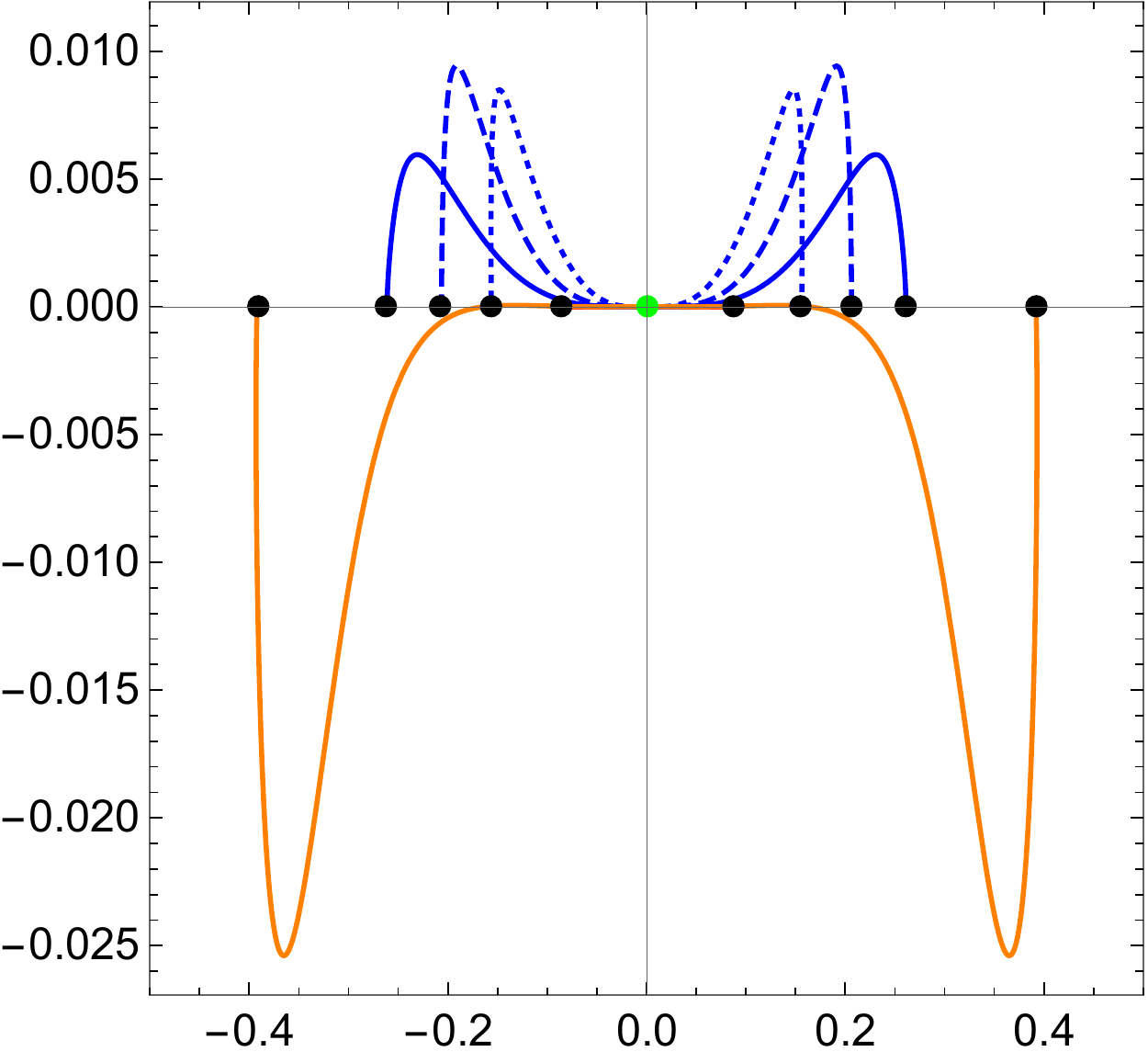} \hspace{10mm} \includegraphics[width=2.6in]{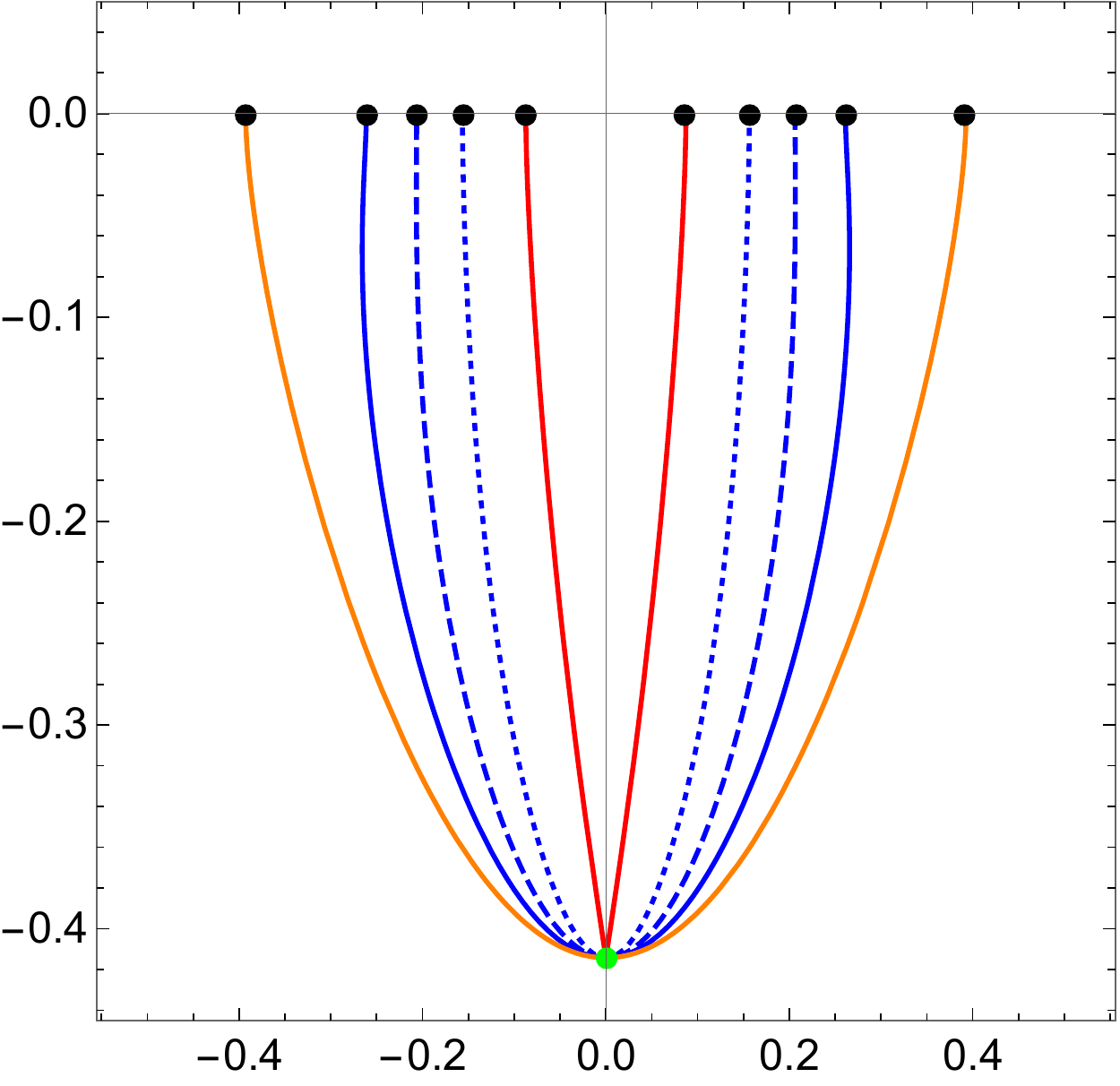}
\put(-315,-15){$\tilde{z}_{4}$}\put(-90,-15){$\tilde{z}_{5}$}}
\caption{{\it Samples of regular flows within the $SU(2)$-invariant sector using the boundary conditions (\ref{boundary_SU2_z123}) for the complex scalar fields $\,\tilde{z}_{1}\,$ (top-left), $\,\tilde{z}_{2}\,$ (top-right), $\,\tilde{z}_{4}\,$ (bottom-left) and $\,\tilde{z}_{5}\,$ (bottom-right). The $\mathcal{N}= 2\,\&\,U(2)$ $AdS_{4}$ vacuum and the D3-brane asymptotics are denoted by green and black dots, respectively. The grey lines in the left column plots correspond to the boundary values $\,|\tilde{z}_{1}|=1\,$. The red flow corresponds to the choice $\,\epsilon \approx 0\,$ whereas the orange one corresponds to $\,\epsilon \approx \epsilon_{\textrm{crit}}\,$. Black dots in the bottom plots are compatible with SO(6) invariance, \textit{i.e.} $\,\tilde{z}_{4}=\tilde{z}_{5}\,$ are real-valued, and correspond to the values $\,\pm\Phi_{0}\,$ of the type IIB dilaton in the D3-brane asymptotics at $\,r \rightarrow \pm \infty\,$.}}
\label{Fig:N=2_plots_1}
\end{center}
\end{figure}

\subsection{$SU(2)$-invariant flows}

The $(SU(2) \cap \mathbb{Z}_{2}^{3})$-invariant sector of the theory is obtained upon identifying the complex scalars as
\begin{equation}
\label{SU(2)_sector_scalars}
z_1 = z_3 \qquad , \qquad z_2\qquad , \qquad z_4\,=\,z_6\qquad , \qquad z_5\,=\,z_7 \ ,
\end{equation}
so the resulting effective model involves four chiral fields. The holomorphic superpotential in (\ref{Superpotential}) simplifies to
\begin{equation}
\label{Superpotential_SU(2)}
\mathcal{V}\,=\,2 \, g\Big( 4 \, z_1 \, z_4 \, z_5 + z_2 \, \left(z_4^2+z_5^2\right)\Big) + 2\, g \, c \, \left(1-z_4^2 \, z_5^2\right)\, .
\end{equation}
The identifications in (\ref{SU(2)_sector_scalars}) are compatible with the Family II of $AdS_{4}$ vacua in (\ref{solution_family_2}) provided
\begin{equation}
\label{cond_axions_U2}
\textrm{Re}z_1 = 0 \ .
\end{equation}
For generic choices of boundary conditions, the algebraic condition (\ref{cond_axions_U2}) will not be preserved along the flows we will construct. However, as we will see, a specific choice of boundary conditions will preserve the condition (\ref{cond_axions_U2}).

\begin{figure}[t!]
\begin{center}
\scalebox{0.75}{
\includegraphics[width=2.8in]{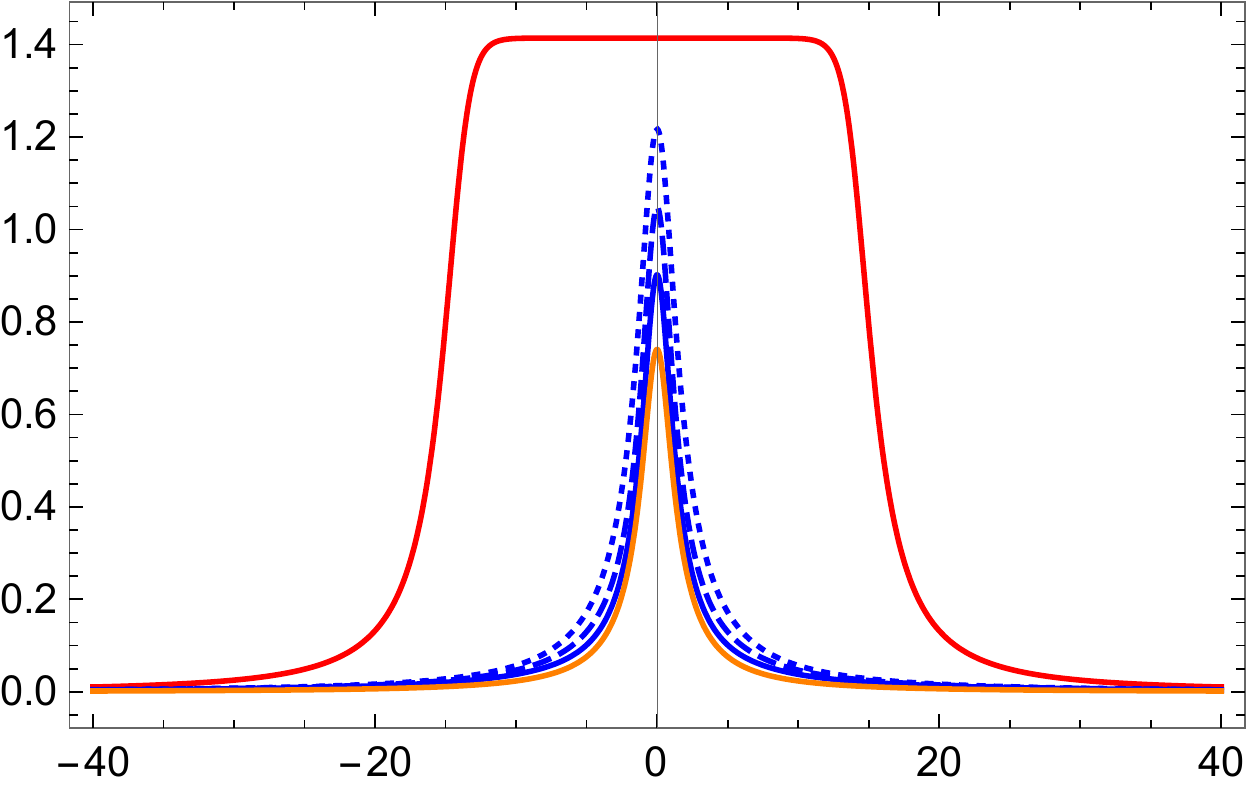}  \hspace{10mm} \includegraphics[width=2.8in]{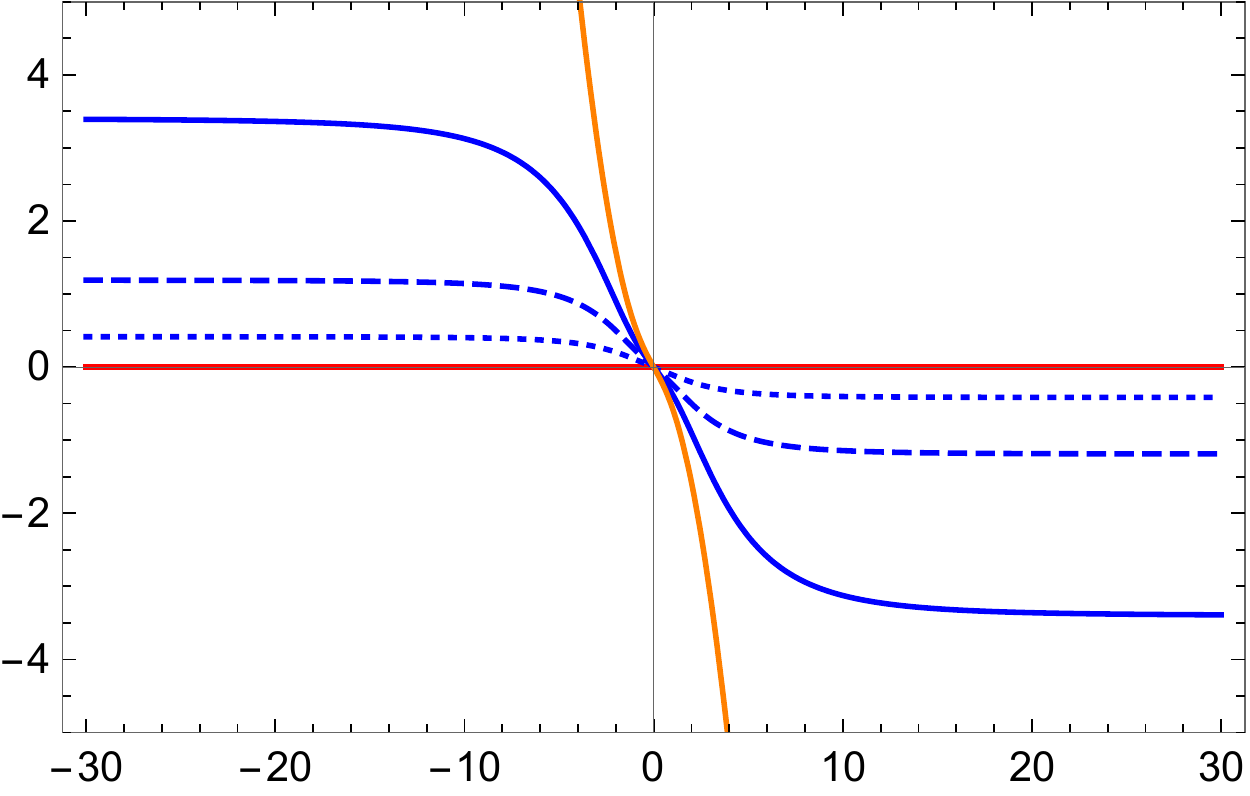} 
\put(-462,50){\rotatebox{90}{$\left(\textrm{Im}z_{1}\right)^{-1}$}}
\put(-218,57){\rotatebox{90}{$\textrm{Re}z_{1}$}}}
\\[4mm]
\scalebox{0.75}{\includegraphics[width=2.8in]{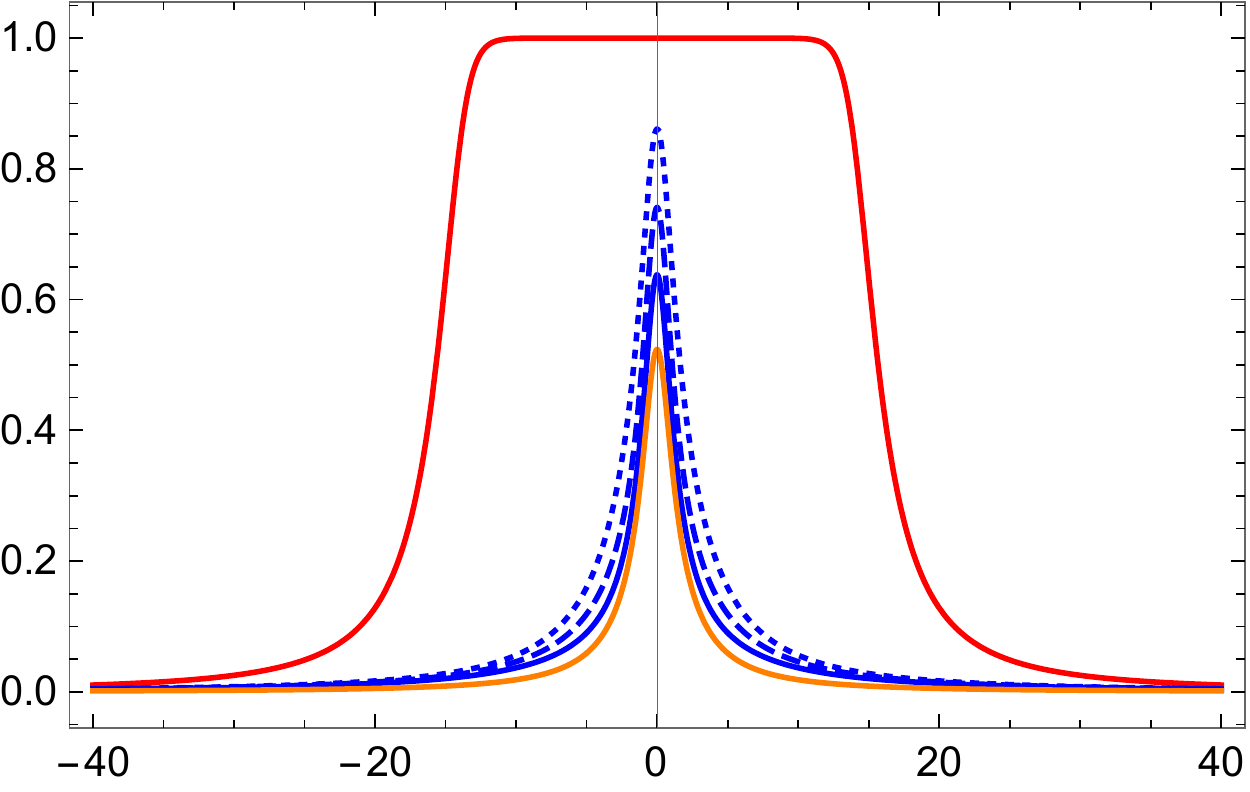}  \hspace{10mm} \includegraphics[width=2.8in]{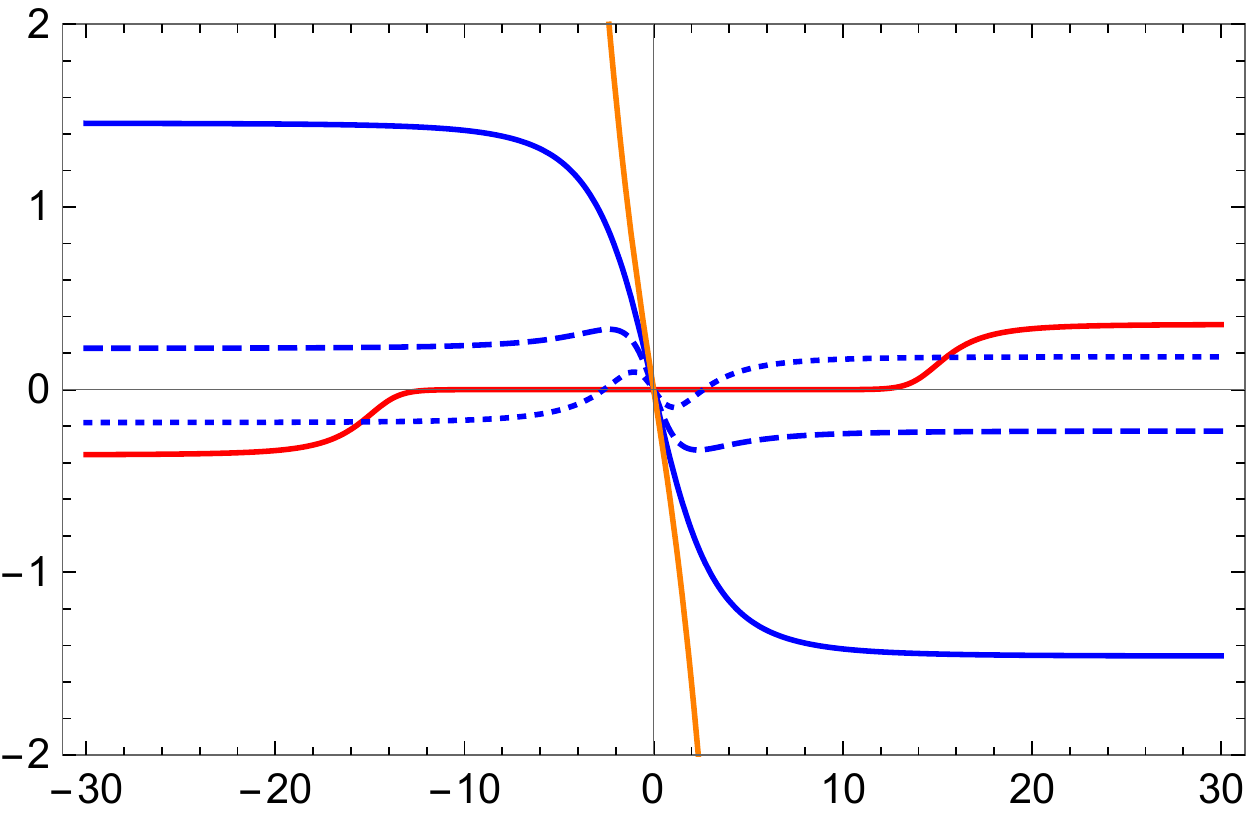} 
\put(-462,50){\rotatebox{90}{$\left(\textrm{Im}z_{2}\right)^{-1}$}}
\put(-218,57){\rotatebox{90}{$\textrm{Re}z_{2}$}}}
\\[4mm]
\scalebox{0.75}{\includegraphics[width=2.8in]{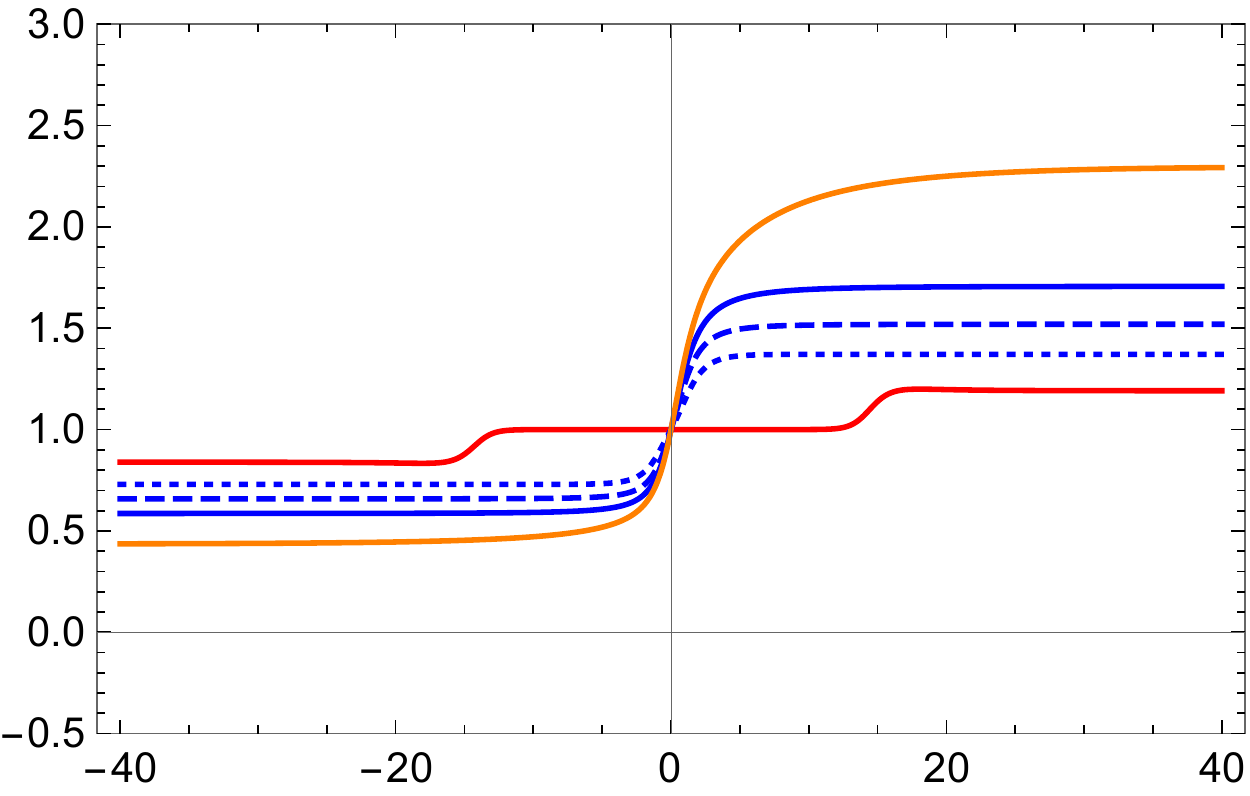}  \hspace{10mm} \includegraphics[width=2.8in]{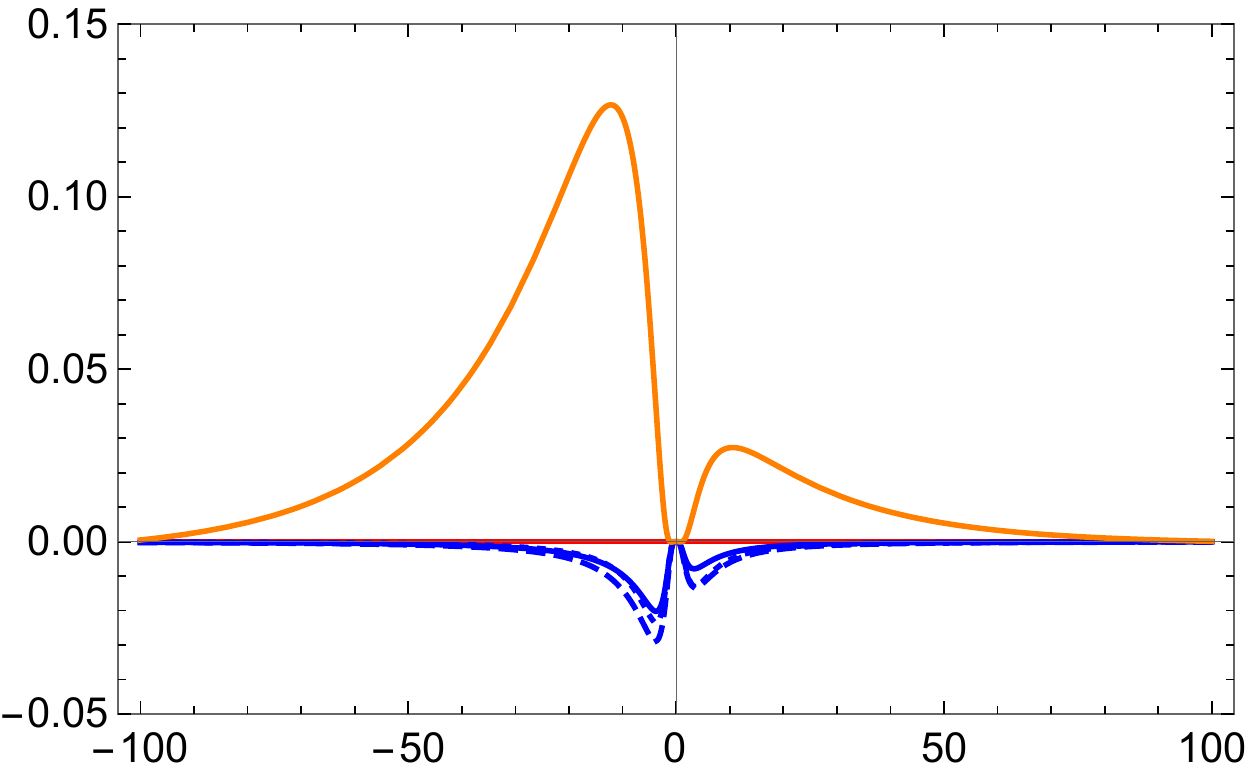} 
\put(-462,50){\rotatebox{90}{$\left(\textrm{Im}z_{4}\right)^{-1}$}}
\put(-218,57){\rotatebox{90}{$\textrm{Re}z_{4}$}}}
\\[2mm]
\scalebox{0.75}{\includegraphics[width=2.8in]{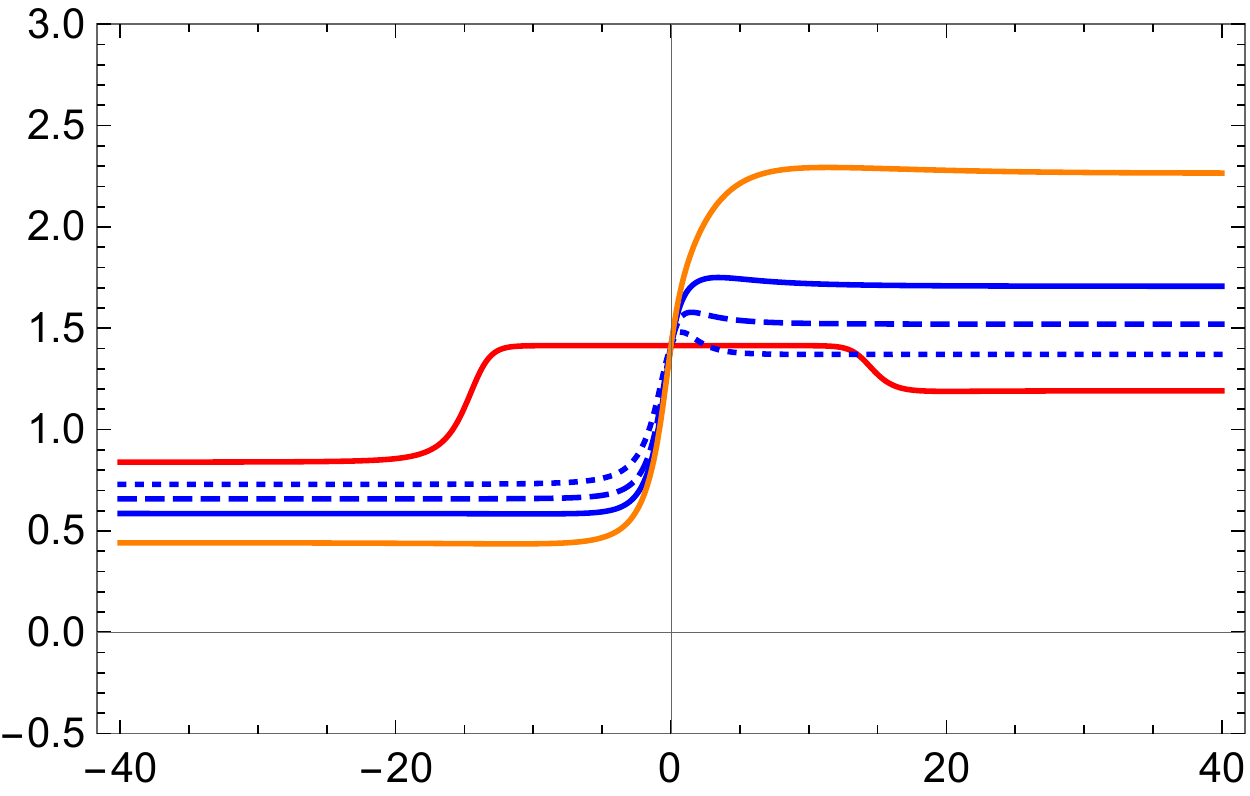}  \hspace{10mm} \includegraphics[width=2.8in]{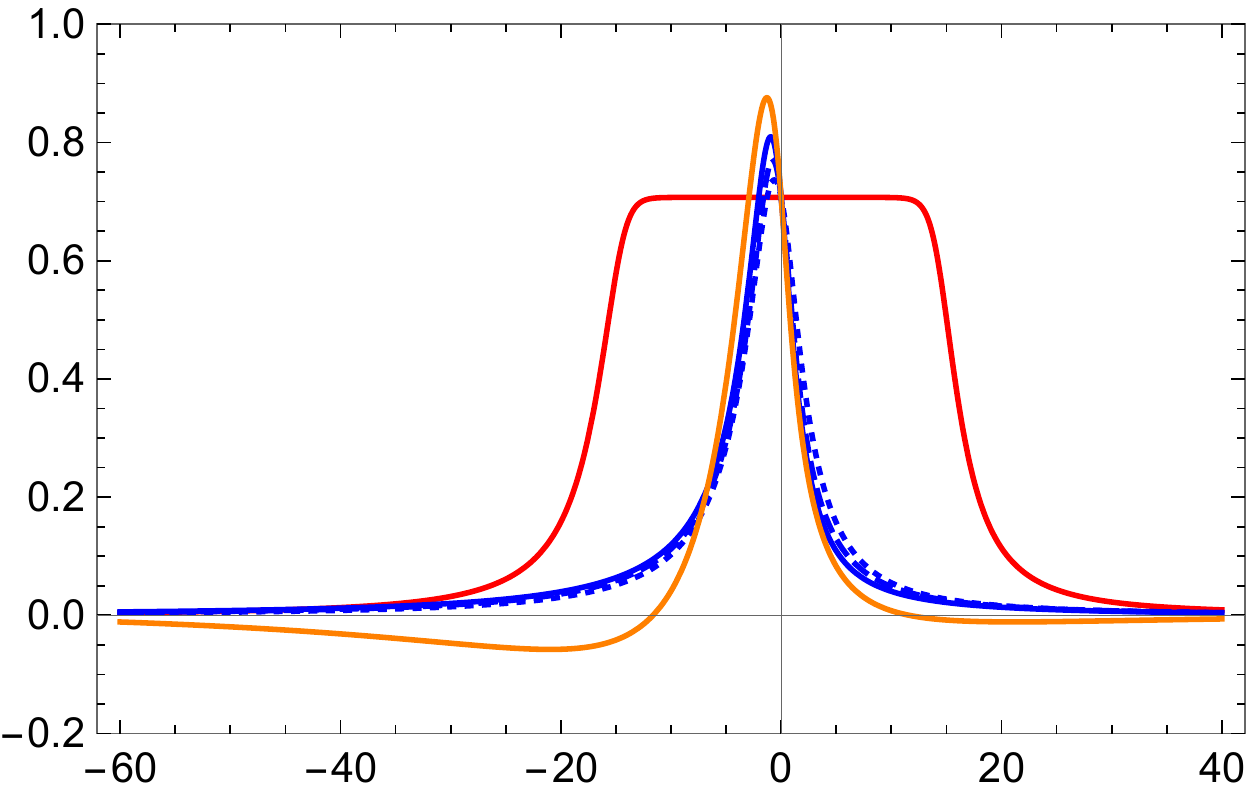} 
\put(-462,50){\rotatebox{90}{$\left(\textrm{Im}z_{5}\right)^{-1}$}}
\put(-218,57){\rotatebox{90}{$\textrm{Re}z_{5}$}}}
\\[4mm]
\scalebox{0.75}{\includegraphics[width=2.8in]{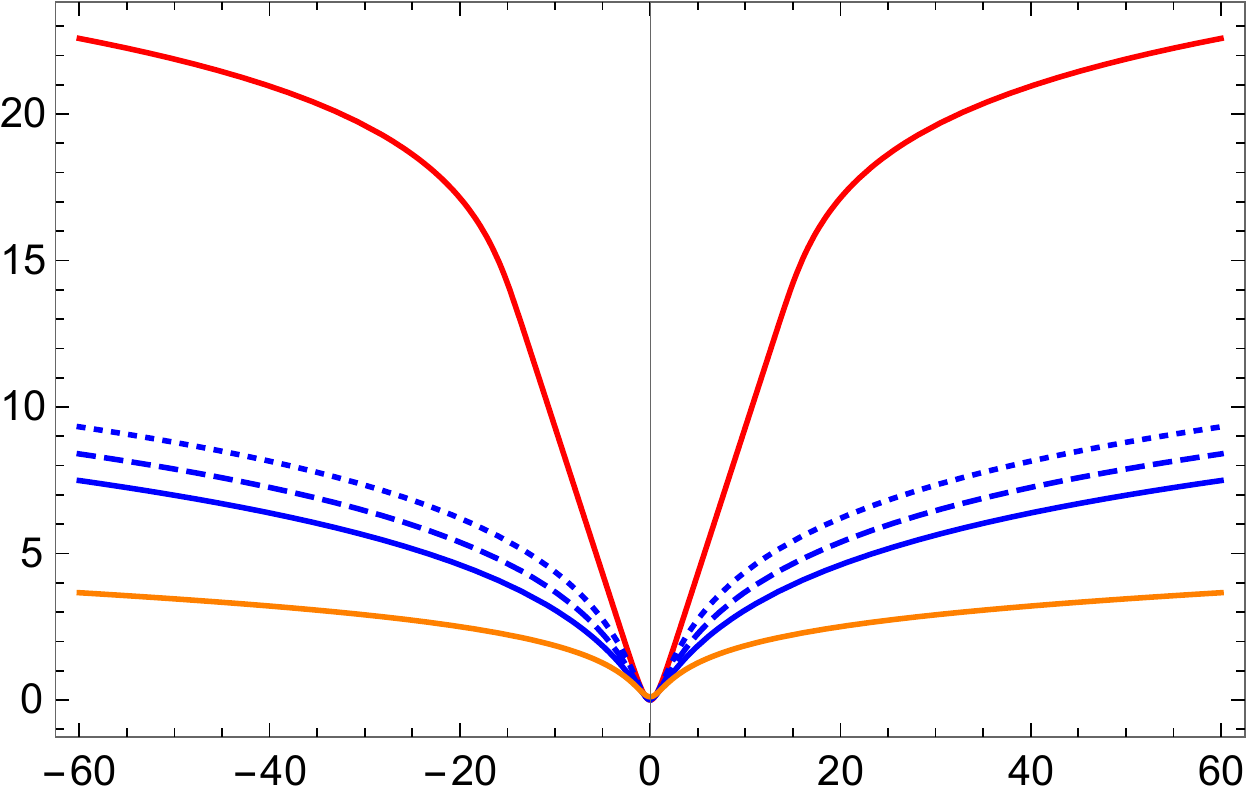}  \hspace{10mm} \includegraphics[width=2.8in]{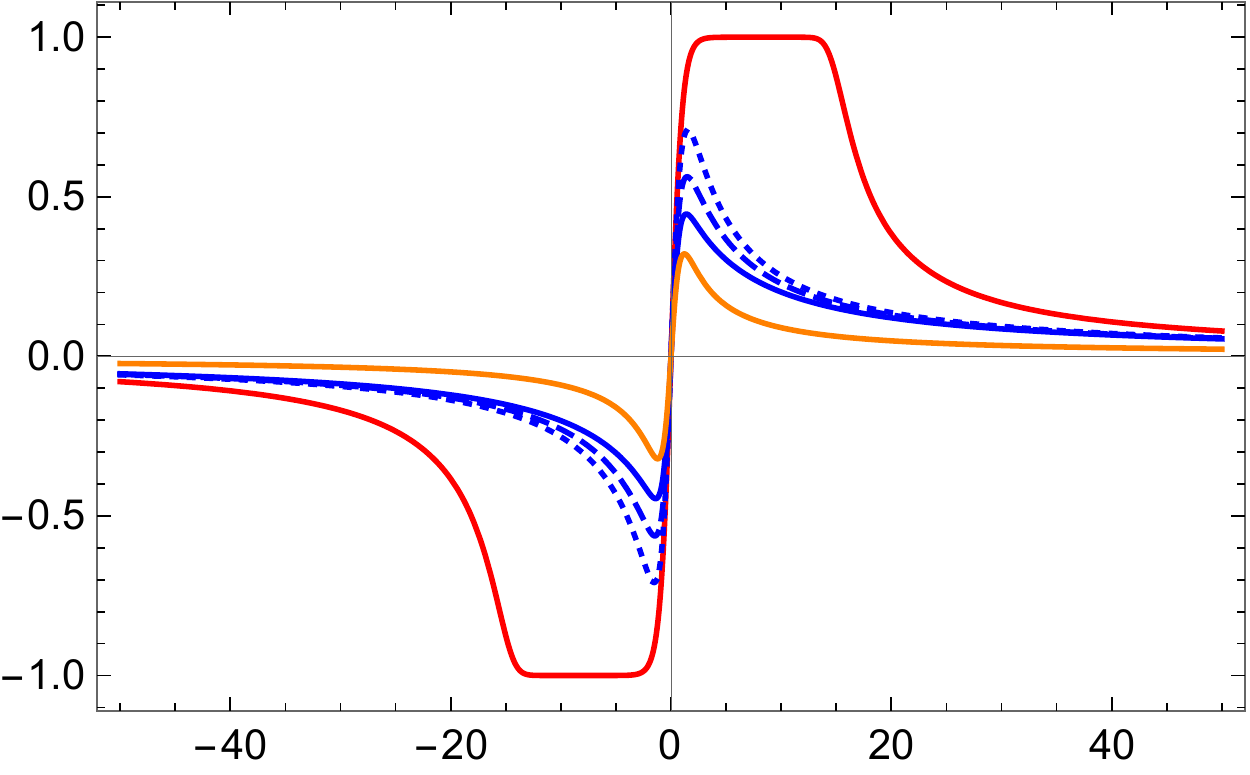} 
\put(-462,65){\rotatebox{0}{$A$}}
\put(-218,65){\rotatebox{0}{$A'$}}
\put(-339,-15){$r$}\put(-96,-15){$r$}}
\caption{{\it Samples of regular flows within the $SU(2)$-invariant sector using the boundary conditions (\ref{boundary_SU2_z123}) for the complex scalar fields $\,z_{1}\,$, $\,z_{2}\,$, $\,z_{4}\,$ and $\,z_{5}\,$.}}
\label{Fig:N=2_plots_2}
\end{center}
\end{figure}

\subsubsection*{Numerical study}

The $SU(2)$ invariant sector we are considering involves four complex scalars and, therefore, it is hard to perform an exhaustive study of numerical flows. Still the location of the $\mathcal{N}= 2\,\&\,U(2)$ $AdS_{4}$ vacuum and the D3-brane attractor point (see Figure~\ref{Fig:N=2_plots_1}) suggest again various natural choices of boundary conditions.

The first choice of boundary conditions consists in distributing the turning points of the numerical flows (pink points in Figure~\ref{Fig:N=2_plots_1}) along the $\,\textrm{Re}\tilde{z}_{1}\,$ and $\,\textrm{Re}\tilde{z}_{2}\,$ axes, while keeping $\,\tilde{z}_{4}\,$ and $\,\tilde{z}_{5}\,$ fixed at their values in the $\mathcal{N}= 2\,\&\,U(2)$ $AdS_{4}$ vacuum. Moreover, the identification $\,z_{1,2,3}\,$ in the D3-brane solution (\ref{D3-brane_4D_scalars}) further motivates the symmetric choice
\begin{equation}
\label{boundary_SU2_z123}
z_{1}(0) = -\chi_{1}^{(*)} + i \,  e^{- \left( \varphi_{1}^{(*)}  - \,  \epsilon \right)}
\,\,\,\,\, , \,\,\,\,\, 
z_{2}(0) =  -\chi_{2}^{(*)} + i \,  e^{- \left( \varphi_{2}^{(*)}  - \,  \epsilon \right)}
\,\,\,\,\, , \,\,\,\,\, 
z_{4}(0) =  z_{4}^{(*)} 
\,\,\,\,\, , \,\,\,\,\, 
z_{5}(0) = z_{5}^{(*)}  \ ,
\end{equation}
where the asterisk $\,^{(*)}\,$ denotes again the expectation values of the scalars at the $\,\mathcal{N}= 2\,\&\,U(2)\,$ $AdS_{4}$ vacuum. As before, regular flows only exist for $\,0 < \epsilon <  \epsilon_{\textrm{crit}}\,$ with $\,\epsilon_{\textrm{crit}} \approx 0.64778 \,$. The profiles for the fields are displayed in Figure~\ref{Fig:N=2_plots_2}. We find again two limiting/bounding flows associated with the boundary values of the parameter $\,\epsilon \approx 0\,$ and $\,\epsilon \approx \epsilon_{\textrm{crit}}\,$, which respectively correspond to the solid red and orange lines in Figure \ref{Fig:N=2_plots_1} and Figure~\ref{Fig:N=2_plots_2}. The flow with $\,\epsilon \approx 0\,$ (red solid lines) develops an $AdS_{4}$ intermediate behaviour with constant scalars around the turning point at $\,r=0$. This $AdS_{4}$ regime uplifts to the $\mathcal{N}= 2\,\&\,U(2)$ type IIB S-fold of \cite{Guarino:2020gfe}. In addition, we observe the presence of a crossing point with $\,\textrm{Im}\tilde{z}_{2}=\textrm{Re}z_{2}=0\,$ at finite $\,r = \pm r_{0}\,$ in all the flows having $\, 0 <\epsilon < \epsilon_{*}\,$ with $\,\epsilon_{*} \approx 0.23428\,$. An example of this is the flow with $\,\epsilon=0.15 < \epsilon_{*}\,$ (blue dotted line) depicted in Figure \ref{Fig:N=2_plots_1} and Figure~\ref{Fig:N=2_plots_2}.\footnote{A crossing point is also present in the red flow with $\,\epsilon \approx 0\,$ although it cannot be appreciated in the figures due to its small size and the fact that $\,r_{0} \rightarrow \pm 0\,$ in this limiting case.} On the contrary, note that, for example, the crossing point is not present in the flow with $\,\epsilon=0.30 > \epsilon_{*}\,$ (blue dashed line) in the same figures. Lastly, the flow at the special value $\,\epsilon \approx \epsilon_{*}\,$ has $\,\textrm{Im}\tilde{z}_{2}=\textrm{Re}z_{2}=0\,$ at $\,r_{0} \rightarrow \pm\infty$, namely, the crossing point has been pushed to the endpoints of the flow.

\begin{figure}[t!]
\begin{center}
\scalebox{0.8}{\includegraphics[width=2.9in]{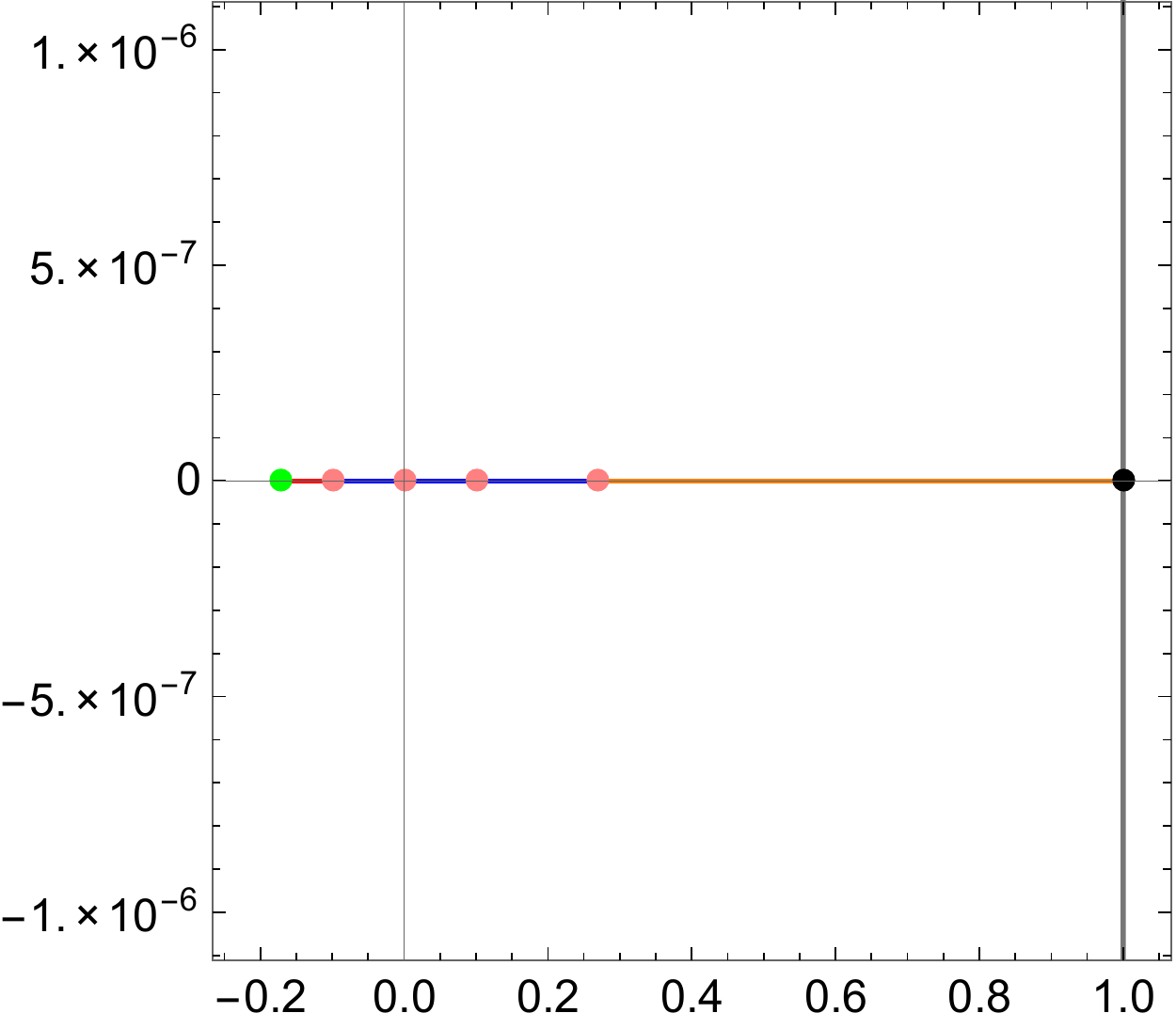} \hspace{10mm} \includegraphics[width=2.6in]{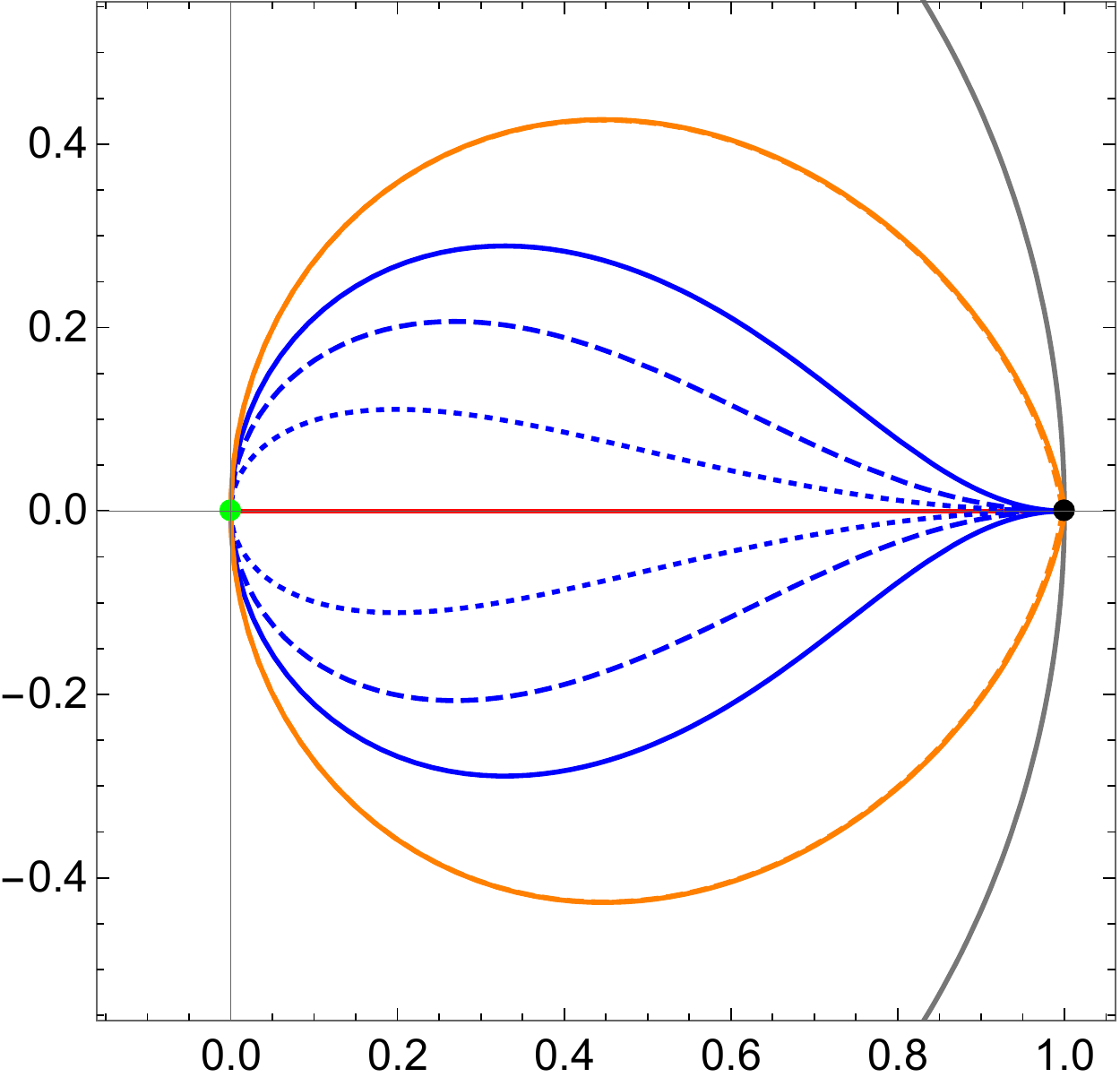}
\put(-315,-15){$\tilde{z}_{1}$}\put(-90,-15){$\tilde{z}_{2}$}} \\[2mm]
\scalebox{0.8}{\includegraphics[width=2.9in]{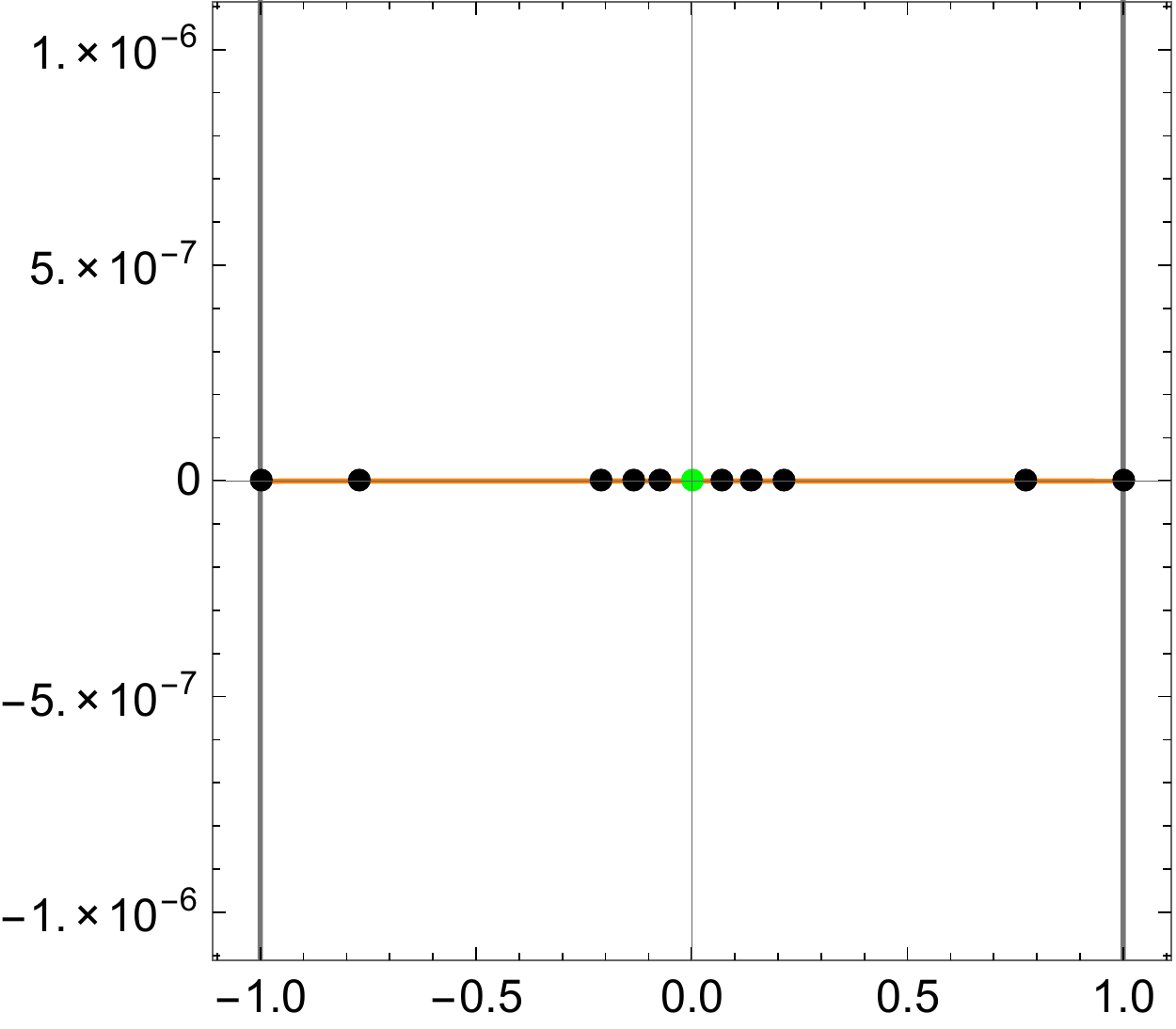} \hspace{10mm} \includegraphics[width=2.6in]{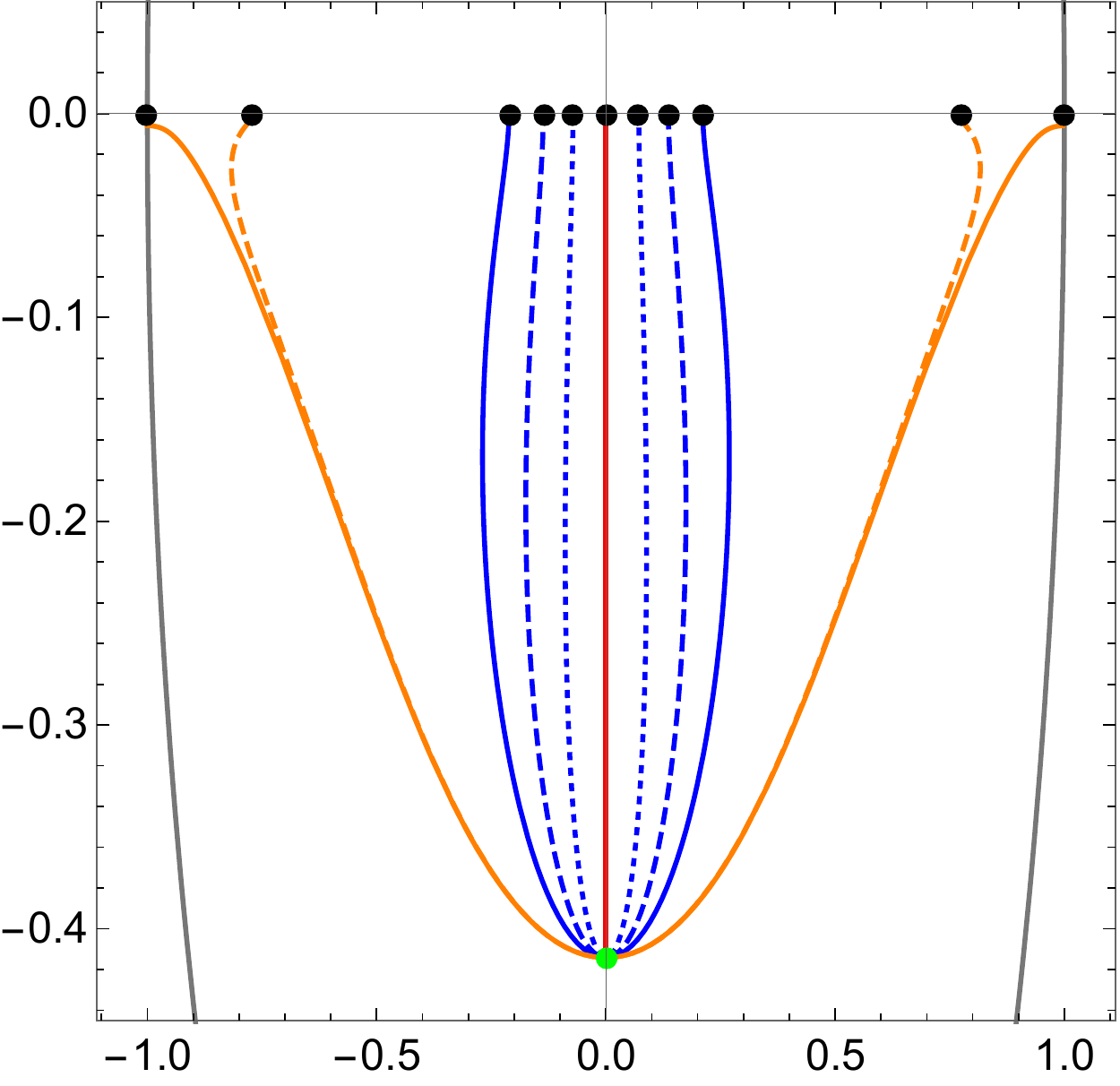}
\put(-315,-15){$\tilde{z}_{4}$}\put(-90,-15){$\tilde{z}_{5}$}}
\caption{{\it Samples of regular flows within the $SU(2)$-invariant sector using the boundary conditions (\ref{boundary_SU2_z13}) for the complex scalar fields $\,\tilde{z}_{1}\,$ (top-left), $\,\tilde{z}_{2}\,$ (top-right), $\,\tilde{z}_{4}\,$ (bottom-left) and $\,\tilde{z}_{5}\,$ (bottom-right). The $\mathcal{N}= 2\,\&\,U(2)$ $AdS_{4}$ vacuum and the D3-brane asymptotics are denoted by green and black dots, respectively. The grey lines in the left column plots correspond to the boundary values $\,|\tilde{z}_{1}|=1\,$. The red flow corresponds to the choice $\,\epsilon \approx 0\,$ whereas the orange one corresponds to $\,\epsilon \approx \epsilon_{\textrm{crit}}\,$. Black dots in the bottom plots are compatible with SO(6) invariance, \textit{i.e.} $\,\tilde{z}_{4}=\tilde{z}_{5}\,$ are real-valued, and correspond to the values $\,\pm\Phi_{0}\,$ of the type IIB dilaton in the D3-brane asymptotics at $\,r \rightarrow \pm \infty\,$.}}
\label{Fig:N=2_plots_3}
\end{center}
\end{figure}

One could also think of relaxing the $\,z_{1,3}(0)=z_{2}(0)\,$ symmetry of the boundary conditions in (\ref{boundary_SU2_z123}). For instance, we can distribute the turning points of the numerical flows (pink points in Figure~\ref{Fig:N=2_plots_3}) only along the $\,\textrm{Re}\tilde{z}_{1}\,$ real axis while keeping $\,\tilde{z}_{2}\,$, $\,\tilde{z}_{4}\,$ and $\,\tilde{z}_{5}\,$ fixed at their values in the $AdS_{4}$ vacuum. This is
\begin{equation}
\label{boundary_SU2_z13}
\tilde{z}_{1}(0) = \tilde{z}_{1}^{(*)} + \epsilon
\quad , \quad 
\tilde{z}_{2}(0) = \tilde{z}_{2}^{(*)}
\quad , \quad 
\tilde{z}_{4}(0) = \tilde{z}_{4}^{(*)}
\quad , \quad 
\tilde{z}_{5}(0) = \tilde{z}_{5}^{(*)} \ .
\end{equation}
This choice of boundary conditions produces regular flows only within the range $\,0 < \epsilon <  \epsilon_{\textrm{crit}}\,$ with $\,\epsilon_{\textrm{crit}} \approx 0.44262 \,$. Furthermore, as it can be seen from Figure~\ref{Fig:N=2_plots_3}, the boundary conditions  (\ref{boundary_SU2_z13}) localise the flows in the real axis both for $\,\tilde{z}_{1}\,$ and $\,\tilde{z}_{4}\,$. In our numerical scanning, we could not find regular flows with boundary conditions different from (\ref{boundary_SU2_z123}) and (\ref{boundary_SU2_z13}).

\subsection{$SO(3)$-invariant flows}

\begin{figure}[t!]
\begin{center}
\includegraphics[width=2in]{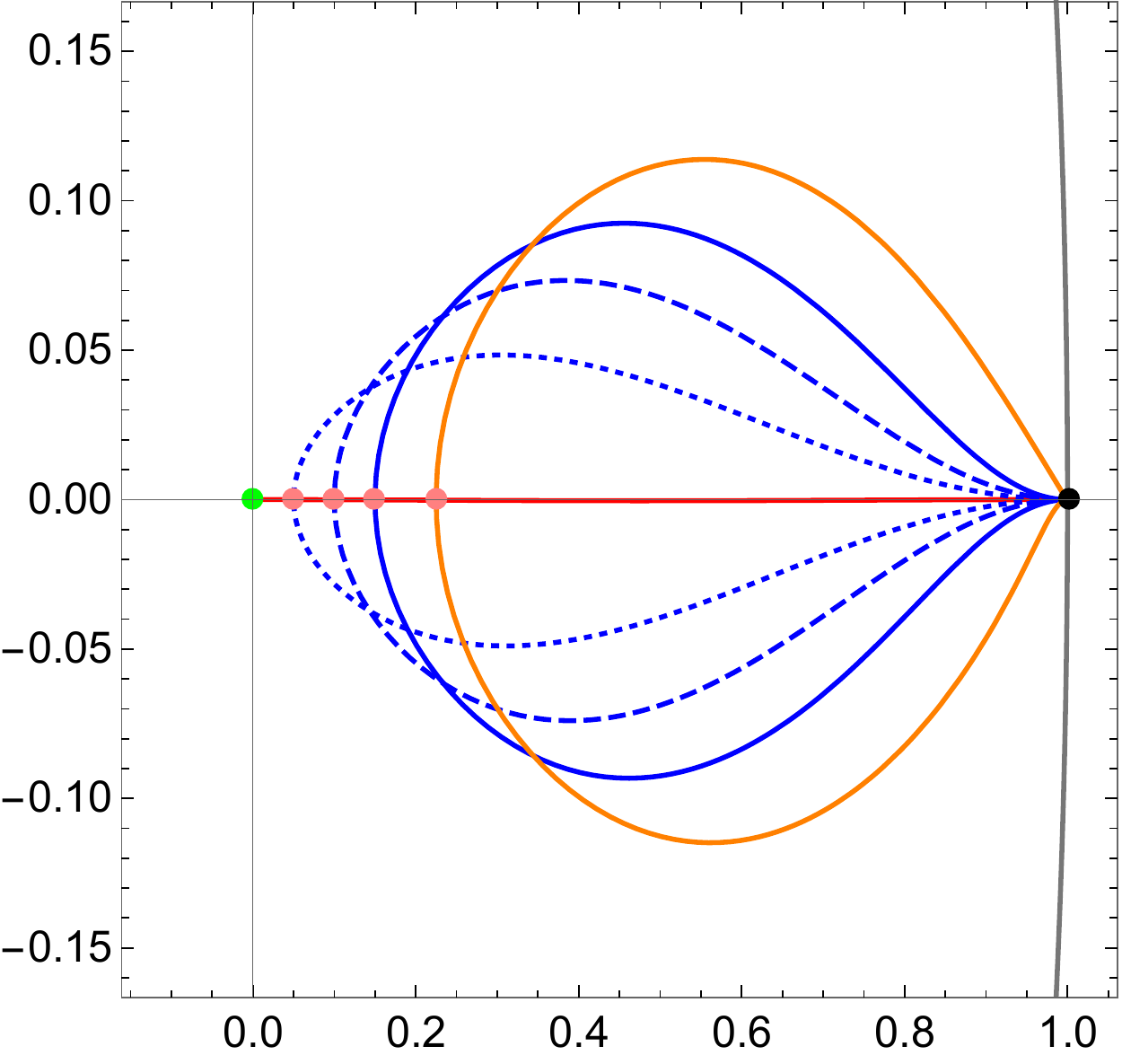} \hspace{5mm} \includegraphics[width=2in]{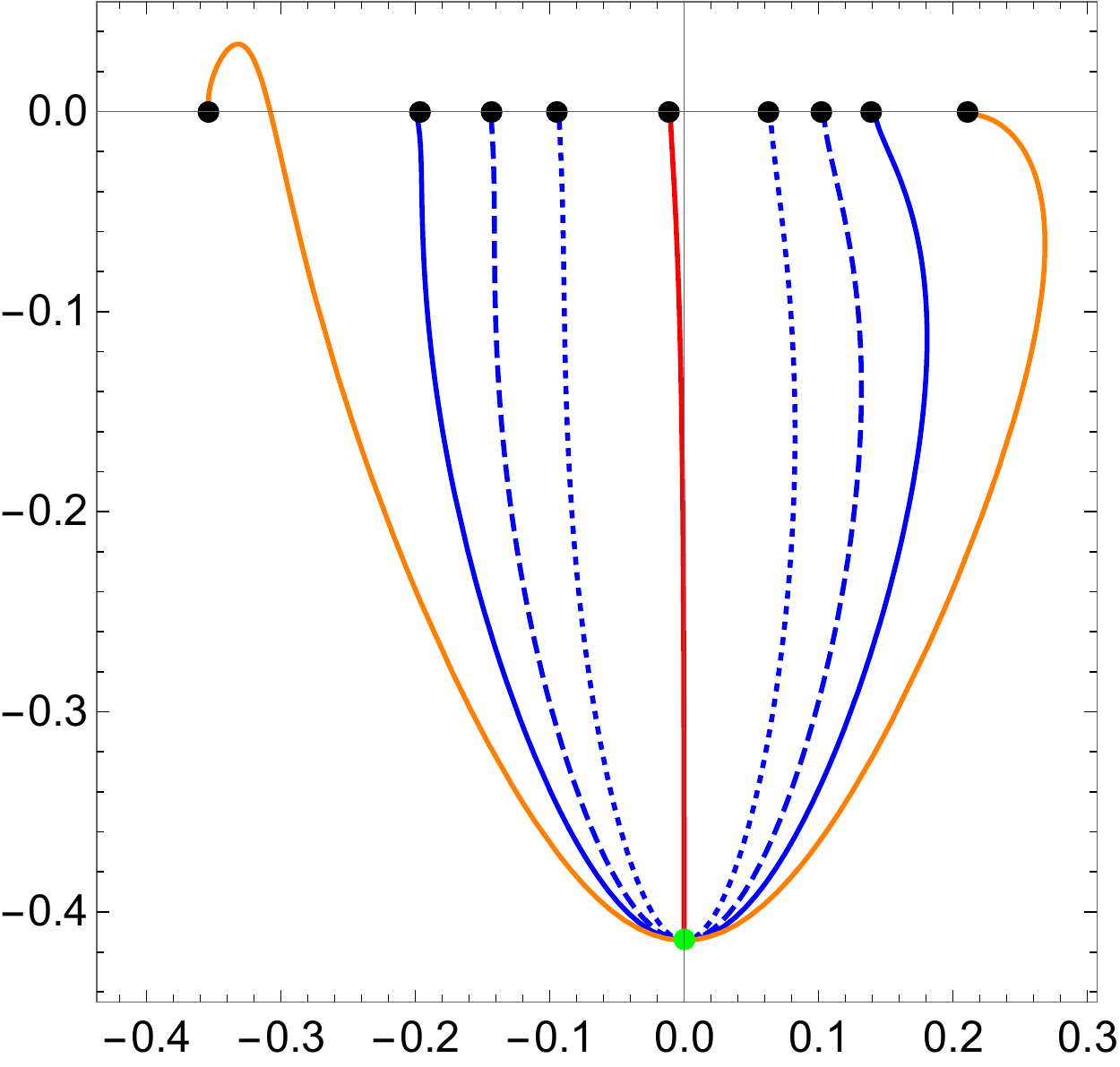} \hspace{5mm} \includegraphics[width=2in]{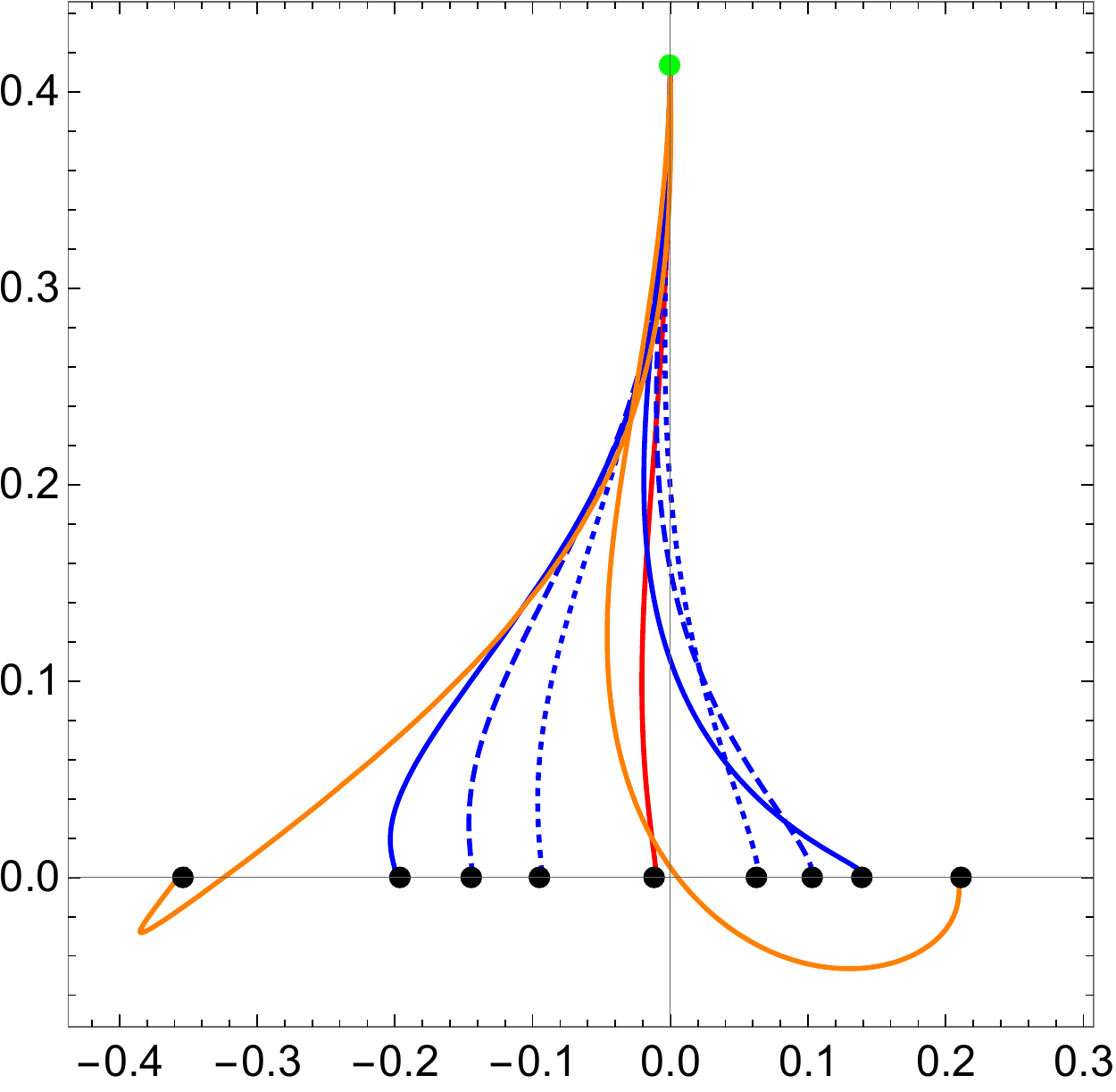}
\put(-400,-15){$\tilde{z}_{1}$}\put(-237,-15){$\tilde{z}_{4}$}\put(-75,-15){$\tilde{z}_{7}$}
\caption{{\it Samples of regular flows within the $SO(3)$-invariant sector using the boundary conditions (\ref{boundary_SO3_z123}) for the complex scalar fields $\,\tilde{z}_{1}\,$ (left), $\,\tilde{z}_{4}\,$ (middle) and $\,\tilde{z}_{7}\,$ (right). The $\mathcal{N}= 4\,\&\,SO(4)$ $AdS_{4}$ vacuum and the D3-brane asymptotics are denoted by green and black dots, respectively. The grey line in the left plot corresponds to the boundary values $\,|\tilde{z}_{1}|=1\,$. The red flow corresponds to the choice $\,\epsilon \approx 0\,$ whereas the orange one corresponds to $\,\epsilon \approx \epsilon_{\textrm{crit}}\,$. Black dots in the middle and right plots are compatible with SO(6) invariance, \textit{i.e.} $\,\tilde{z}_{4}=\tilde{z}_{7}\,$ are real-valued, and correspond to the values $\,\pm\Phi_{0}\,$ of the type IIB dilaton in the D3-brane asymptotics at $\,r \rightarrow \pm \infty\,$.}}
\label{Fig:N=4_plots_1}
\end{center}
\end{figure}

\begin{figure}[t!]
\begin{center}
\scalebox{0.8}{
\includegraphics[width=2.8in]{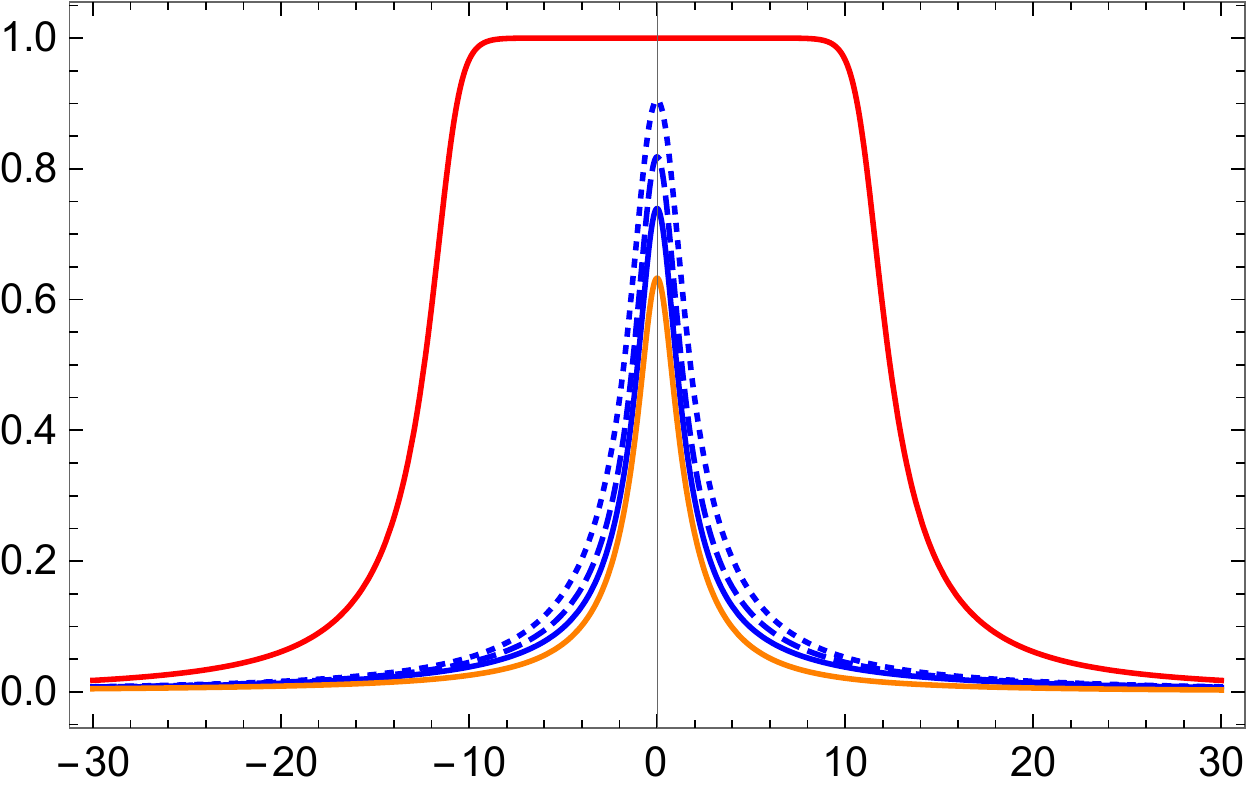}  \hspace{10mm} \includegraphics[width=2.8in]{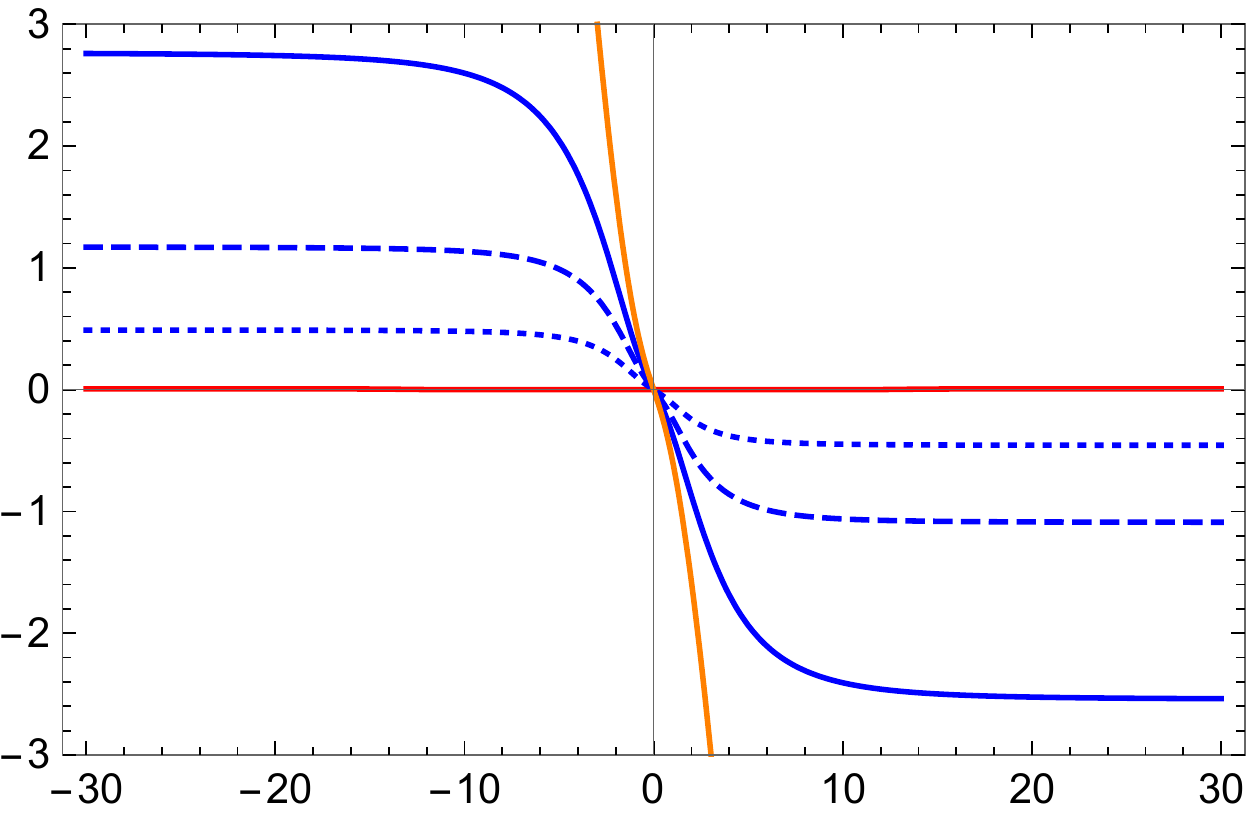} 
\put(-462,50){\rotatebox{90}{$\left(\textrm{Im}z_{1}\right)^{-1}$}}
\put(-218,57){\rotatebox{90}{$\textrm{Re}z_{1}$}}}
\\[4mm]
\scalebox{0.8}{\includegraphics[width=2.8in]{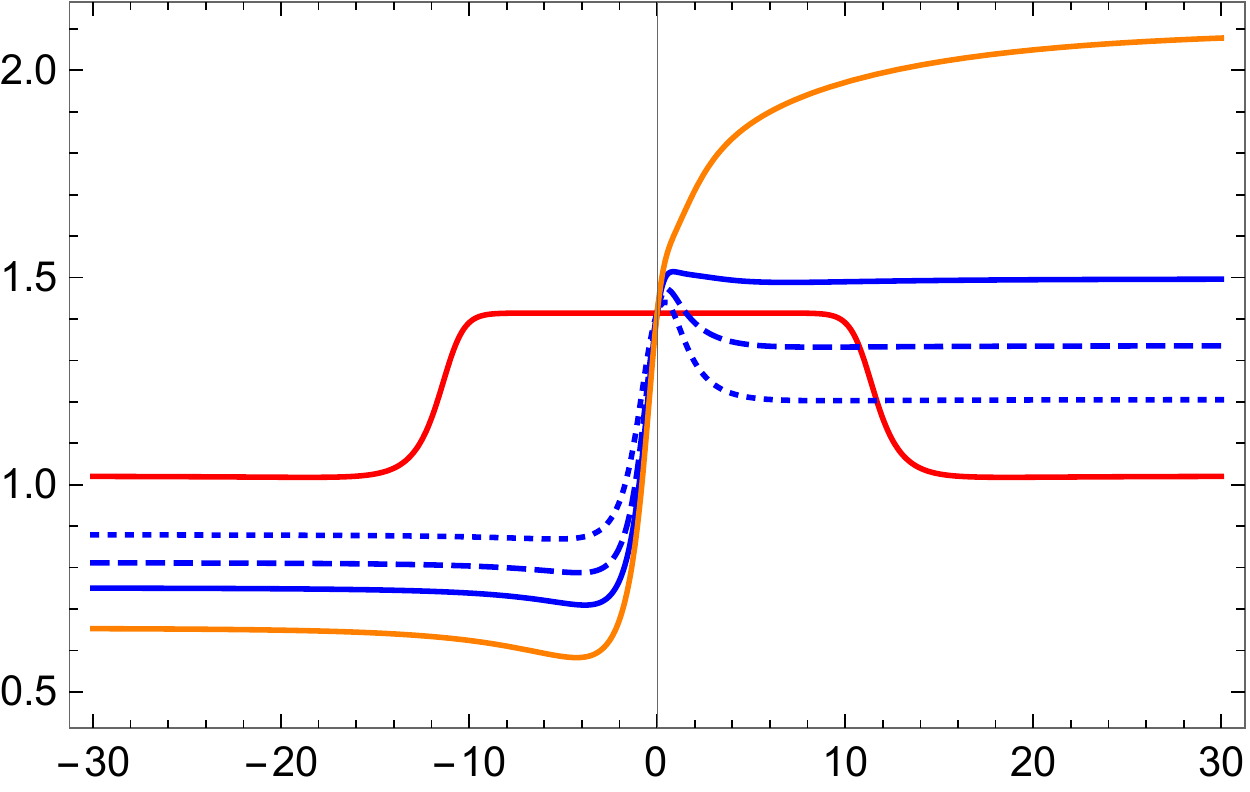}  \hspace{10mm} \includegraphics[width=2.8in]{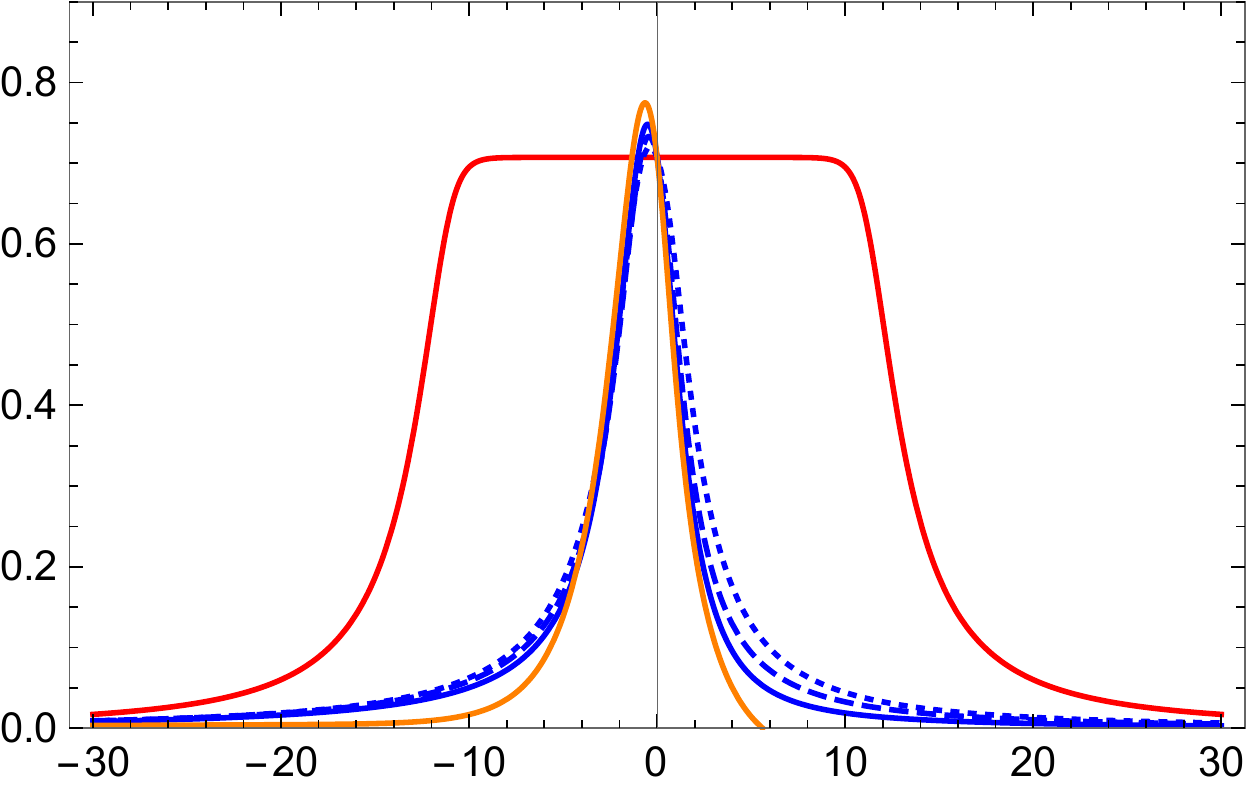} 
\put(-462,50){\rotatebox{90}{$\left(\textrm{Im}z_{4}\right)^{-1}$}}
\put(-218,57){\rotatebox{90}{$\textrm{Re}z_{4}$}}}
\\[4mm]
\scalebox{0.8}{\includegraphics[width=2.8in]{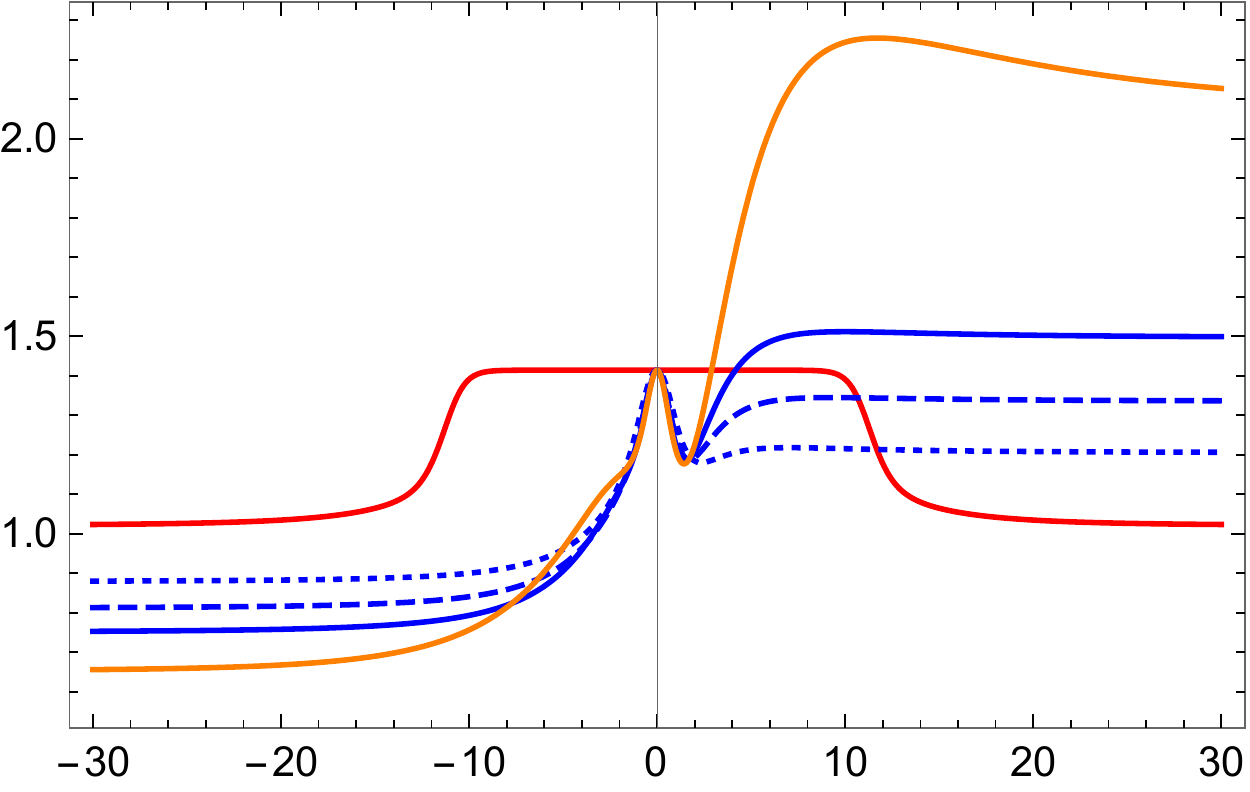}  \hspace{10mm} \includegraphics[width=2.8in]{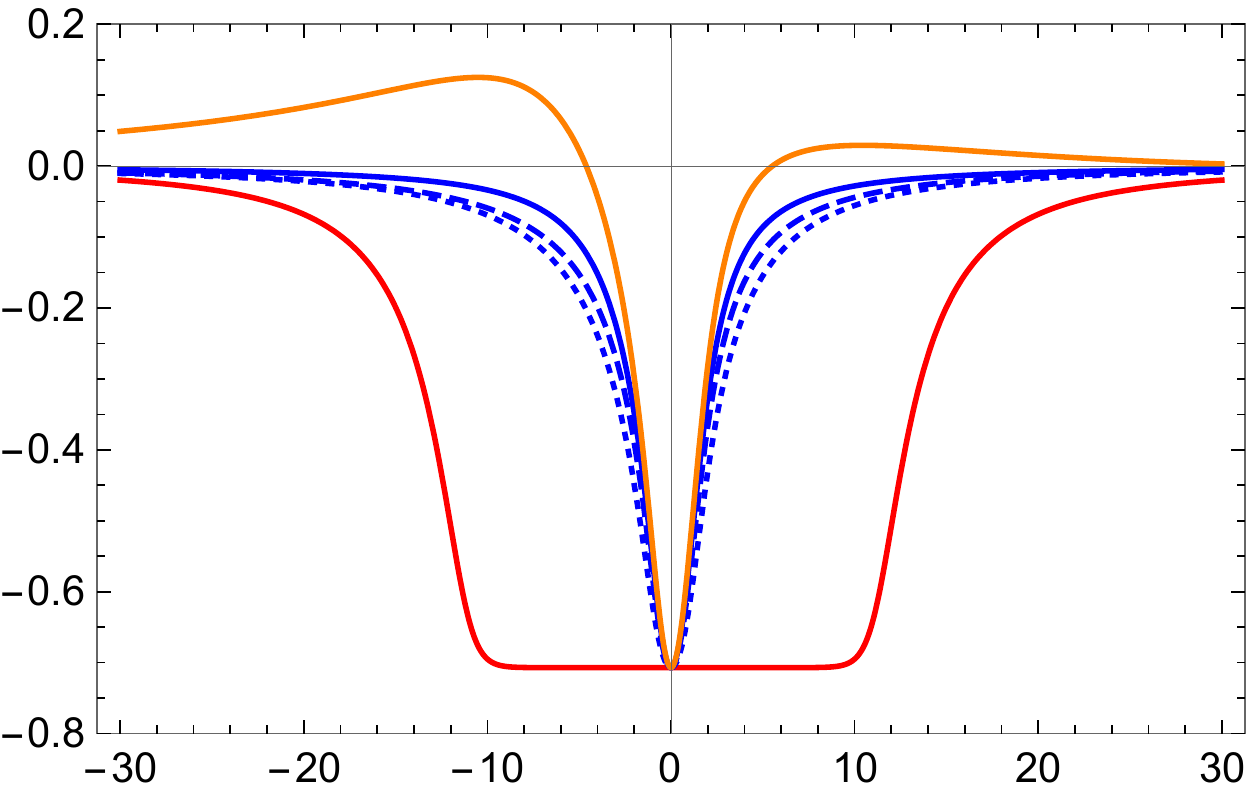} 
\put(-462,50){\rotatebox{90}{$\left(\textrm{Im}z_{7}\right)^{-1}$}}
\put(-218,57){\rotatebox{90}{$\textrm{Re}z_{7}$}}}
\\[4mm]
\scalebox{0.8}{\includegraphics[width=2.8in]{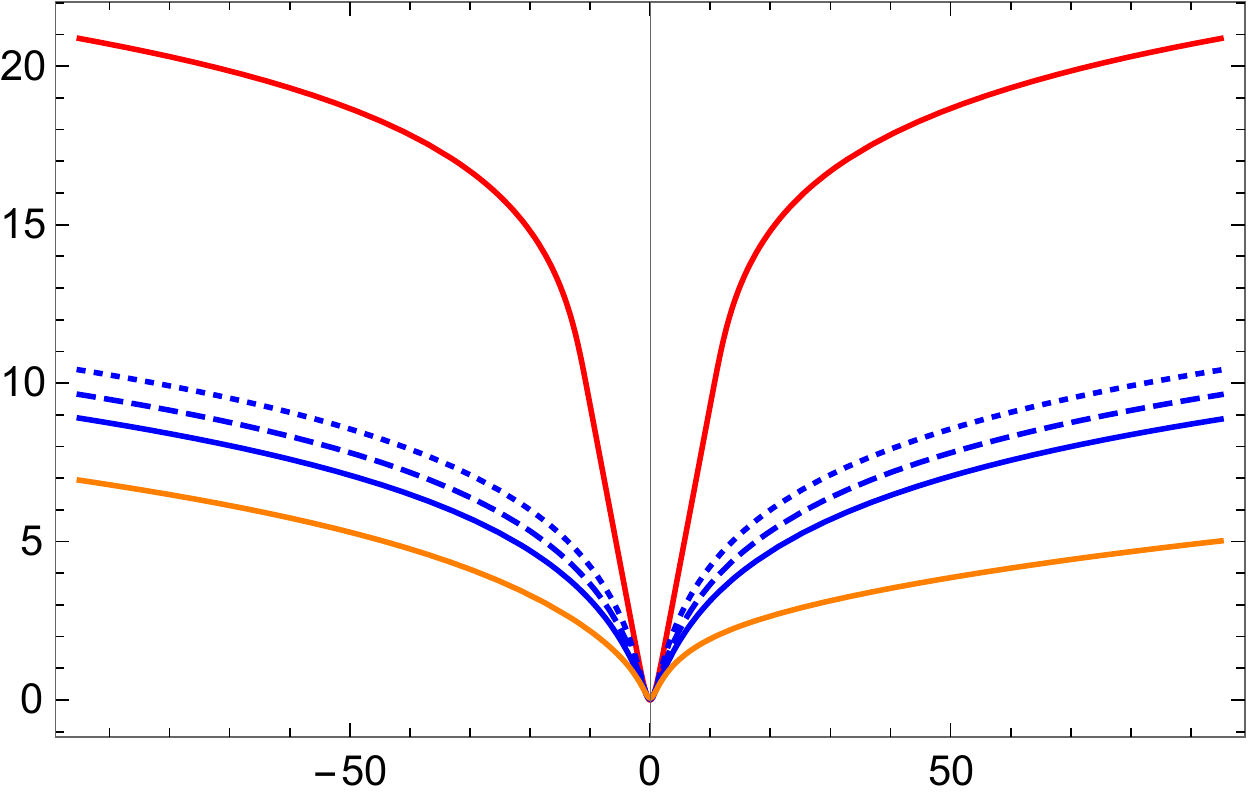}  \hspace{10mm} \includegraphics[width=2.8in]{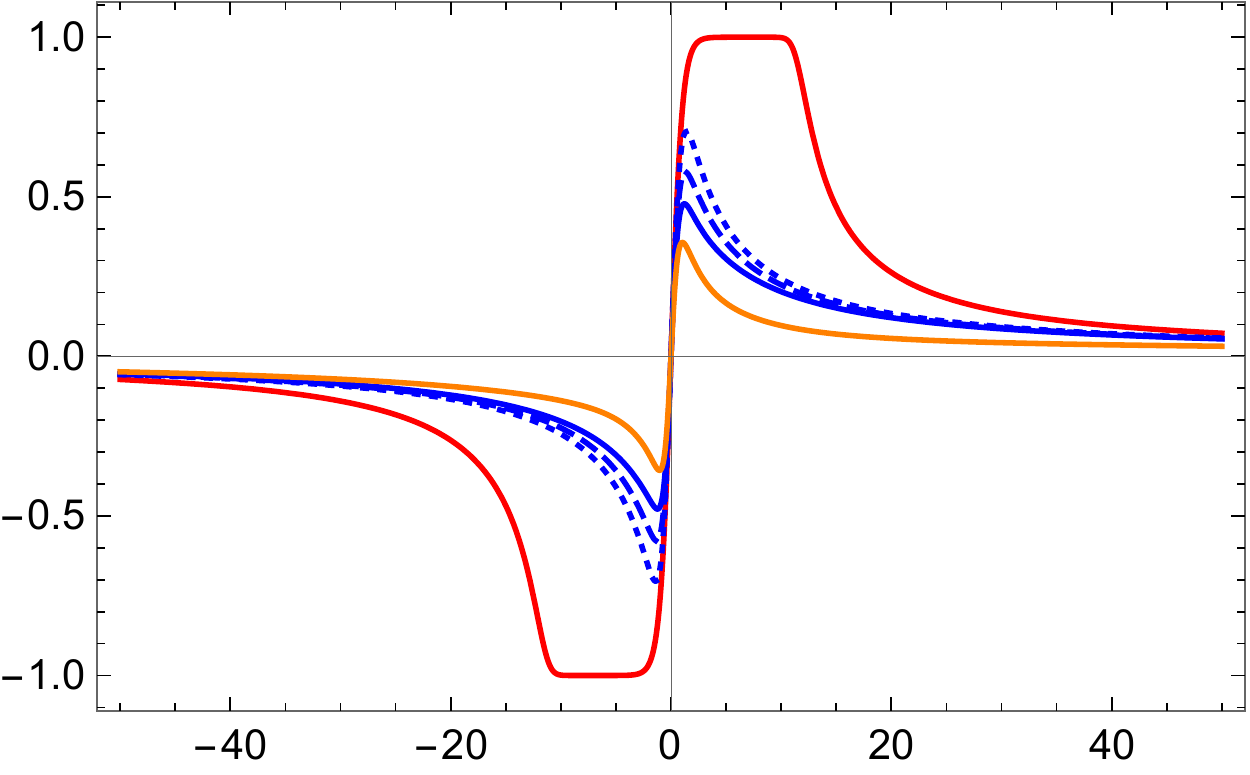} 
\put(-462,65){\rotatebox{0}{$A$}}
\put(-218,65){\rotatebox{0}{$A'$}}
\put(-339,-15){$r$}\put(-96,-15){$r$}}
\caption{{\it Samples of regular flows within the $SO(3)$-invariant sector using the boundary conditions (\ref{boundary_SO3_z123}) for the complex scalar fields $\,z_{1}\,$, $\,z_{4}\,$ and $\,z_{7}\,$.}}
\label{Fig:N=4_plots_2}
\end{center}
\end{figure}

\begin{figure}[t!]
\begin{center}
\includegraphics[width=2in]{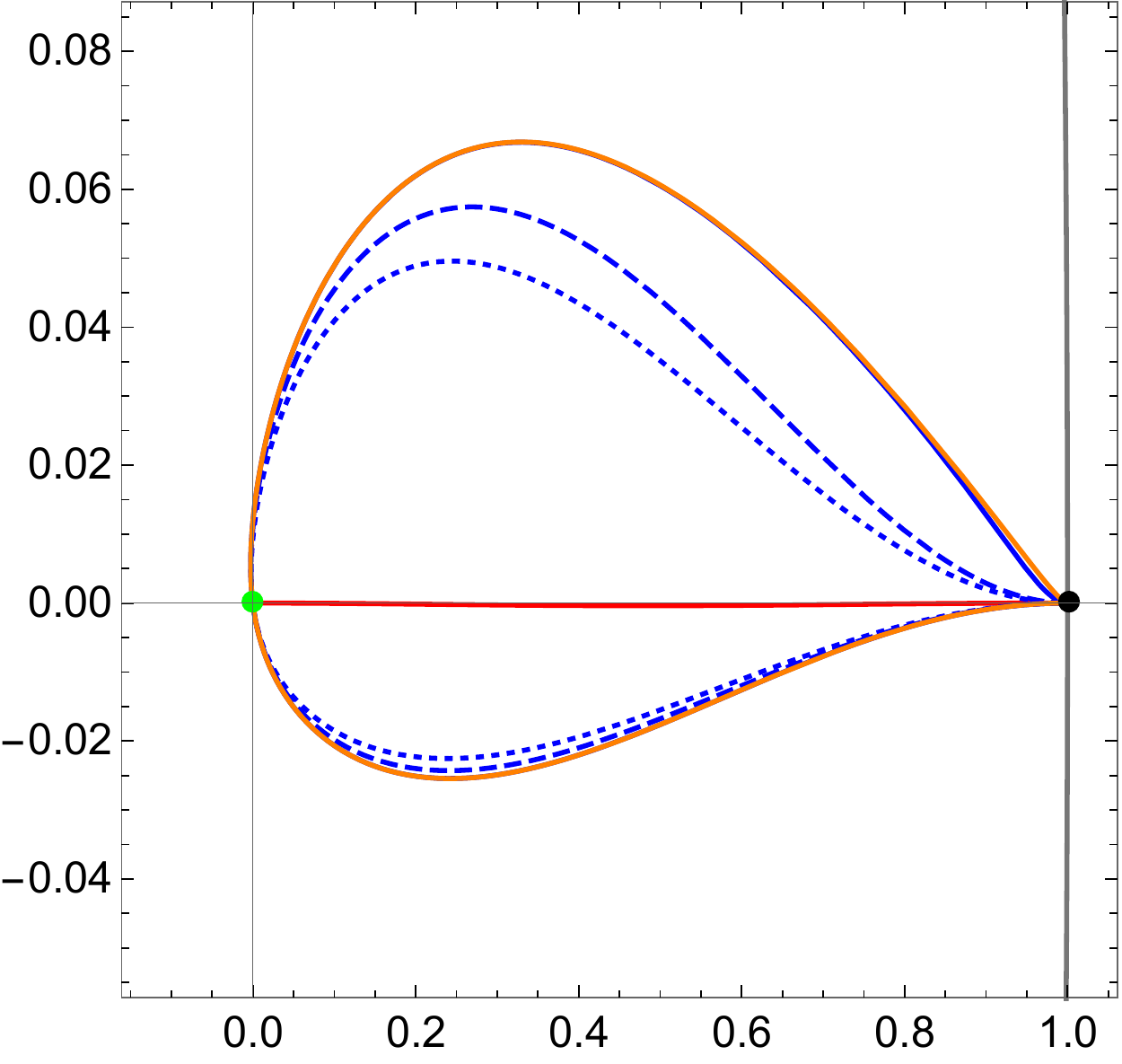} \hspace{5mm} \includegraphics[width=2in]{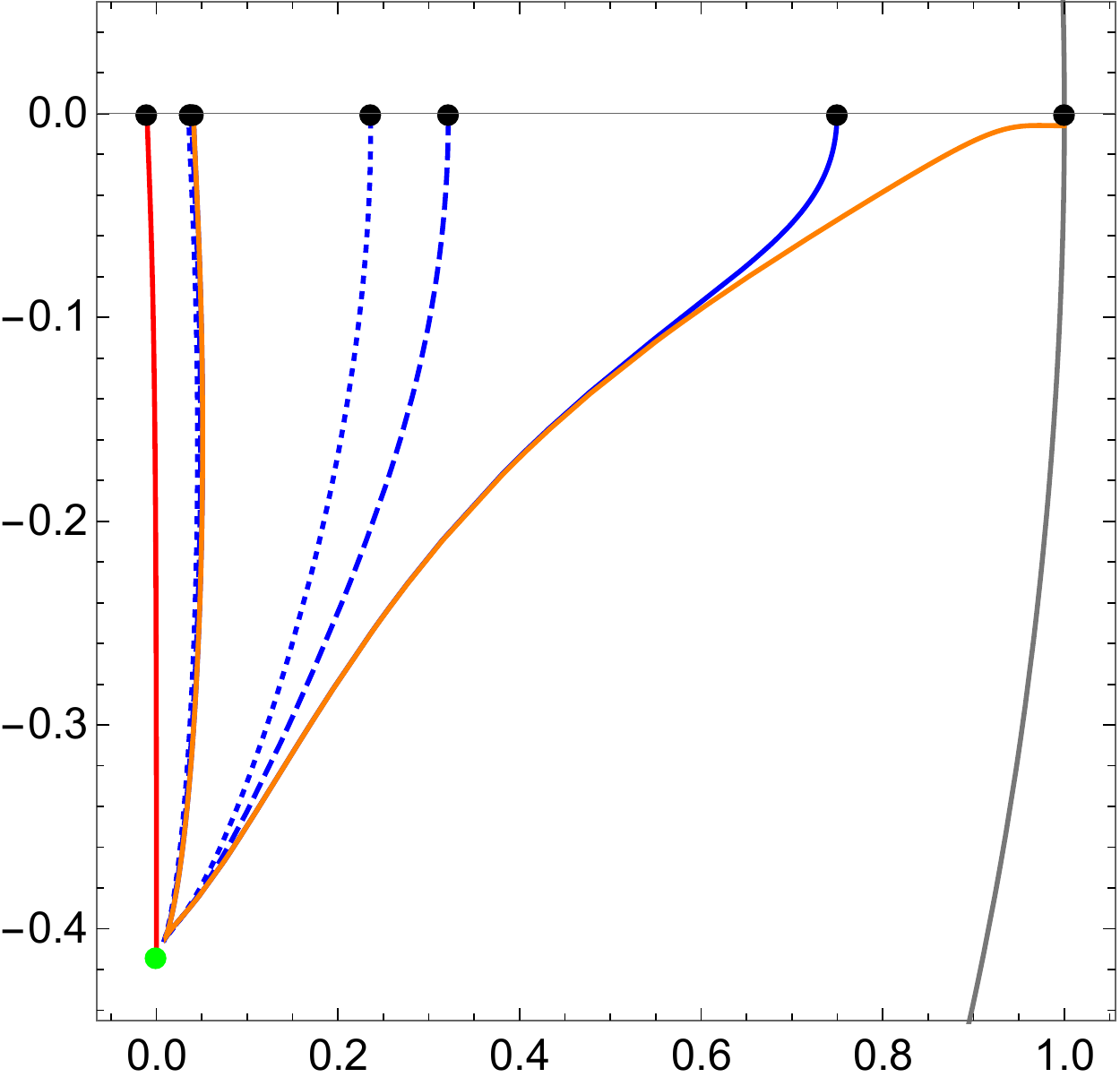} \hspace{5mm} \includegraphics[width=2in]{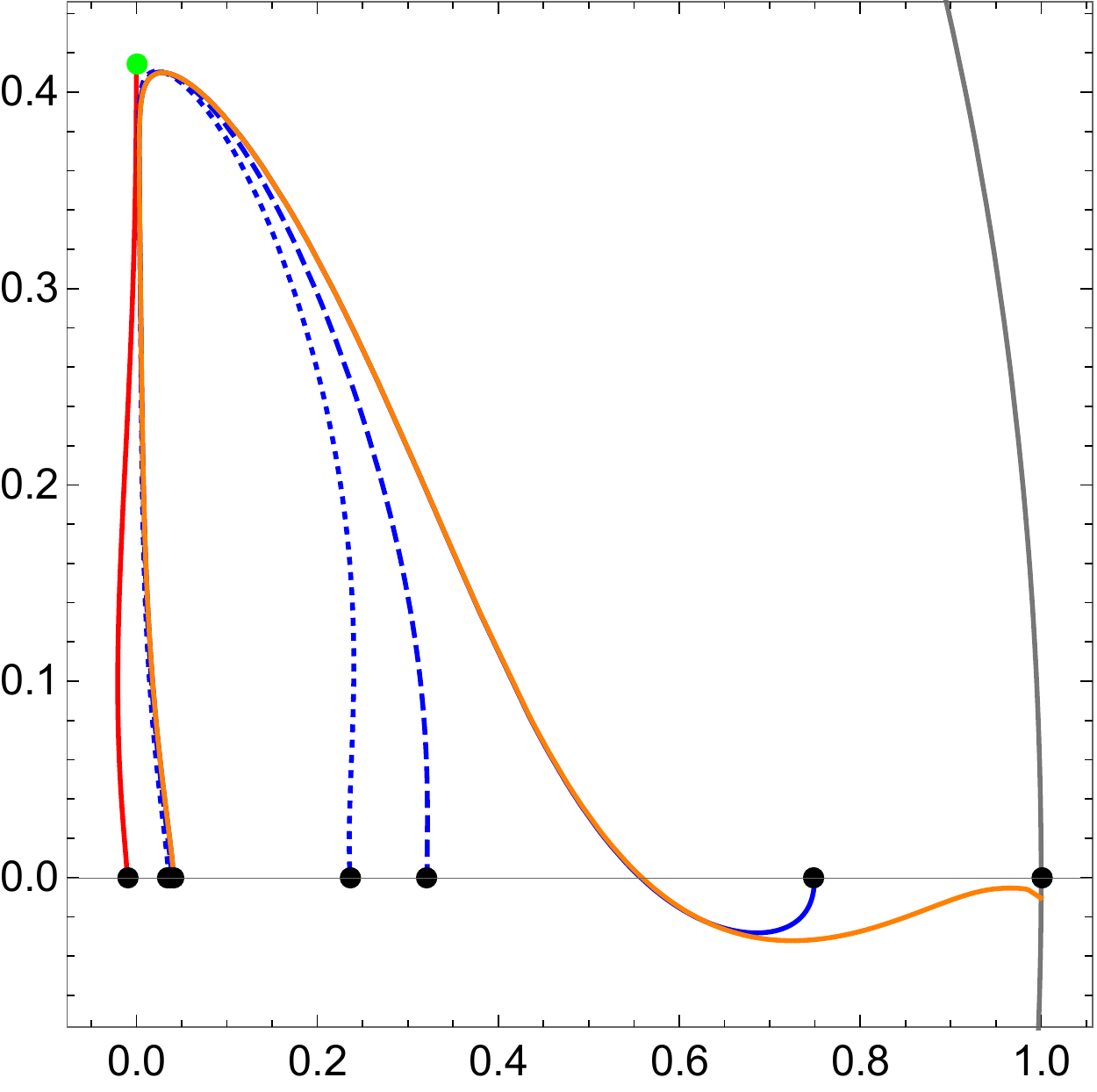}
\put(-400,-15){$\tilde{z}_{1}$}\put(-237,-15){$\tilde{z}_{4}$}\put(-75,-15){$\tilde{z}_{7}$}
\caption{{\it Samples of regular flows within the $SO(3)$-invariant sector using the boundary conditions (\ref{boundary_SO3_z4567}) for the complex scalar fields $\,\tilde{z}_{1}\,$ (left), $\,\tilde{z}_{4}\,$ (middle) and $\,\tilde{z}_{7}\,$ (right). The $\mathcal{N}= 4\,\&\,SO(4)$ $AdS_{4}$ vacuum and the D3-brane asymptotics are denoted by green and black dots, respectively. The grey line in the left plot corresponds to the boundary values $\,|\tilde{z}_{1}|=1\,$. The red flow corresponds to the choice $\,\epsilon \approx 0\,$ whereas the orange one corresponds to $\,\epsilon \approx \epsilon_{\textrm{crit}}\,$. Black dots in the middle and right plots are compatible with SO(6) invariance, \textit{i.e.} $\,\tilde{z}_{4}=\tilde{z}_{7}\,$ are real-valued, and correspond to the values $\,\pm\Phi_{0}\,$ of the type IIB dilaton in the D3-brane asymptotics at $\,r \rightarrow \pm \infty\,$.}}
\label{Fig:N=4_plots_3}
\end{center}
\end{figure}

The $(SO(3) \cap \mathbb{Z}_{2}^{3})$-invariant sector of the theory is a three field model that is obtained upon the identifications
\begin{equation}
\label{SO(3)_sector_scalars}
z_1\,=\,z_2\,=\,z_3 \quad , \quad z_4\,=\,z_5\,=\,z_6 \quad , \quad z_7 \,.
\end{equation}
The holomorphic superpotential (\ref{Superpotential}) reduces in this case to
\begin{equation}
\label{Superpotential_SO(3)}
\mathcal{V}\,=\,6 \, g \, z_{1} \, z_{4} \, \left(z_{4}+z_7\right) + 2 \, g \, c \left(1-z_{4}   z_7^3\right) \ .
\end{equation}
The identification in (\ref{SO(3)_sector_scalars}) is compatible with the $\mathcal{N}\,=\,4 \, \&\, SO(4)$ invariant vacuum in (\ref{solution_family_4}) which has $\,\textrm{Im}z_{1}=0\,$. As in the previous cases, this condition will no longer hold along the numerical flows when approaching the D3-brane behaviour describing the endpoints of the generic flows.

\subsubsection*{Numerical study}

First we will consider the simplest case in which the turning points of the numerical flows (pink points in Figure~\ref{Fig:N=4_plots_1}) are distributed along the $\,\textrm{Re}\tilde{z}_{1}\,$ real axis while $\,\tilde{z}_{4}\,$ and $\,\tilde{z}_{7}\,$ are kept fixed at their values in the $AdS_{4}$ vacuum. This amounts to set
\begin{equation}
\label{boundary_SO3_z123}
\tilde{z}_{1}(0) = \tilde{z}_{1}^{(*)}  + \epsilon
\qquad , \qquad 
\tilde{z}_{4}(0) = \tilde{z}_{4}^{(*)}
\qquad , \qquad 
\tilde{z}_{7}(0) = \tilde{z}_{7}^{(*)}  \ ,
\end{equation}
where $\,\tilde{z}_{1,4,7}^{(*)}\,$ denote the expectation values of the scalars at the $\,\mathcal{N}= 4\,\&\,SO(4)\,$ $AdS_{4}$ vacuum. We find a structure of flows similar to the cases previously analysed. Namely, regular flows only exist within the numerical window $\,0 < \epsilon <  \epsilon_{\textrm{crit}}\,$ with $\,{\epsilon_{\textrm{crit}} \approx 0.22592}\,$. These flows lie in between two limiting/bounding flows at $\,\epsilon \approx 0\,$ and $\,\epsilon \approx \epsilon_{\textrm{crit}}\,$ which respectively correspond to the solid red and orange lines in Figures \ref{Fig:N=4_plots_1} and \ref{Fig:N=4_plots_2}. As in the other cases, the flow with $\,\epsilon \approx 0\,$ (red solid lines) features an $AdS_{4}$ intermediate behaviour with constant scalars around the turning point at $\,r=0$. This $AdS_{4}$ regime uplifts this time to the $\mathcal{N}= 4\,\&\,SO(4)$ type IIB S-fold of \cite{Inverso:2016eet}.

More general boundary conditions $\,\lbrace  \tilde{z}_{1}(0)  \,,\, \tilde{z}_{4}(0)  \,,\, \tilde{z}_{7}(0) \rbrace\,$ can be specified which also give rise to regular flows. A choice that is motivated by the D3-brane solution (\ref{D3-brane_4D_scalars}) is given by
\begin{equation}
\label{boundary_SO3_z4567}
z_{1}(0) = z_{1}^{(*)}  
\qquad , \qquad 
z_{4}(0) = -\chi_{4}^{(*)} + i \,  e^{- \left( \varphi_{4}^{(*)}  - \,  \epsilon \right)}
\qquad , \qquad 
z_{7}(0) = -\chi_{7}^{(*)} + i \,  e^{- \left( \varphi_{7}^{(*)}  - \,  \epsilon \right)}\ .
\end{equation}
The above choice of boundary conditions produces regular flows only within the range $\,{0 < \epsilon <  \epsilon_{\textrm{crit}}}\,$ with $\,\epsilon_{\textrm{crit}} \approx 0.03472 \,$. Some examples are depicted in Figure~\ref{Fig:N=4_plots_3}. Performing an exhaustive study of the entire six-dimensional parameter space goes beyond the scope of this paper.

\section{Final remarks}

In this work we have investigated a class of Janus-type solutions in type IIB supergravity, some of them accommodating an S-fold intermediate regime at their core. Instead of addressing the problem directly in ten dimensions, we have made use of the effective four-dimensional ${[SO(1,1)\times{S}O(6)]\ltimes\mathbb{R}^{12}}$ gauged supergravity that arises upon a \textit{twisted} compactification of type~IIB supergravity on $S^1 \times S^5$ \cite{Inverso:2016eet}. The implementation of an S-duality twist in the reduction along the $S^1$ is a necessary ingredient for the intermediate S-fold region to exist. This twist totally codifies the simple and analytic dependence of the Janus-type solution on the coordinate $\,\eta\,$ along the $\,S^1\,$ (see Appendix~\ref{app:uplift}). In order to obtain the flow solutions along the radial coordinate $r$, we have resorted to numerical methods and solved the BPS equations in the four-dimensional ${[SO(1,1)\times{S}O(6)]\ltimes\mathbb{R}^{12}}$ gauged supergravity imposing an $AdS_{3}$-sliced domain-wall ansatz.

When uplifted back to ten dimensions, the flow solutions are interpreted as type IIB Janus-type solutions that interpolate between a supersymmetric deformation of $AdS_{5} \times S^5$ on each side of the Janus ($r \rightarrow \pm \infty$) and generically display an $\mathbb{R} \times AdS_{3} \times S^{1} \times S^5$ geometry along the flow. However, upon tuning of the boundary conditions for the running scalars when solving the four-dimensional BPS equations, it is possible to obtain a special class of Janus-type solutions for which an S-fold geometry of the form $AdS_{4}  \times S^{1} \times S^{5}$ emerges at the core of the solution ($r=0$). We have constructed such special classes of Janus-type solutions for the three largest symmetric S-folds preserving $\,{\mathcal{N}=1 \,\& \,SU(3)}$ \cite{Guarino:2019oct}, $\mathcal{N}=2 \,\& \, U(2)\,$ \cite{Guarino:2020gfe} and $\,{\mathcal{N}=4 \,\& \, SO(4)}\,$ \cite{Inverso:2016eet} symmetry. A natural extension is to consider Janus-type solutions involving less symmetric S-folds obtained, for example, by turning on the so-called flat deformations dual to marginal deformations in the dual S-fold CFT's \cite{Guarino:2021hrc,Guarino:2022tlw}.

One may wonder whether the solutions presented in this work could have been found in the much more studied Einstein-scalar models describing consistent subsectors of the $SO(6)$ gauged supergravity in five dimensions \cite{Gunaydin:1984qu, Gunaydin:1985cu,Pernici:1985ju}. However, this is not possible because the axions (pseudo-scalars) $\textrm{Im}\tilde{z}_{1,2,3}$ in four dimensions originate from Wilson lines of five-dimensional vector fields along the $\,S^1\,$ \cite{Guarino:2021kyp}. And even if these axions are set to zero at the intermediate S-fold regime of the Janus-type solutions, they are generically activated away from the core of the solutions. Nonetheless, there are direct flows in which  $\textrm{Im}\tilde{z}_{1,2,3}=0$ all along the Janus-type solutions. These particular flows therefore stand a chance of being found analytically in four, five and ten dimensions. It would be very interesting to investigate these direct flows in more detail.

On the other hand, sticking to the common characteristic of usual Janus solutions -- $AdS_{d-1}$ solutions with $AdS_{d}$ asymptotics on each side of the Janus --, it would be interesting to construct flows that depart from an $AdS_4$ S-fold and arrive at either the same or a different $AdS_4$ S-fold while preserving the $SO(2,2)$ isometries of $AdS_3$ along the flow. The latter would be the curved counterparts of the holographic RG-flows (with a three-dimensional flat slicing) connecting S-fold CFT's in \cite{Guarino:2021kyp}, and would holographically describe conformal interfaces in three-dimensional S-fold CFT's. In the M-theory context, examples of supersymmetric Janus solutions dual to conformal interfaces in ABJM theory \cite{Aharony:2008ug} (and mass deformed versions thereof) were put forward in \cite{Bobev:2013yra}.

Let us conclude with some remarks on the Janus-type solutions with an S-fold at the core for which the space-time geometry conforms to $\,AdS_{5}-AdS_{3}-AdS_{4}-AdS_{3}-AdS_{5}\,$. Each side of these Janus-type solutions can be viewed as a D3-brane to S-fold flow with a geometry of the form $\,AdS_{5}-AdS_{3}-AdS_{4}\,$, and holography anticipates an interplay between field theories of different dimensionality. Despite the presence of nested $AdS$ factors, the geometry in our solutions is different from the Janus within Janus solutions considered in \cite{Hirano:2006as} describing interfaces of higher codimensions in $\,\mathcal{N}=4\,$ SYM$_{4}$. In our case there is no dependence of the ten-dimensional geometry on the fifth coordinate $\,\eta\,$ on $\,S^1\,$ since the metric is a singlet under S-duality and therefore is not affected by the twisted reduction (see eq.(\ref{10D_metric_SU(3)_lambda})). The setup here also differs from the Janus on the brane construction of \cite{Gutperle:2020gez} realising interfaces on surface defects in $\,\mathcal{N}=4\,$ SYM$_{4}$. This construction considers an $\,AdS_{2} \times \mathbb{R}\,$ slicing on the $\,AdS_{3} \subset AdS_{3} \times S^1\,$ parameterising the worldvolume of the D3-brane. It would therefore be interesting to make progress in understanding the holography of the Janus-type solutions presented here in light of the AdS/CFT correspondence. We leave this and other related questions for the future.

\section*{Acknowledgements}

\noindent AG would like to thank Colin Sterckx for collaboration on related works. The work of AG is partially supported by the AEI through the Spanish grant PGC2018-096894-B-100 and by the FICYT through the Asturian grant SV-PA-21-AYUD/2021/52177. MS is supported by the National Research Foundation of Korea under the grant NRF-2020R1A2C1008497.

\appendix

\section{Type IIB uplift of the $SU(3)$-invariant sector}
\label{app:uplift}

The $SU(3)$-invariant sector of the theory was uplifted to type IIB supergravity in Section~$4.3.1$ of \cite{Guarino:2021kyp}. We employ the results there to uplift some of the solutions presented in the main text. For example, the non-$AdS_{4}$ exact flow at $c=0$ given in (\ref{D3-brane_4D_scalars})-(\ref{D3-brane_4D_metric}) that uplifts to the D3-brane solution (\ref{D3-brane_10D_1})-(\ref{D3-brane_10D_2}).

\subsection*{Ten-dimensional metric}

A Sasaki-Einstein metric can be locally written as 
\begin{equation}
ds^2(SE_5)\,=\,ds^2(KE_4)+\eta_1\otimes\eta_1 \ ,
\end{equation}
where $ds^2(KE_4)$ is a local K\"ahler-Einstein metric with positive curvature and $\eta_1$ is a globally well-defined one-form dual to the Reeb Killing vector. There are also a globally well-defined K\"ahler two-form, $J_2$, and a $(2,0)$-form complex structure, $\Omega_2$, and they satisfy
\begin{align}
d\eta_1\,=&\,2J_2\ , \notag \\
d\Omega_2\,=&\,3i\eta_1\wedge\Omega_2\ .
\end{align}

The uplift of the $SU(3)$-invariant sector of the $[SO(1,1)\times{S}O(6)]\ltimes\mathbb{R}^{12}$ dyonically-gauged maximal supergravity is the following. The ten-dimensional metric is given by
\begin{equation}
\label{10D_metric_SU(3)}
ds_{10}^2\,=\,\frac{1}{2\Delta}ds_{4}^2+\frac{1}{g^2c^2}Hd\eta^2+\frac{1}{g^2}F\left(ds_{\mathbb{CP}_2}^2+\frac{1}{F^2} \, \eta_1\otimes\eta_1\right) \ ,
\end{equation}
where we have defined the metric functions
\begin{equation}
\label{def:F_H_Delta}
F\,=\,\frac{\text{Im}z_{4,5,6,7}}{|z_{4,5,6,7}|}\,, \qquad H\,=\,\frac{1}{F}\text{Im}[z_{1,2,3}]^2\,, \qquad \Delta\,=\,F\,\text{Im}z_{1,2,3} \ ,
\end{equation}
and $ds_{4}^2$ is the four-dimensional metric in \eqref{4met}. The metric on $\mathbb{CP}_2$ is given by
\begin{equation}
ds_{\mathbb{CP}_2}\,=\,d\alpha^2+\frac{1}{4}\sin^2\alpha\left(\sigma_1^2+\sigma_2^2+\cos^2\alpha \, \sigma_3^2\right) \ ,
\end{equation}
where the left-invariant one-forms are
\begin{align}
\sigma_1\,=&\,-\sin\psi{d}\theta+\cos\psi\sin\theta{d}\theta\,, \notag \\
\sigma_2\,=&\,\cos\psi{d}\theta+\sin\psi\sin\theta{d}\theta\,, \notag \\
\sigma_3\,=&\,d\psi+\cos\theta{d}\phi \ .
\end{align}
As a non-trivial monodromy on the internal manifold is induced by the set of non-zero constant axion $\text{Re}z_{1,2,3}$ the real one-form is given by
\begin{equation}
\label{U(1)-fibration}
\eta_1\,=\,d\beta+A_1+\text{Re}z_{1,2,3} \, d\eta \ ,
\end{equation}
where the one-form potential on $\mathbb{CP}_2$ is 
\begin{equation}
A_1\,=\,\frac{1}{2}\sin^2\alpha \, \sigma_3 \ .
\end{equation}
They satisfy the torsion conditions for the $SU(2)$-structure
\begin{equation}
d\left(\eta_1-\text{Re}z_{1,2,3} \, d\eta\right)\,=\,2J_2\,, \qquad dJ_2\,=\,0\,, \qquad d\Omega_2\,=\,3i\left(\eta_1-\text{Re}z_{1,2,3} \, d\eta\right)\wedge\Omega_2 \ .
\end{equation}

To understand the geometry along the flows, it proves convenient to perform a change of radial coordinate of the form
\begin{equation}
\label{dlambda_appendix}
d\lambda = \dfrac{g}{\sqrt{2}} \,  \textrm{Im}[z_{1,2,3}]^{-\frac{1}{2}} \, F^{-\frac{1}{2}} \, dr \ ,
\end{equation}
so that the metric in (\ref{10D_metric_SU(3)}) becomes
\begin{equation}
\label{10D_metric_SU(3)_lambda}
\begin{array}{rll}
ds_{10}^2 &=& g^{-2}  \Big[d\lambda^2+f_{1}(\lambda)\,ds_{AdS_3}^2 + f_{2}(\lambda)\,d\eta^2+ F(\lambda) \, ds_{\mathbb{CP}_2}^2 + F(\lambda)^{-1} \, \eta_1\otimes\eta_1 \Big] \ , 
\end{array}
\end{equation}
with
\begin{equation}
\label{f1f2_funcs_appendix}
f_{1}(\lambda) = \tfrac{1}{2} \, g^2 \, F^{-1}  \, e^{2 A} \, \textrm{Im}[z_{1,2,3}]^{-1} 
\hspace{8mm} \textrm{ and } \hspace{8mm}
f_{2}(\lambda) = c^{-2} \, F^{-1}   \, \textrm{Im}[z_{1,2,3}]^2\ .
\end{equation}
Note that, at the core of the $SU(3)$-invariant flow  with $\,\epsilon \approx 0\,$ (red solid line) in Figure~\ref{Fig:N=1_plots_2}, the scalars undergo a constant behaviour so that $\,f_{1}(\lambda) \propto e^{2A} \approx \left(\frac{L}{l}\right)^2 \cosh^2\left(\frac{\lambda}{L} \right)\,$ and $\,f_{2}(\lambda) \approx \textrm{cst}\,$. As a result, the five-dimensional space-time metric conforms to $\,AdS_{4} \times S^1\,$ as required by the S-fold regime. Note also that, at finite $\,c\,$, the function $\,f_{2}\,$ in (\ref{f1f2_funcs_appendix}) vanishes whenever $\,\textrm{Im}[z_{1,2,3}]=0\,$ ($F\,$ in (\ref{def:F_H_Delta}) is bounded both above and bellow). Having $\,\textrm{Im}[z_{1,2,3}]=0\,$ implies $\,|\tilde{z}_{1,2,3}|=1$. Therefore, when this occurs, $\,\tilde{z}_{1,2,3}\,$ hits the boundary of the unit-disk and the solution becomes singular: the $\,S^{1}\,$ in (\ref{10D_metric_SU(3)_lambda}) parameterised by $\,\eta\,$ collapses. Flows of this type can be constructed for which the singularity occurs at a finite radial distance, although we are not investigating them in this work. As mentioned in the introduction, these are the counterparts of the flows to Hades investigated in \cite{Bobev:2013yra,Anabalon:2022fti}.

\subsection*{Axion-dilaton}

The axion-dilaton matrix including the ten-dimensional dilaton $\Phi$ and the Ramond--Ramond (RR) axion $C_{0}$ is given by
\begin{equation}
m_{\alpha\beta}\,=\,\left(A^{-t}\right)_\alpha{}^\gamma \, \mathfrak{m}_{\gamma\delta}\, \left(A^{-1}\right)^\delta{}_\beta 
= e^{\Phi} \left(   
\begin{matrix}
e^{-2 \Phi} + C_{0}^2 & - C_{0} \\
- C_{0} & 1
\end{matrix}
\right) \ ,
\end{equation}
where
\begin{equation}
\mathfrak{m}_{\gamma\delta}\,=\,\left(
\begin{array}{ll}
 |z_{4,5,6,7}|^2 & 0 \\
 0 & |z_{4,5,6,7}|^{-2}
\end{array}
\right)\ .
\end{equation}
The $SL(2)$ hyperbolic twist matrix induced by the electromagnetic parameter (we are setting $c=1$) reads
\begin{equation}
A^\alpha{}_\beta\,=\,\left(
\begin{array}{ll}
 \cosh\eta & \sinh\eta \\
 \sinh\eta & \cosh\eta
\end{array}
\right) \quad , \quad
\left(A^{-1}\right)^\alpha{}_\beta\,=\,\left(
\begin{array}{ll}
 \,\,\,\,\, \cosh\eta & -\sinh\eta \\
 -\sinh\eta & \,\,\,\,\, \cosh\eta
\end{array}
\right)\ .
\end{equation}
Importantly, the twist matrix trivialises to $A^\alpha{}_\beta=\delta^\alpha{}_\beta$ when having a purely electric gauging with $c=0$. This is the case studied in Section~\ref{sec:D3-brane_c=0}.

\subsection*{Three-form fluxes}

The type IIB two-form potentials $\mathbb{B}^\alpha=(B_{2},C_{2})$ are given by
\begin{equation}
\label{mathbbB}
\mathbb{B}^\alpha\,=\,A^\alpha{}_\beta \, \mathfrak{b}^\beta\,.
\end{equation}
In terms of the complex combination of potentials one finds
\begin{equation}
\begin{array}{lll}
\mathfrak{b}^2+i \, |z_{4,5,6,7}|^2 \, \mathfrak{b}^1 &=&\dfrac{i}{g^2}  \, \text{Re}z_{4,5,6,7}\, \Omega_2 \\[4mm]
&=& \dfrac{i}{g^2} \, e^{3i\beta} \, \text{Re}z_{4,5,6,7} \left(d\alpha+\frac{i}{4}\sin(2\alpha)\sigma_3\right)\wedge\left(\frac{1}{2}\sin\alpha\left(\sigma_1-\sigma_2\right)\right)\,,
\end{array}
\end{equation}
or, equivalently,
\begin{equation}
\label{mathfrakb1b2}
\mathfrak{b}^1\,=\,-\frac{1}{g^2}\frac{\text{Re}z_{4,5,6,7}}{|z_{4,5,6,7}|^2} \, \text{Re}[\Omega_2]\,, \qquad \mathfrak{b}^2\,=\,\frac{1}{g^2}\text{Re} z_{4,5,6,7}  \, \text{Im}[\Omega_2]\,.
\end{equation}
Their three-form field strengths are readily computed as
\begin{equation}
\mathbb{H}^\alpha\,=\,d\mathbb{B}^\alpha\,=\,\left(H_3 \, , F_3\right) \ ,
\end{equation}
and we obtain
\begin{equation}
\label{H_tensor_SU(3)}
\mathbb{H}^\alpha\,=\,A^\alpha\,_\beta\left(d\eta_1\wedge\mathfrak{b}^\gamma\theta_\gamma\,^\beta+d\mathfrak{b}^\beta\right) \ ,
\end{equation}
with
\begin{align}
d\mathfrak{b}^1\,=&\,-\frac{1}{g^2}d\left(\frac{\text{Re}z_{4,5,6,7}}{|z_{4,5,6,7}|^2}\right)\wedge\text{Re}[\Omega_2]+\frac{3}{g^2}\frac{\text{Re}z_{4,5,6,7}}{|z_{4,5,6,7}|^2}\left(\eta_1-\text{Re}z_{1,2,3} \, d\eta\right)\wedge \text{Im}[\Omega_2] \ , \notag \\
d\mathfrak{b}^2\,=&\,\frac{1}{g^2}d\text{Re}z_{4,5,6,7} \wedge\text{Im}[\Omega_2]+\frac{3}{g^2}\text{Re}z_{4,5,6,7}\left(\eta_1-\text{Re}z_{1,2,3} \, d\eta\right)\wedge\text{Re}[\Omega_2] \ .
\end{align}
In (\ref{H_tensor_SU(3)}) we introduced the constant matrix
\begin{equation}
\theta_\gamma\,^\beta\,=\,\left(
\begin{array}{ll}
 0 & 1 \\
 1 & 0
\end{array}
\right) \ .
\end{equation}

\subsection*{Five-form flux}

The self-dual five-form field strength is given by
\begin{align}
\widetilde{F}_5\,= \,&\,g\left(1+*\right)\Big[\Big(4-6\left(1-F^2\right)\Big)vol_{\mathbb{CP}_2}\wedge{d}\beta \notag \\
+&\left(4 \, \text{Re}z_{1,2,3} +\text{Re}[z_{4,5,6,7}]^2\left(1-\frac{1}{|z_{4,5,6,7}|^4}\right)\right) vol_{\mathbb{CP}_2}\wedge{d}\eta \notag \\
-& d\text{Re}z_{1,2,3} \wedge d\eta\wedge\Big(d\alpha \, \wedge\frac{1}{2}\sin\alpha \, \sigma_1 \wedge{A_1}+2 \, J_2 \wedge d\beta \Big)\Big] \ .
\end{align}

\bibliography{references}


\end{document}